\documentclass[twocolumn]{aastex631}

\usepackage{textcomp, gensymb}
\usepackage{graphicx}
\usepackage{lipsum}       
\usepackage{float}
\usepackage{xspace}
\usepackage{soul}
\usepackage{xurl}
\usepackage{silence}
\usepackage{morefloats}

\newcommand\um{\ensuremath{\mu\mathrm{m}}\xspace}

\shorttitle{NIR Survey of W51A}
\shortauthors{Dawson et al.}

\begin{document}

\title{How Massive Star Clusters Form and Evolve: A Near-IR Survey of the W51 Complex}

\author{Aden Dawson}
\affiliation{Department of Astronomy, University of Florida, Gainesville, FL 32601, USA}

\author{Adam Ginsburg}
\affiliation{Department of Astronomy, University of Florida, P.O. Box 112055, Gainesville, FL 32601, USA}

\author{Carlos G. Rom\'{a}n-Z\'{u}\~{n}iga}
\affiliation{Universidad Nacional Aut\'{o}noma de M\'{e}xico, Instituto de Astronom\'{i}a, AP 106, Ensenada 22800, BC, M\'{e}xico}

\begin{abstract}
We present near-infrared \textit{JHK\textsubscript{s}} and narrow-band \textit{H$_2$(1-0)} photometric observations of the W51A region, obtained with GTC EMIR, aiming to characterize its young stellar population, and provide mass estimates for individual cluster members and the proto-clusters. Our observations reveal over 3000 new sources, out of which 88 are located in the proto-clusters, W51 IRS2 and W51 Main. The average extinction ($A_{V}$), measured from the $J-H$ color, of sources is $19 \space A_{V}$ in W51 IRS2 and $14 \space A_{V}$ in W51 Main. We document 17 new instances of H\textsubscript{2} emission in the region by utilizing observations from the \textit{H$_2$(1-0)} narrow-band filter. Despite limited completeness, we estimated masses for each cluster member and estimated the total cluster mass to be in the range of $900-4700M_\odot$ for W51 IRS2 and $500-2700M_\odot$ for W51 Main, using an assumed age range of 1-3 Myr. We measured the initial mass function (IMF) in the proto-clusters assuming a range of ages from 1-3 Myr and found that the IMF slopes for both proto-clusters are consistent with the Salpeter IMF in the mass range $M\gtrsim8M_\odot$ within $1\sigma-2\sigma$.
 
\end{abstract}

\section{Introduction}

The W51 star-forming complex is among the most massive and luminous in the Galaxy, containing $\gtrsim10^6 M\textsubscript{\(\odot\)}$ of molecular gas \citep{Combes1991,Carpenter1998} and exhibiting $\sim10^7 L\textsubscript{\(\odot\)}$ in far-infrared luminosity \citep{Harvey1986}. W51A (centered on (\textit{l}, \textit{b}) $\sim$ 49.5, -0.4) is the most studied region in W51. W51A hosts two proto-clusters, W51 IRS2 and W51 Main \citep{Ginsburg2012} at its most active sites of star formation. There is also a distributed population of main-sequence OB stars dispersed in the surroundings of W51A \citep{Bik2019}. The stellar population of W51A has been studied with infrared observations that cover both large scales \citep{Okumura2000,Kumar2004} and small scales with adaptive optics \citep{Figueredo2008}. However, the past large-scale observations were unable to detect most stars in the crowded, embedded clusters in the region, and had to rely on making inferences about W51A based only on the brightest members of the stellar population. 

W51 has been observed in the near-infrared (NIR) several times.
\citet{Okumura2000} observed W51A (referred to as G49.5-0.4 in their work) using the OASIS camera on the National Astronomical Observatory of Japan 1.88m Telescope at Okayama Astrophysical Observatory with the intent to measure the stellar masses, ages, and spatial distributions of stars in order to derive the massive star initial mass function (IMF). They took images in the NIR \textit{JHK'} and the \textit{Br$\gamma$} filters. They measured an IMF slope of $\Gamma$ = 1.8, a total stellar mass of 8200 $M\textsubscript{\(\odot\)}$, and age $\lesssim$ 1 Myr for Region 3 (the area of their observations containing the W51A proto-clusters). \citet{Kumar2004} also took NIR \textit{JHK} images of W51 using the UFTI imager on the 3.8m United Kingdom Infrared Telescope at Maunakea Observatory, but expanded their observations to cover six fields in the W51 giant molecular cloud (GMC). The authors estimated a total mass of 9200 $M\textsubscript{\(\odot\)}$ for G49.5-0.4 but did not attempt to measure the stellar IMF. \citet{Figueredo2008} took \textit{JHK\textsubscript{s}} images (using the infrared camera ISPI at the CTIO Blanco 4m telescope) and \textit{K}-band spectroscopy (using NIRI on Gemini-North) of W51A with the primary intent to better constrain the distance to W51. More relevant to the work presented here, the authors also provide analysis of high-resolution images taken of W51 IRS2 by the adaptive optics NIR camera NAOS CONICA (NACO) at the ESO VLT UT4. These images allow us to see many stars in the central region of W51 IRS2 that seeing-limited NIR observations are unable to detect.


A crucial piece of information in the analysis of a star cluster is the cluster's age, and W51A has a complicated star formation history.  \textit{Karl G. Jansky} Very Large Array observations from \citet{Ginsburg2016} reveal a cluster of ultra and hypercompact HII regions (U/HCHII), indicating the presence of proto-OB stars, deeply embedded in W51A that are not observed in the NIR. Spectroscopic studies of the distributed population of OB stars \citep{Bik2019,Barbosa2022}, and ultracompact HII regions (originally observed by \citet{Mehringer1994}) conclude that the age of the observed stars/UCHII regions is $\lesssim4$ Myr.  The W51 IRS2 and W51 Main star clusters we discuss here are likely younger, indicating that these star clusters have formed in the environment of a broader OB association. Unlike bound star clusters, OB associations are observed to have age spreads of several million years \citep{Pecaut2012,Bik2012}, while young star clusters with high stellar densities, such as NGC3603 and Westerlund 1, tend to consist of a single age stellar population \citep{Kudryavtseva2012}. It is likely that the ages for the distributed stellar population do not reflect the age of the embedded proto-clusters. 



The age and mass function of a population can be determined through photometric studies by first determining the distribution of \textit{K\textsubscript{s}}-band magnitudes of the cluster members, commonly referred to as the \textit{K\textsubscript{s}}-band luminosity function (KLF). The KLF is a powerful tool in constraining the age of the cluster. An approximate age range can be determined by comparing the observed KLF to either model KLFs or synthetic stellar populations \citep[e.g.]{RomanZ+2015,Yasui2016B,Yasui2016A,Rawat2024}. Having a tight constraint on the age gives insight on the overall evolutionary state of the cluster and its stellar population. Knowing the age of the cluster allows for determination of stellar mass. Then, the IMF can be constructed with the mass estimates for individual cluster members. The existence of the proto-clusters offers a unique opportunity to measure the stellar IMF of a cluster which has not lost any members due to stellar evolution or mass segregation. With knowledge of the shape of the stellar IMF, we can estimate the total mass of the clusters. 


In this work, we present new GTC EMIR observations of the W51A region in the \textit{J, H, K\textsubscript{s}} and the narrow band 2.12 $\mu$m filters, with the goal of measuring the aforementioned properties for the W51A proto-clusters, as well as any other cluster in the field.
In Section \ref{sec:observations}, we describe the observations and data reduction.
In Section \ref{sec:photometry}, we describe the photometric measurements, how foreground contaminants were removed, and how individual source extinctions were estimated.
In Section \ref{results_and_discussion}, we discuss characteristics of the newly discovered stellar population observed in W51A. We also present the results of our clustering analysis and our age estimations for both proto-clusters, how we estimated the masses for each cluster member, and present the IMFs for the proto-clusters.
We summarize our conclusions in Section \ref{conclusion}.

\section{Observations and Data Reduction}
\label{sec:observations}
\subsection{Observations}
NIR \textit{J-, H-, K\textsubscript{s}-}band filters, and the narrow-band, H$_2$ filter observations of the W51 GMC were conducted on 2020 June 7th, and 2020 July 12th. 
The observations were carried out using the NIR imaging camera and spectrograph EMIR\footnote{\href{http://www.gtc.iac.es/instruments/emir/media/EMIR_USERMANUAL.pdf}{\texttt{EMIR User Manual}}}, installed on the Gran Telescopio Canarias at the Roque de los Muchachos Observatory in the Canaries, Spain.
The detector in EMIR's imaging mode is a $2048 \times 2048$ Teledyne HAWAII-2 HgCdTe NIR optimized chip, yielding a field of view of $6.67' \times 6.67'$ and a plate scale of 0.1942" pixel\textsuperscript{-1}. 
Individual frames in all filters had exposure times of 10 seconds. Table \ref{obs_table} summarizes important observational details of the W51A observing run. 

\begin{figure*}
    \centering
    \includegraphics[width=1.0\linewidth]{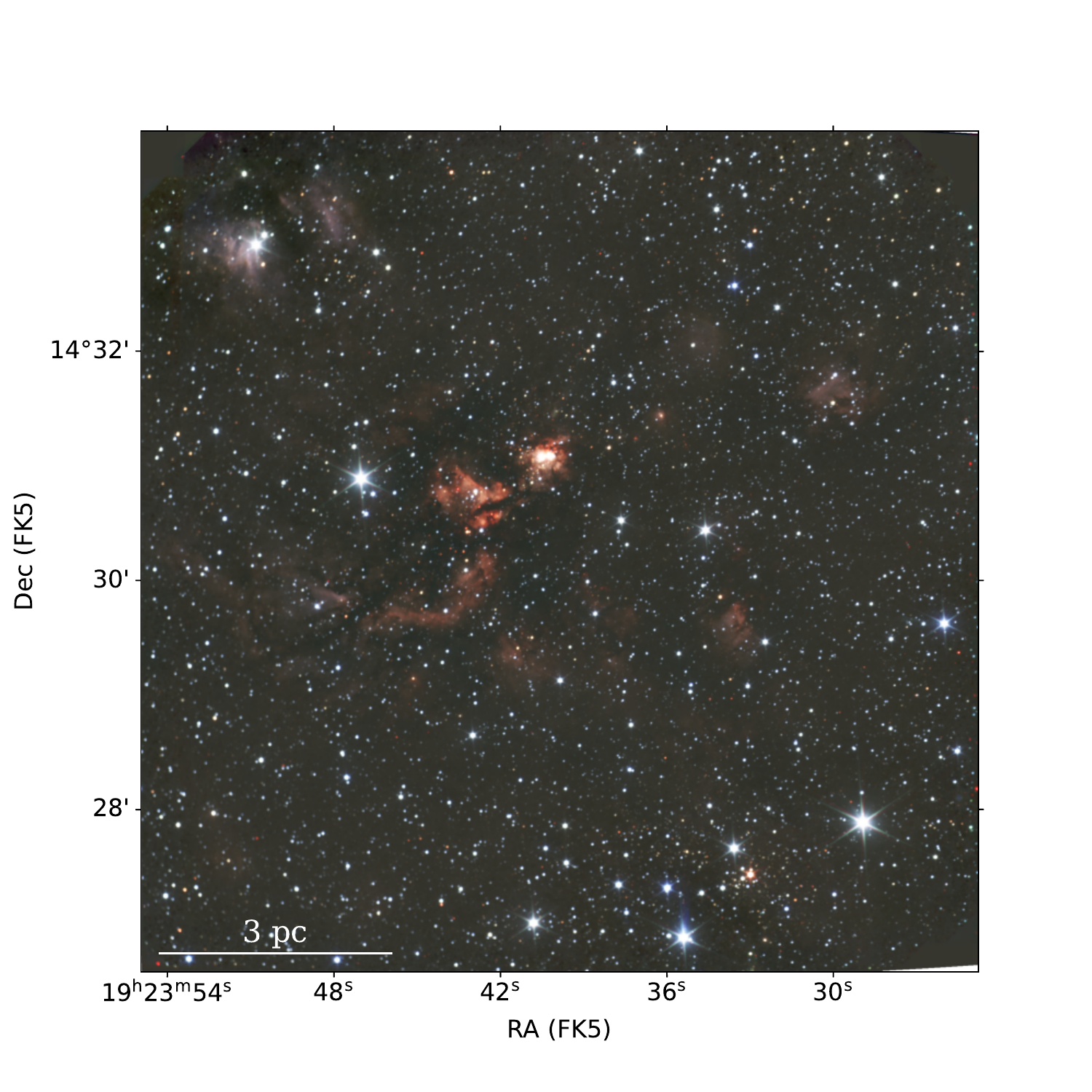}
    \caption{RGB (\textit{K\textsubscript{s}HJ}) image of W51A created from our observations.}
    \label{fig:3ci}
\end{figure*}

Three science fields and one control field were observed in total. However, we will only be discussing the results of one science field in this work. This field is centered on W51A, with right ascension $\alpha$ = 290.916\degree, and declination $\delta$ = 14.505\degree. Figure \ref{fig:3ci} shows a three-color image we created of the region from our data. We utilized the control field (centered on $\alpha$ = 290.81\degree, and declination $\delta$ = 14.63\degree) in our extinction estimation.

\begin{table}
\begin{center}
\caption{Observing details for the W51A observations.}
\label{obs_table}
\begin{tabular}{c c c c}
 \hline\hline
 Filter & Observing Date & Exposure Time (s) & Seeing (") \\ 
 \textit{J} & 2020-07-13 & 900 & 1.0 \\ 
 \textit{H} & 2020-07-13 & 900 & 1.2 \\
 \textit{K\textsubscript{s}} & 2020-08-07 & 900 & 1.1 \\
 \textit{$H_{2}$(1-0)} & 2020-08-07 & 900 & 1.2 \\
 \hline
\end{tabular}
\textbf{Note.} Reported seeing is the average seeing in each filter.
\end{center}
\end{table}

\subsection{Data Reduction}
Data reduction was done using the PyEmir\footnote{\url{https://pyemir.readthedocs.io/en/latest/}} Python package. A median-combined, normalized flat field image for each filter, a bias image, and a bad pixel mask were supplied to the data reduction pipeline. The individual data frames were reduced using these components.  

The reduced images had incorrect world coordinate system (WCS) information in the metadata contained in the FITS header, causing the centroid of stars to be slightly offset from their true positions. To fix this issue, the location of the bright stars from each frame were compared to their on-sky location measured by GAIA (GAIA eDR3, \citet{GaiaeDR32021}) and the WCS was shifted accordingly to provide a closer match. We discuss specifics about the WCS correction process in Appendix \ref{app_wcs}. 

Once the reduced images had this first-order WCS correction, we applied a re-projection to combine the frames. We ran the reduced images through the \texttt{find\_optimal\_celestial\_wcs} function to find the WCS frame that overlaps with all the images. We then re-projected the reduced images to the target WCS using the \texttt{reproject\_interp} function which re-projects the images onto a new array (having the same shape as the inputted shape) using the inputted WCS reference frame. The re-projected images were averaged to produce our final image.


The UKIDSS galactic plane survey (GPS) \citep{Lucas2008} catalog was used to measure the astrometric accuracy once the final mosaics were created. We found that Field 1 has an rms value of $<$ 0.15". Specifics about the astrometric accuracy are also discussed more in Appendix \ref{app_wcs}. 

\section{Photometry}
\label{sec:photometry}
\subsection{Point Spread Function Photometry}
\begin{figure*}
\gridline{\fig{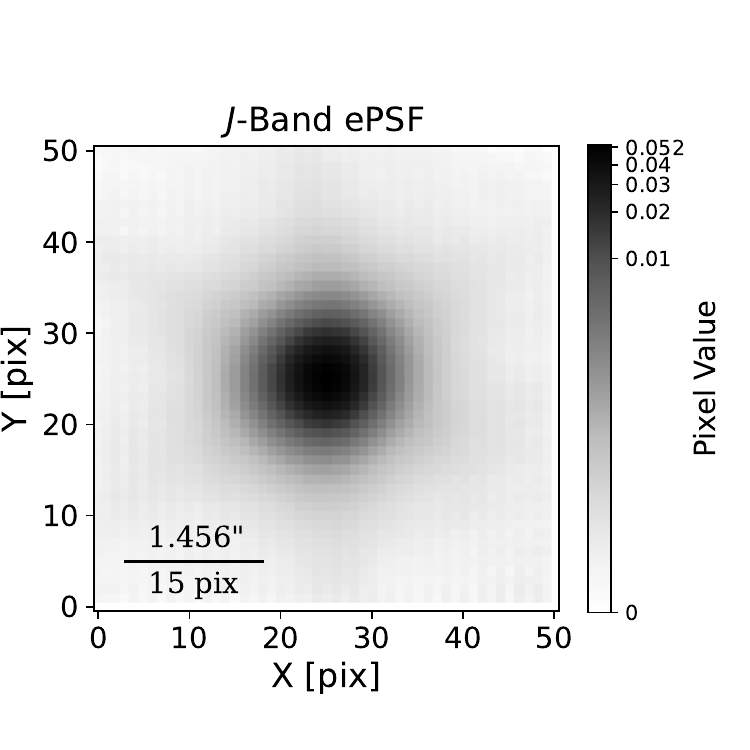}{0.32\textwidth}{}
          \fig{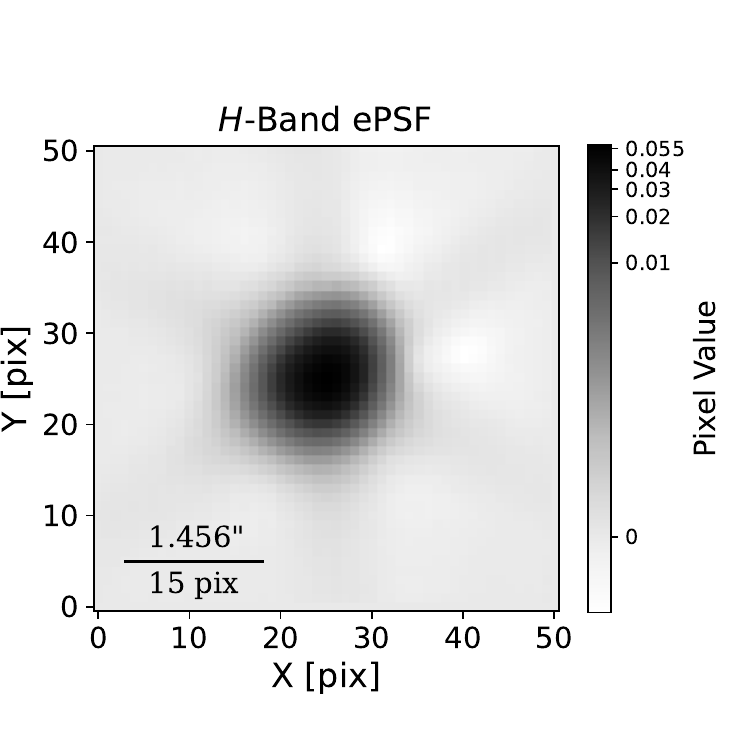}{0.32\textwidth}{}
          \fig{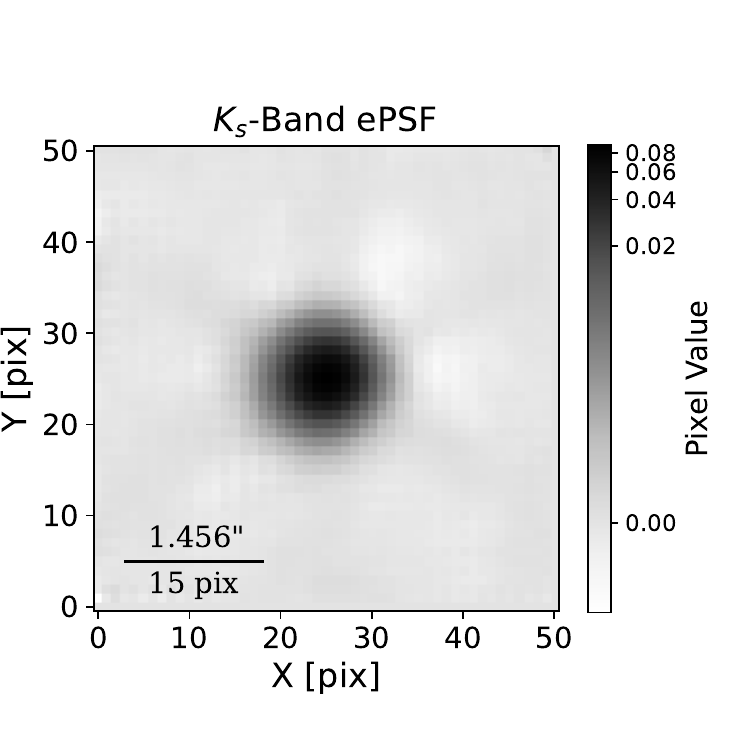}{0.32\textwidth}{}}
\caption{The \textit{J, H,} and \textit{K\textsubscript{s}} ePSFs used in the star-finding algorithm with a FWHM of 1.13", 0.74", and 0.55" respectively. The pixel size in each figure is 0.097".}
\label{fig:epsf_models}
\end{figure*}

Point spread function (PSF) photometry was carried out using the DAOPHOT \citep{Stetson1987} star finding algorithm. We used the EPSFBuilder class provided in the \texttt{photutils.psf} package\footnote{\url{https://photutils.readthedocs.io/en/stable/index.html}} to create an effective PSF for each band. Our ePSF arguments differed from the defaults as follows: \texttt{oversampling = 2}, \texttt{norm\_radius = 10.5}, \texttt{smoothing kernel = ‘quadratic’}. To improve ePSF quality in the $H$- and $K_{s}$-bands, we removed 400 pixels from each image edge to exclude regions of poor image quality and subtracted off the diffuse emission using a $10\times10$ 2D median filter (\texttt{scipy.ndimage}). We used stars from these altered images to build the ePSF for the $H$- and $K_{s}$-bands. We found that these corrections were not necessary to create a high quality ePSF for the \textit{J}-band due to better image quality throughout the FOV and the lack of diffuse emission. The effective PSFs for each filter are shown in Figure \ref{fig:epsf_models}.


We used the \texttt{DAOStarFinder} class from \texttt{photutils.detection} (which is an implementation of the DAOFIND algorithm \citep{Stetson1987}) with a threshold of 5.0 times the background RMS of each image for source detection. This value was calculated using \texttt{MADStdBackgroundRMS}, from \texttt{photutils.background}, over the entire image. The diffuse background had to be removed prior to running the photometry routine to reduce the background flux contamination in the stellar flux measurements. We implemented an iterative background subtraction into photometry method to accomplish this. 


To begin, we ran the PSF photometry routine on the background-subtracted image used to create the ePSFs. We visually inspected four-panel images showing each source in the $J$, $H$, $K_{s}$, and $\mathrm{H}_2$ bands. Circular masks (diameter of 5 pixels) were placed over sources we deemed real in our examination, then using a custom 2D median filter (size $15\times15$ box, \texttt{generic\_filter} from \texttt{scipy.ndimage} capable of processing nan pixel values) we created the filtered image estimating the diffuse emission in our images. These images were used to subtract off the diffuse background in the science images, and we re-ran the photometry routine on the background-subtracted science image. This process of masking, estimating background emission (with a $10\times10$ custom median filter instead), and then photometry was iterated until no new sources were recovered.


To validate the final catalog, we again visually inspected each detection in the proto-cluster region (this was the only region in the images that had a significant amount of false detections) in all four bands to ensure no false detections were included. We kept sources with a clear starlike appearance in at least one filter, and rejected those that were un-starlike. Through visual inspection, we removed 343 false detections on extended emission. This is $\sim3\%$ of the total sources detected, and we assume this percentage as an upper limit on the uncertainty associated with the visual inspection. We also omitted any detection that was to the left of the solid black line in Figure \ref{fig:sdm_inset} because we could not recover accurate photometry for sources in this region due to the poor image quality.

We calibrated the broad-band magnitudes obtained from PSF photometry by direct comparison to the measured magnitudes in the UKIDSS GPS \citep{Lucas2008} catalog, using the ``jAperMag3", ``hAperMag3", and ``k\_1AperMag3" for the $J$, $H$, and $K_{s}$-band respectively. Stars which were saturated in our images ($H < 13$, $K_{s} < 12$) had their measured magnitudes replaced by their respective UKIDSS magnitudes. 



A potential issue with adopting the UKIDSS magnitudes for saturated sources located in the HII regions is that background flux may be contaminating the star's measured magnitude. We utilized the $K_{s}$-band, adaptive optics images taken by the NAOS+CONICA (NACO) \citep{Lenzen2003,Rousset2003} instrument at the ESO VLT UT4, and measured the $K_{s}$-band magnitudes of stars in these images (see Appendix \ref{naco_photometry} for a detailed description of the photometry process). If a saturated star was within the NACO FOV, we opted to replace its magnitude with the magnitude measured from the NACO image instead to reduce potential flux contamination from diffuse emission. There are also 24 saturated sources which had inconsistent UKIDSS magnitudes compared to 2MASS or \citet{Bik2019}. We opted to use the 2MASS magnitudes to avoid incorrect extinction estimates for these foreground sources.

\begin{figure*}\centering
\gridline{\fig{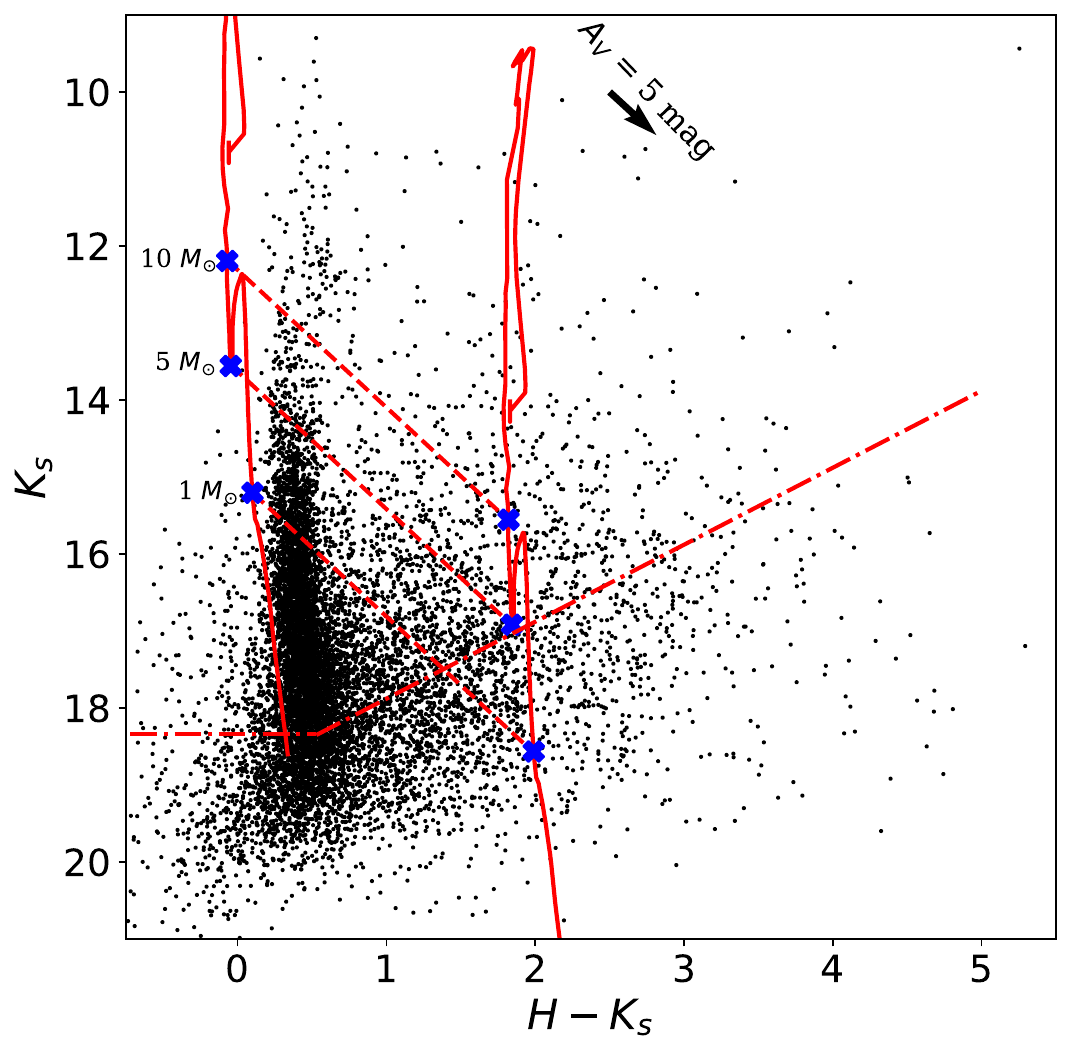}{0.5\textwidth}{(a)}
          \fig{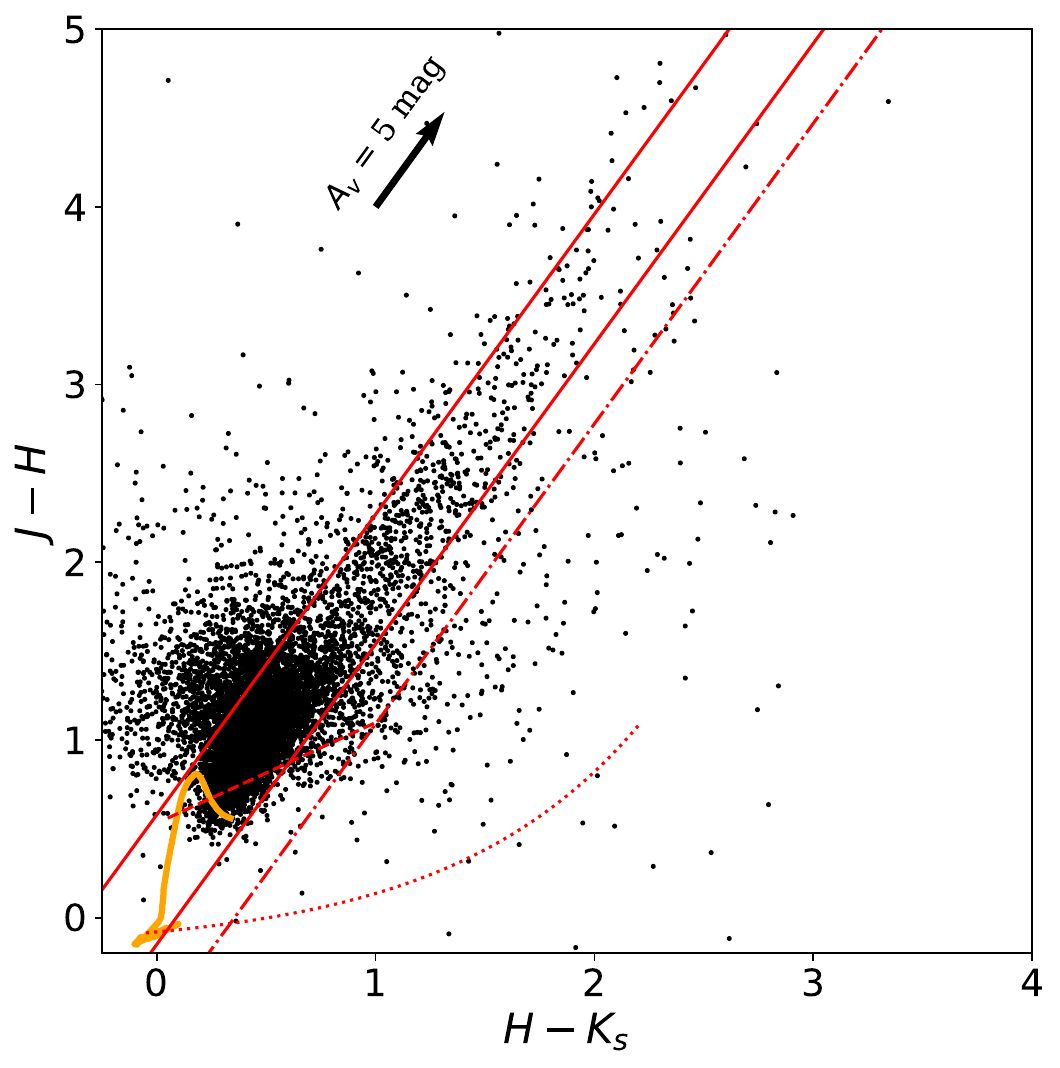}{0.5\textwidth}{(b)}}
\caption{(a) \textit{H-K\textsubscript{s}} vs \textit{K\textsubscript{s}} color-magnitude diagram of the region with 1 Myr isochrones reddened by $A_{V}=0$ and $A_{V}=30$ over-plotted. We specify the locations of 1, 5, and 10 $M_\odot$ stars on the $A_{V}=0$ isochrone and the reddening line connecting to their locations on the $A_{V}=30$ isochrone. The estimated 90\% completeness limit is show as the segmented dash-dotted line. (b) \textit{H-K\textsubscript{s}} vs \textit{J-H} color-color diagram of the region with an un-extincted, 1 Myr isochrone over-plotted in orange. The two solid, parallel lines encompassing the isochrone is the main-sequence reddening zone (reddening slope from \citet{RL85}; See Section \ref{ext}.). The dashed line indicates the locus of classical T-Tauri stars (CTTS) \citep{Meyer1997}, which is encompassed  by the left-most solid line and the dashed-dotted line, showing the boundaries of the CTTS locus reddening zone. The dotted line represents the Herbig AeBe star locus \citep{Lada1992}.}
\label{fig:cmd_and_ccd}
\end{figure*}

The $J-H$ vs. $H-K_{s}$ color-color diagram (CCD) and the $K_{s}$ vs. $H-K_{s}$ color-magnitude diagram (CMD) for the entire population of observed stars are shown in Figure \ref{fig:cmd_and_ccd}. The CMD and CCD are over-plotted with 1 Myr PARSEC-COLIBRI isochrones \citep{Bressan2012,Chen2014,Chen2015,Tang2014,Marigo2017,Pastorelli2019,Pastorelli2020}.

\subsection{Extinction}\label{ext}
We measure the extinction toward each source based on their observed colors and an extinction law.
We tested 4 galactic extinction laws from \citet{RL85}, \citet{Cardelli1989}, \citet{Indebetouw2005}, and \citet{Nishiyama2009}. We determined that the extinction law of \citet{RL85} provided the best fit to our data, because the reddening slope of this extinction law aligned well with the reddening slope observed in the CCD. We discuss this with more detail in Appendix \ref{ext_law_comp}. This extinction law was used for all extinction conversions between different bands. 
 
We chose to use the Near Infrared Color Excess (NICE) method \citep{Lada1994} for measuring extinction for individual sources, with a small change to the process. Instead of using the $H-K_{s}$ color to estimate extinction, we use the $J-H$ color to minimize the impact $K_{s}$-band excess on our estimates. This method uses the observed $J-H$ color of a source in the science field and the average, intrinsic $J-H$ color of sources in a nearby control field to find $E(J-H)$ for each source in the science field. $E(J-H)$ is then converted to $A_{V}$ using the determined extinction law.

Many sources in our observations lack a $J$-band detection, and some sources (particularly in the proto-clusters) are only observed in the $K_{s}$-band. We estimated the extinction for these sources using two different methods. The first method estimated their extinction by averaging the extinction of the five nearest neighbors to that source. Background contamination may impact extinction estimates for stars that are not in the proto-cluster regions, but this is less of a concern in the central HII region as the light from background stars would be unable to penetrate the cloud due to the high extinction. The second method assigned a “brightest possible” magnitude in undetected bands based on the 50\% completeness limits from Section \ref{measuring_completeness}, using values specific to the region (W51 IRS2: $J=22.15$, $H=20.23$; W51 Main: $J=22.19$, $H=20.33$; and a region with no diffuse emission: $J=22.5$, $H=20.43$). The extinction value for these sources was calculated using the NICE method described prior. We acknowledge that these approaches may lead to inaccurate extinction estimates for some sources and have marked them in the catalog.


These two methods provide a range of possible extinction values ($A_{V min}$ and $A_{V max}$) for sources lacking a $J$- or $H$-band magnitude. This propagates into other properties of these sources (like the absolute magnitude/mass). We then calculated the absolute $K_{s}$-band magnitude (denoted as $M_{K_S}$ hereafter) for each source by subtracting the estimated extinction value and the distance modulus to W51 IRS2 from their apparent \textit{K\textsubscript{s}}-band magnitude. Sources with a range of possible $A_{V}$ values are represented by a range of $M_{K_S}$ values. In Appendix \ref{sample_populations}, we discuss how we generate sample populaions from these ranges to use in our analysis.

\begin{figure}[H]
    \centering
    \includegraphics[width=1.0\linewidth]{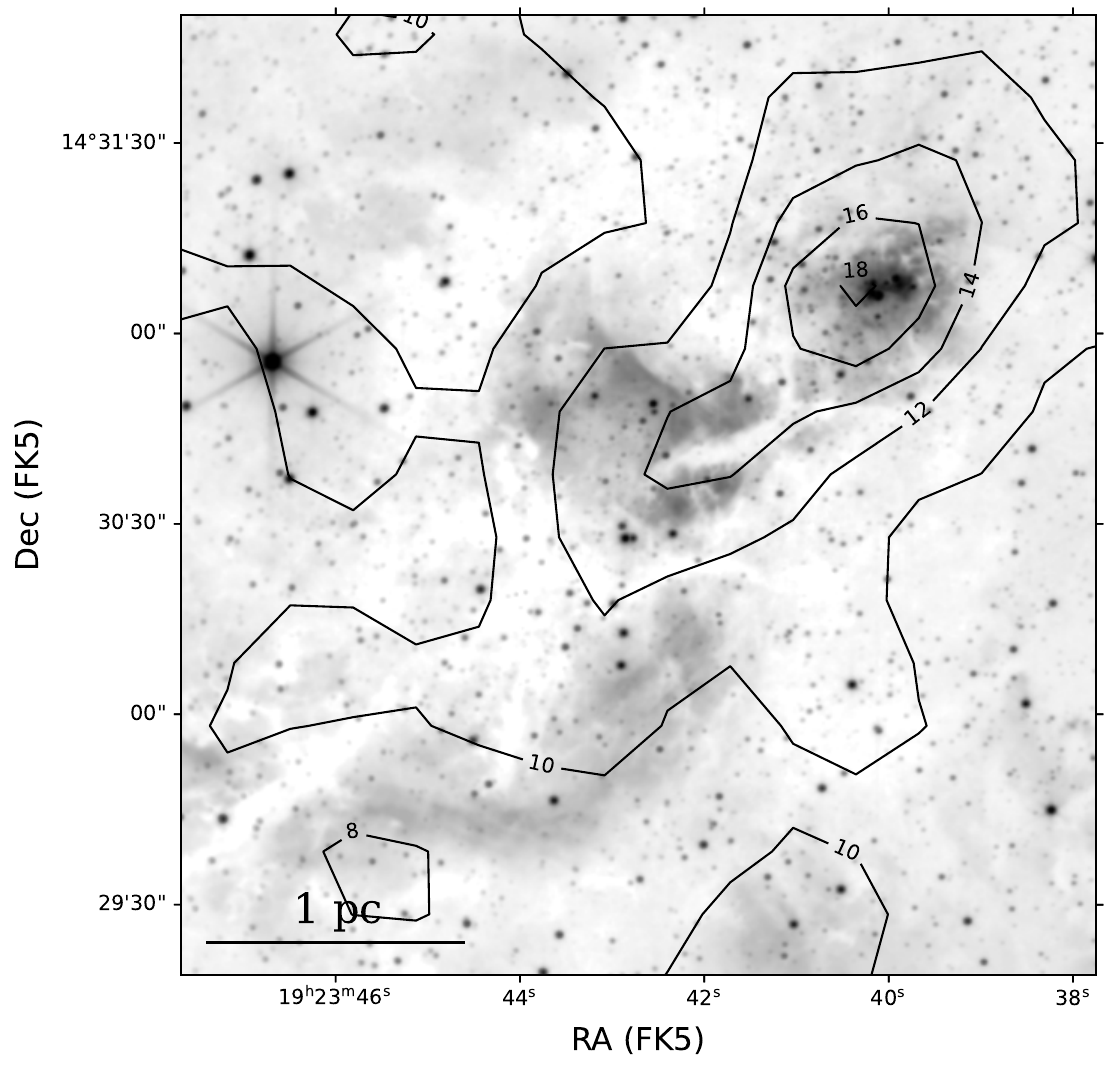}
    \caption{Extinction contours over-plotted on the W51A proto-clusters. Contour labels are given in visual extinction ($A_{V}$). The contours were created with 10"$\times$10" bins, and smoothed using a 2D Gaussian kernel with an 8-pixel width.}
    \label{ext_con}
\end{figure}

\begin{figure}
    \centering
    \includegraphics[width=1.0\linewidth]{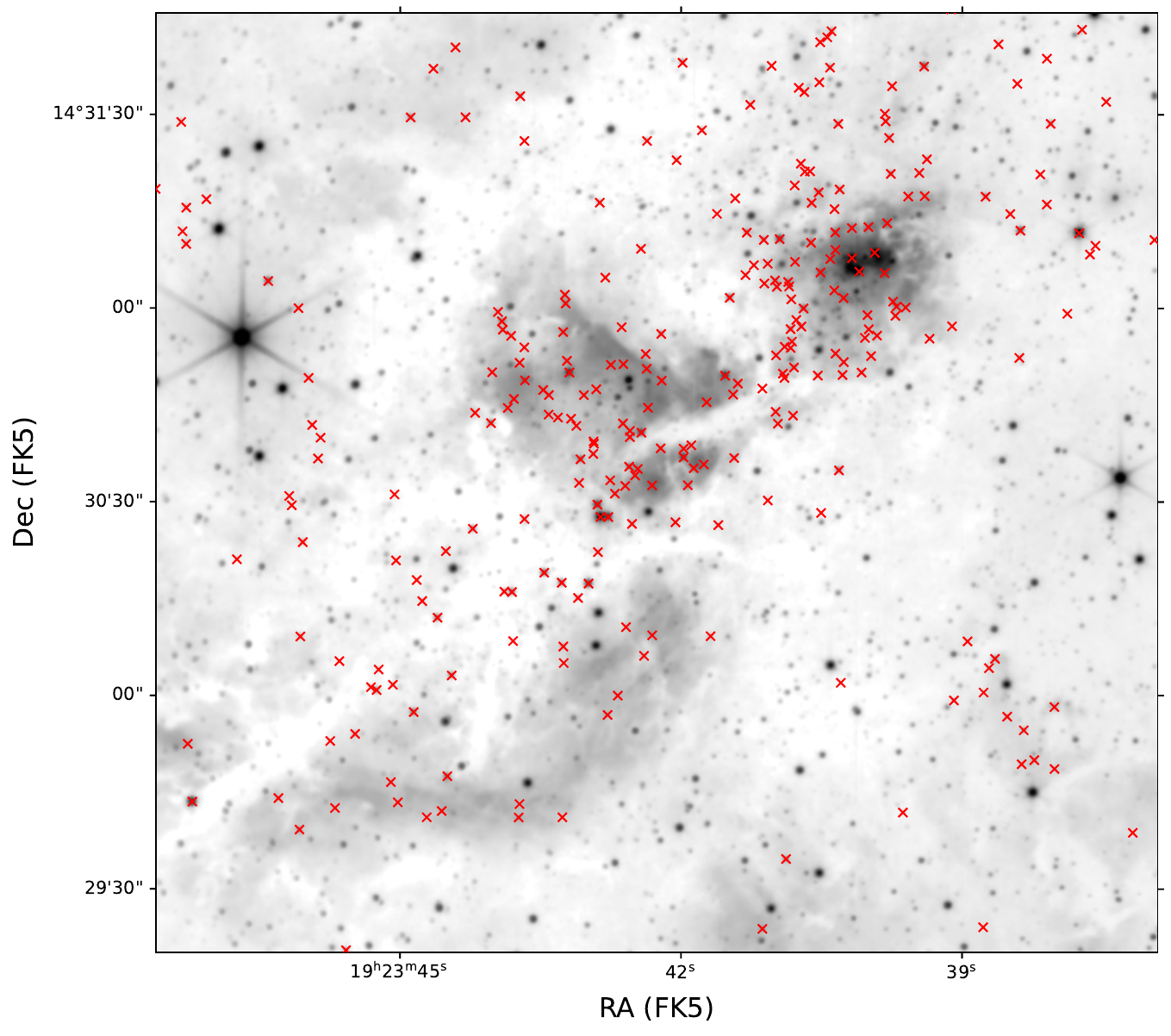}
    \caption{\textit{K\textsubscript{s}}-band image focused on the central HII region of W51A. The red X's show the location of sources with an \textit{H}-\textit{K\textsubscript{s}} color greater than 2.}
    \label{excess_sources_protocluster}
\end{figure}

Figure \ref{ext_con} shows the W51A proto-clusters over-plotted with extinction contours. From this figure, it can be seen that the sources in W51 IRS2 suffer from more extinction (average $A_{V} \sim 19$) than sources in W51 Main (average $A_{V} \sim 14$). Both clusters have a similar number of sources with \textit{H-K\textsubscript{s}} colors $>$ 2.0 ($A_{V}$ $\gtrsim$ 25, assuming no $K_{s}$-band excess), with W51 IRS2 having 56 sources and W51 Main having 62. The distribution of these sources in the region can be seen in Figure \ref{excess_sources_protocluster}. Both clusters suffer from differential extinction, with  individual source values of $6<A_V<44$ mag  in W51 IRS2, and $5<A_V<53$ mag in W51 Main.

\subsection{Foreground Contamination}
\label{sec:foreground}
We classified stars as foreground contaminants if they exhibited extinction less than the minimum expected for the W51 cloud. To determine that minimum extinction, we used the 3D dust maps created from the Pan-STARRS 1, Two-micron All Sky Survey (2MASS), and Gaia surveys by \citet{Green2019}. The authors host web-based plots showing distance modulus vs. E(g-r)\footnote{The plots for W51A can be found \href{http://argonaut.skymaps.info/query?lon=49.8&lat=-0.4&coordsys=gal&mapname=bayestar2019}{\texttt{here}}.}. Converting the plot to distance vs. $A_{V}$, we find that the minimum extinction to W51A is $A_{V}$ $\sim$ 5. The distance modulus towards W51A (13.5, from a distance of 5.1\raisebox{0.2ex}{$\begin{array}{c}  2.9 \\  -1.4 \end{array}$} kpc \citep{Xu2009}) falls outside the range of reliable distance moduli (7.25-12.58) of these dust maps, so this can be interpreted as the absolute minimum extinction towards W51A. Using this information, we removed any source from the catalog with $A_{V}<5$ as a first step. For details about individual source extinction estimations see Section \ref{ext}. This removed $\sim1050$ sources from the catalog. 


We searched for additional foreground contaminants by cross-matching the GTC catalog to optical catalogs from GAIA and IPHAS \citep{Barentsen2014}. Since the distance modulus plus extinction to W51 is $>18.5$, we assume that any source with an optical detection in GAIA or IPHAS is a foreground contaminant; only relatively massive ($M > 2 M_\odot$) stars with no local extinction should be detectable at optical wavelengths within W51. To allow inclusion of such sources, we kept those optical detections with a $J-K_{s}$ color $>2.88$ (this is equivalent to $A_{V}\sim15$ using the Milky Way average extinction curve \citep{Gordon2009}). This is the minimum $J-K_{s}$ color of the spectroscopically classified sources from \citet{Bik2019}. The minimum $J-K_{s}$ color of these sources were chosen because we know they are main-sequence stars at the distance of W51A, and they are not embedded within the HII region. Using this process, we removed an additional 870 sources from the GTC catalog. 


\subsection{Measuring Completeness}\label{measuring_completeness}
W51A is a region that varies in stellar density and background emission across the FOV. To account for this, we measured the completeness in two different regions in our images: the regions encompassing the proto-clusters (these regions are defined in Section \ref{pcb}), and the area south of the central HII region with no diffuse emission. The pixel scale for our observations is 0.1942"/pixel, and the average measured PSF in the K$_{s}$-band is $\sim$1.05".

For both regions, we added 10 randomly placed synthetic stars of a given magnitude and then ran the star finder to determine how many synthetic stars were recovered. This process was repeated 10 times per magnitude for a total of 100 synthetic stars per specified magnitude. We did not do visual inspection during the completeness testing, so if a synthetic source was randomly placed very close ($< 1$ pixel) to a false detection, and the synthetic source was not detected, the completeness at that magnitude would be slightly overestimated.

\begin{figure}[H]
    \centering
    \includegraphics[width=1.0\linewidth]{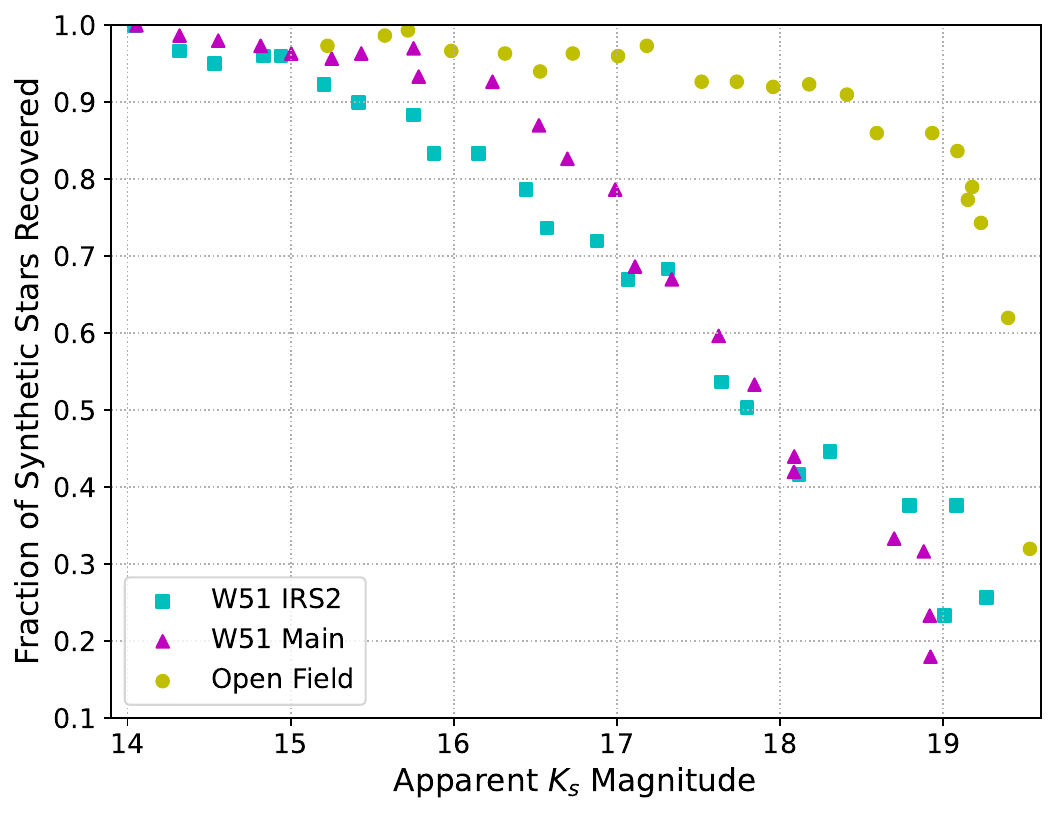}
    \caption{$K_{s}$-band completeness curves for three regions in W51A: W51 IRS2, W51 Main, and the region south of the proto-clusters with no diffuse emission.}
    \label{completeness_curve}
\end{figure}

Figure \ref{completeness_curve} shows the $K_{s}$-band completeness curves for the central HII region and the region south of the proto-clusters (with no diffuse emission). We measure the 90\% completeness limits away from the central HII region to be 19.38, 18.87, and 18.34 in the $J$-, $H$-, and $K_{s}$-bands respectively. Even in the background-subtracted images, the effects of the diffuse emission are significant in the $K_{s}$-band; the 90\% completeness limits in the proto-clusters are 15.45 in W51 IRS2 and 16.36 in W51 Main. We used the completeness curves in Figure \ref{completeness_curve} to exclude sources in the proto-clusters whose magnitudes were fainter than the 50\% completeness limit during the analysis.


\section{Results and Discussion}\label{results_and_discussion}

\subsection{Stellar Surface Density Map and Proto-cluster Boundaries}\label{pcb}
\begin{figure*}
    \centering
    \includegraphics[width=1.0\textwidth]{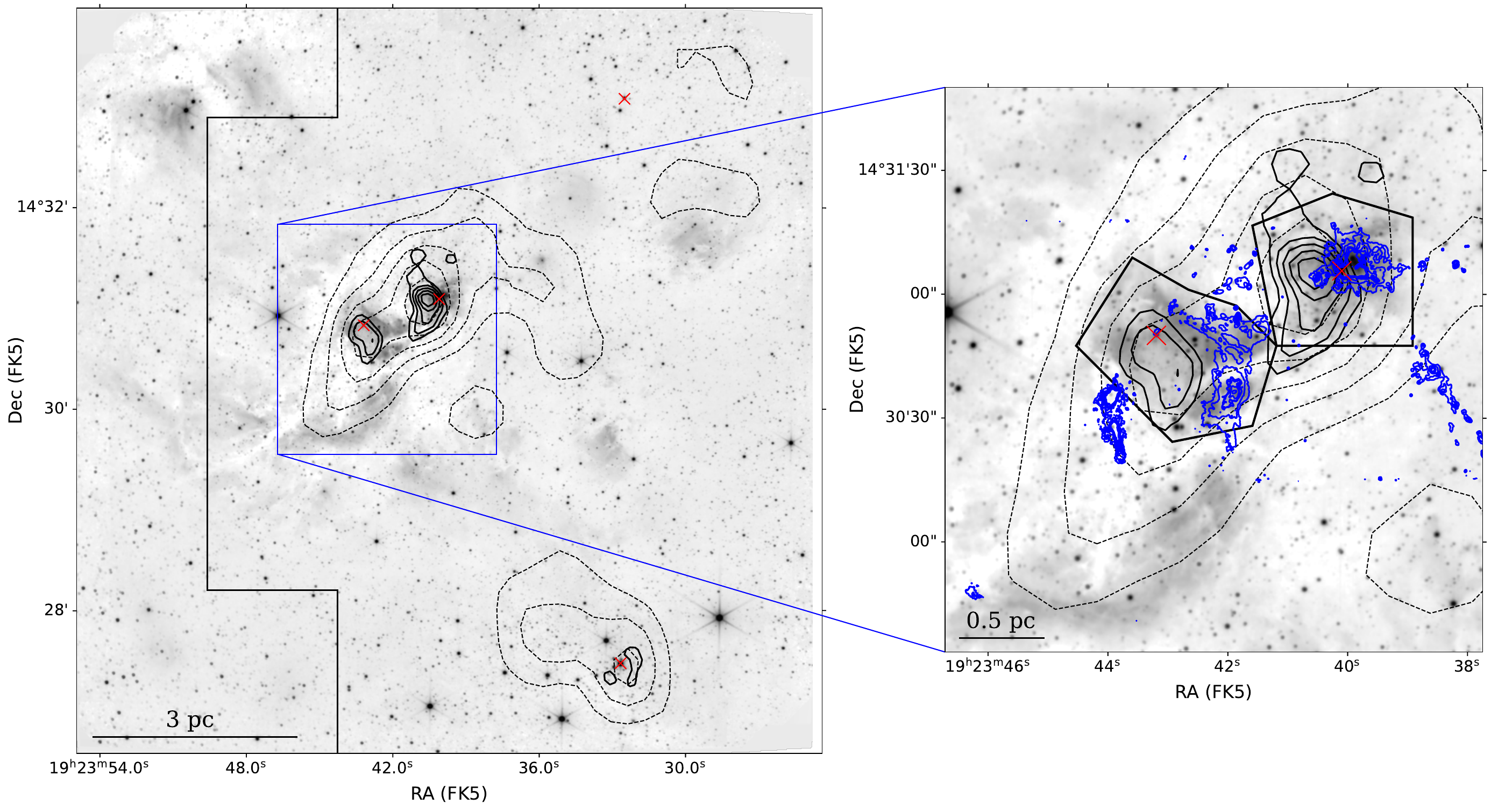}
    \caption{$K_{s}$-band image with a zoom-in of the W51A proto-cluster region showing the stellar density map. Dashed-line contours were created with 10"$\times$10" bins, and the solid black contours were created with 3"$\times$3" bins. The contour levels for both sets of contours start at 50\% of the value of the most populated bin, increasing by 10\% increments for each level. The lowest contour level (shown by the dotted contours) represents 134 stars/pc$^2$ (0.08 stars/arcsec$^2$) and the highest contour level (shown by the solid contours) represents 484 stars/pc$^2$ (0.30 stars/arcsec$^2$). The 1 mm ALMA emission is shown by the blue contours. The defined proto-cluster regions are shown as the black polygons encompassing the areas of both high stellar density, and bright 1 mm emission. The solid line in the full FOV shows the region we have excluded due to poor image quality. Red X's indicate the position of the extremely luminous sources discussed in Section \ref{SoI}.}
    \label{fig:sdm_inset}
\end{figure*}

Creating a stellar surface density map (SDM) is common practice when analyzing the spatial distribution of stars within clusters. They provide quantitative estimates to stellar surface densities and how they change throughout the cluster. We present our SDM of W51A in Figure \ref{fig:sdm_inset}. The SDM was created using 2D histograms of source counts, with two bin sizes: 10"$\times$10" (dotted contours) for large-scale stellar distributions, and 3"$\times$3" (solid contours) to show how the stars are distributed in areas of high stellar density. Both maps were smoothed with an 11-pixel wide 2D Gaussian kernel, using pixel sizes matching each bin scale (10"$\times$10" or 3"$\times$3", depending on the respective SDM being smoothed). This configuration produced the smoothest contours while keeping the peaks of stellar density respective to each proto-cluster separate.


As expected, there is an over-density of stars embedded in the HII region at the location of the proto-clusters, with an average stellar density of 242 stars/pc$^2$ for W51 IRS2 and 215 stars/pc$^2$ for W51 Main, compared to a typical background density of 26 stars/pc$^2$ for the rest of the observed field. However, there is another area of high stellar density located south of the HII region, in close proximity to the source LS2 \citep{Okumura2000}. We discuss this region further in Section \ref{third_dense_region}.

In combination with the SDM, we used K-means clustering to better constrain the outlines on the boundaries of both proto-clusters and to identify any other possible clusters in the field not seen in the SDM. 
The generated clusters were then compared to the stellar density map of W51A to define the proto-cluster boundaries. We placed the regions defining the boundaries of the proto-clusters over the areas of the highest stellar density, while also encompassing the bright HII regions these stars are embedded in, and the areas of the proto-clusters with the brightest gas emission from 1 mm ALMA observations from \citet{Ginsburg2017}. Converting from pixel units to physical units, the sizes of the proto-cluster boundaries are $\sim$0.74 pc$^2$ for W51 IRS2, and $\sim$0.8 pc$^2$ for W51 Main. The stellar density map and the defined proto-cluster regions are shown in Figure \ref{fig:sdm_inset}. There are a total of 192 sources in W51 IRS2 and 198 sources in W51 Main that we can resolve from our data before down-selection based on completeness.

\subsection{Newly Discovered Stars and Sources of Interest}\label{new_stars}
We detect several sources not previously identified in any catalogs of observations at the same wavelength.
The observations we have presented here have comparable seeing in the $K_{s}$-band to the UKIDSS GPS ($\leq$ 1.0", \citet{Lucas2008}) and the photometric catalog from \citet{Bik2019} ($\sim$0.95"), while having worse seeing than the observations from \citet{Kumar2004} (average of 0.5"). We are able to detect more sources in W51A mainly due to our careful removal of the background emission, allowing us to detect more sources on top of the extended background in the central HII region. Another contributor is the slightly increased sensitivity of our observations compared to the past surveys. The 90\% completeness limit in the $K_{s}$-band for the UKIDSS GPS is 18.0 and 17.3 for the observations of \citet{Kumar2004}, while it is 18.34 for our observations.


We compared the GTC catalog to 2MASS, the UKIDSS galactic plane survey, as well as the catalogs of \citet{Kumar2004}, \citet{Bik2019}, and our own photometric catalog covering the NACO images of W51 IRS2. There is a total of 3003 stars in our field that were not documented in any previous NIR catalog of the region, with 88 of these sources being located within the proto-clusters. There are 2265 newly detected sources with corresponding $H$-band detections, and 1298 with corresponding $J$-band detections. 

\begin{figure*}[htbp!]\centering
\gridline{\fig{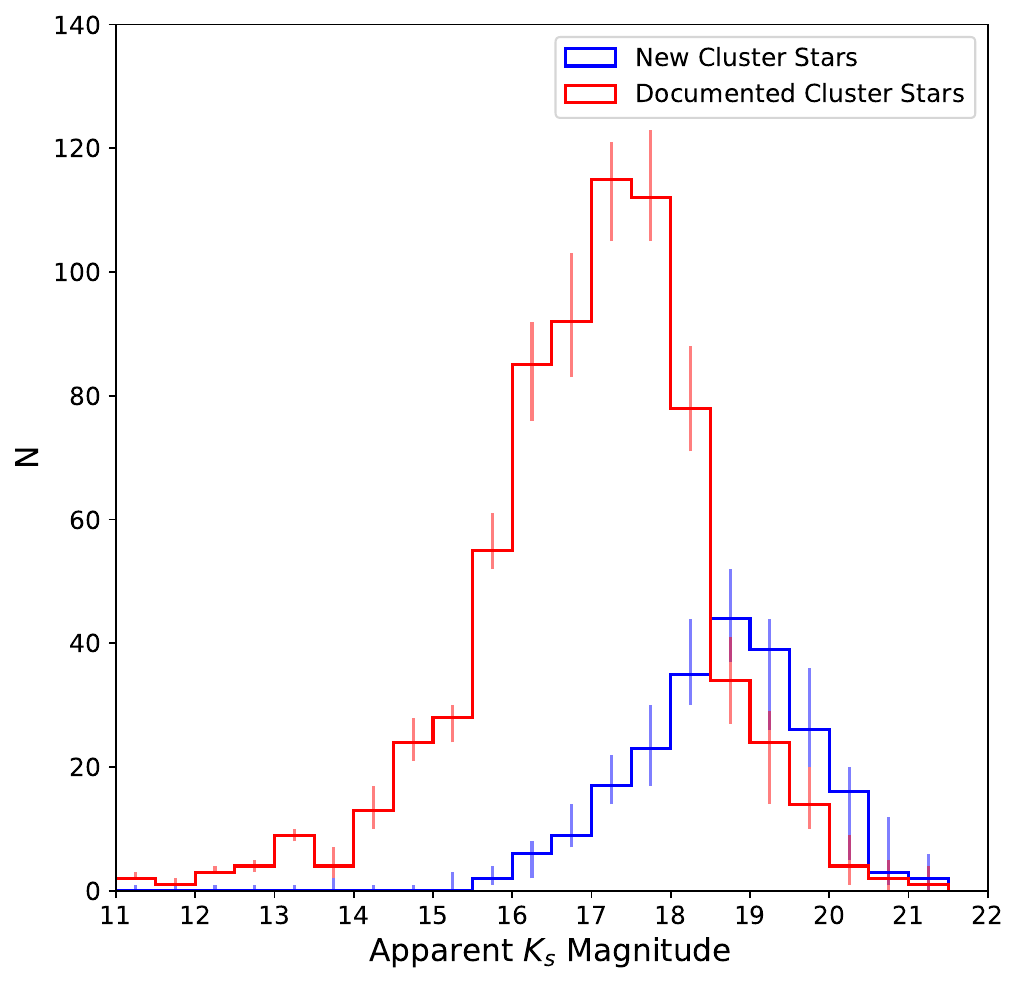}{0.45\textwidth}{(a)}
          \fig{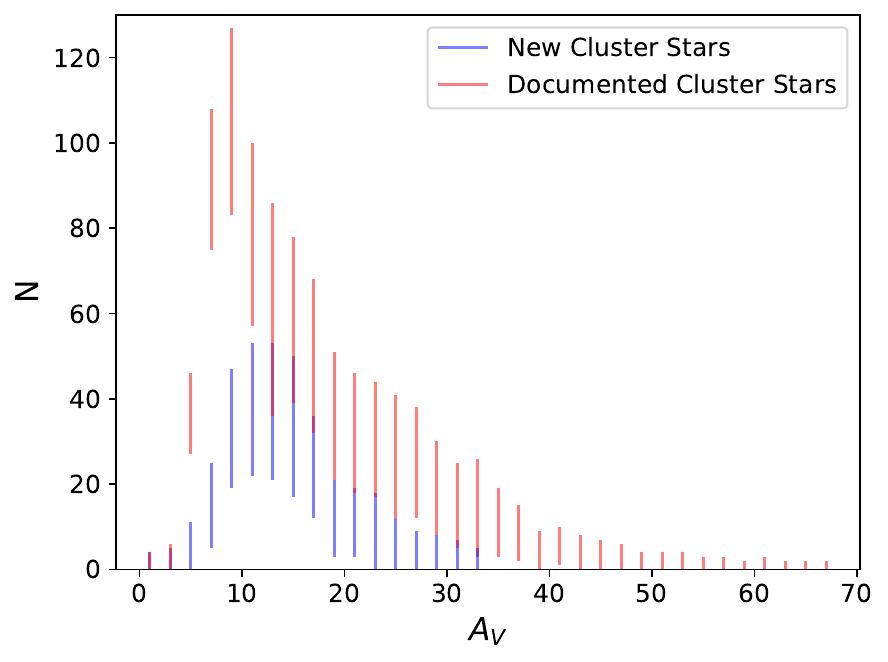}{0.45\textwidth}{(b)}}
\caption{Histograms comparing the distribution of (a) apparent $K_{s}$-band magnitude and (b) extinction ($A_{K_{s}}$) between the newly-discovered and previously-documented stars in the W51A proto-clusters (see Section \ref{new_stars}).}
\label{fig:comparison_between_new_old}
\end{figure*}


These new sources spanned the magnitude range $13.9 < m\textsubscript{$K_{s}$} < 25.5$.
The detection of new bright sources suggests the reason these stars have not been observed by previous authors is because their PSFs are getting confused with either the bright HII region or nearby sources.

When comparing the newly-discovered to the previously-documented cluster stars, we find that the median apparent $K_{s}$-band magnitude of newly-discovered stars is much fainter than the previously-documented stars (18.7 mag compared to 17.1 mag), and the median extinction of the newly discovered and previously documented stars are the same ($A_{V}\sim13$). We present histograms comparing the distributions of apparent $K_{s}$-band magnitude and extinction of the two populations in Figure \ref{fig:comparison_between_new_old}. Panel (b) lacks concrete bin values because we account for the range of possible extinction values for each source, showing only the minimum and maximum bin values from the generated sample populations (See Appendix \ref{sample_populations}). 

Figure \ref{fig:new_stars_protocluster} shows the locations of the newly discovered sources within the proto-clusters. In the figure, some of the newly detected sources are difficult to distinguish from the residual structured background. Sources marked in blue are only observed in the $K_{s}$-band, which introduces some uncertainty regarding the legitimacy of the detection. The sources detected only in the $K_{s}$-band were omitted from the analysis to prevent false detections from affecting the results.




\begin{figure*}
\gridline{\fig{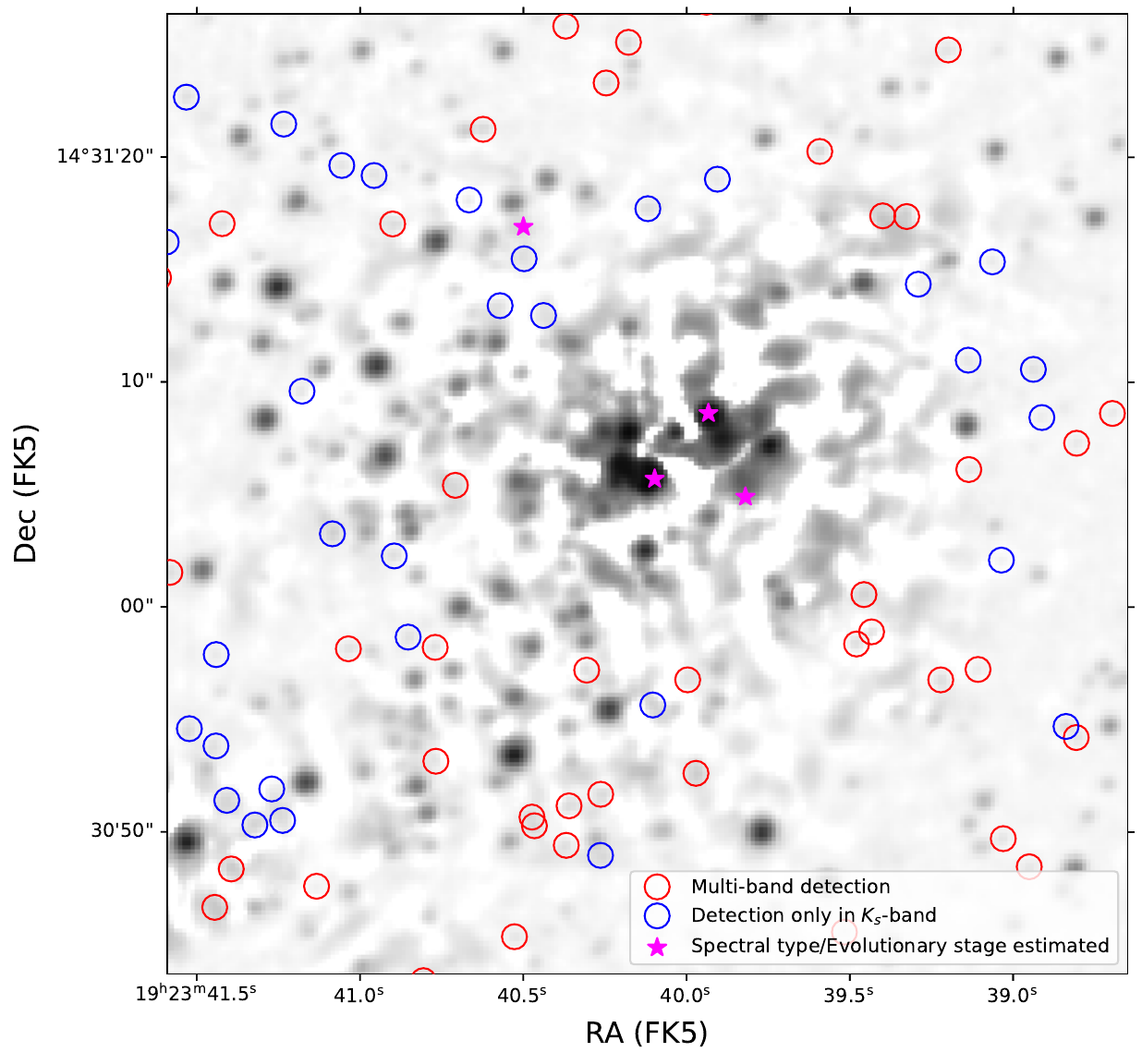}{0.5\textwidth}{}
          \fig{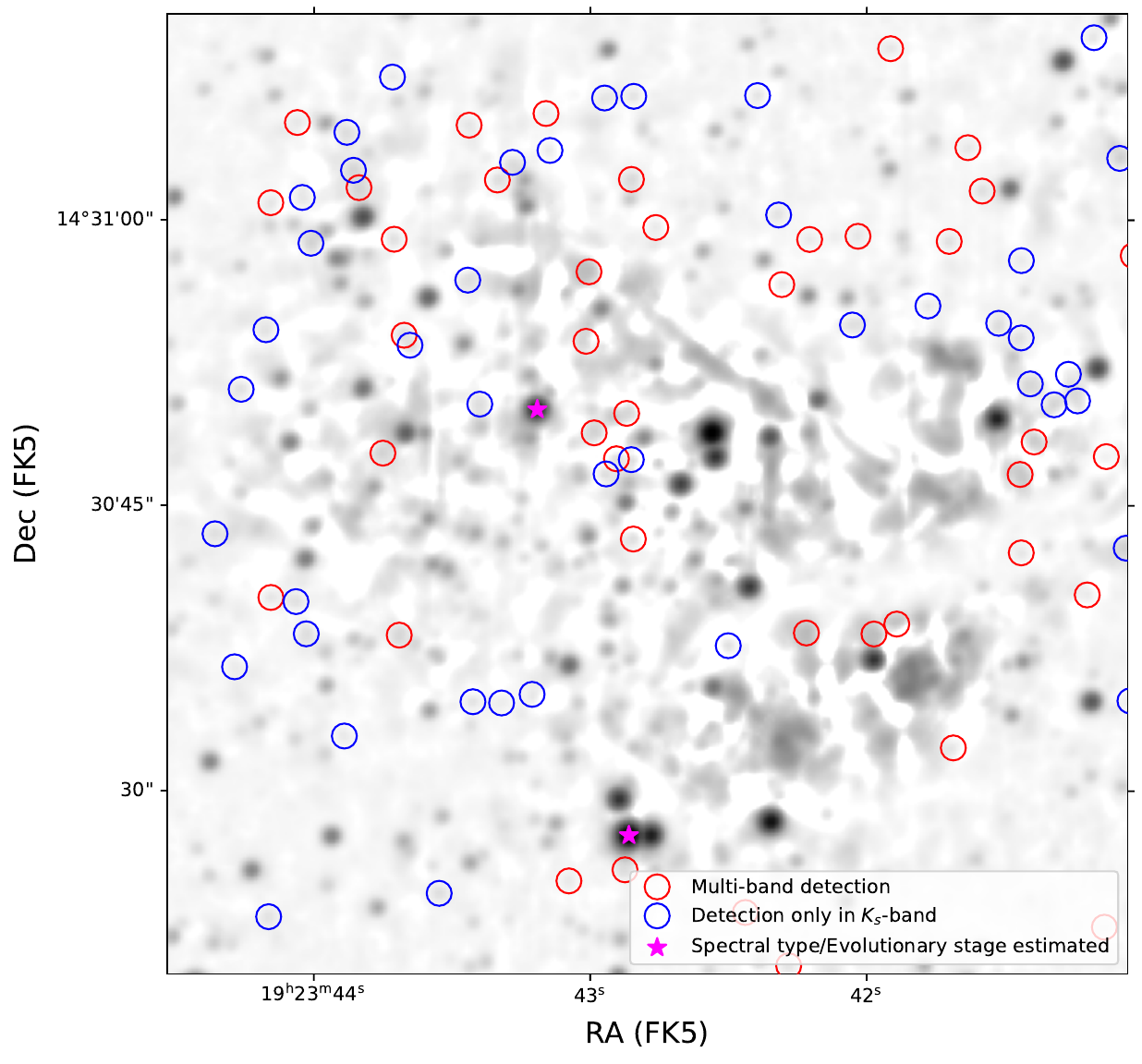}{0.5\textwidth}{}}
\caption{Background-subtracted $K_{s}$-band image showing newly detected sources in W51 IRS2 (left) and W51 Main (right). The magenta star markers show the location of stars which have had their spectral type/evolutionary stage estimated (See Appendix \ref{age_estimation_app}).}
\label{fig:new_stars_protocluster}
\end{figure*}

\subsubsection{X-ray Sources}\label{xray}

\begin{figure*}
    \centering
    \includegraphics[width=1.0\linewidth]{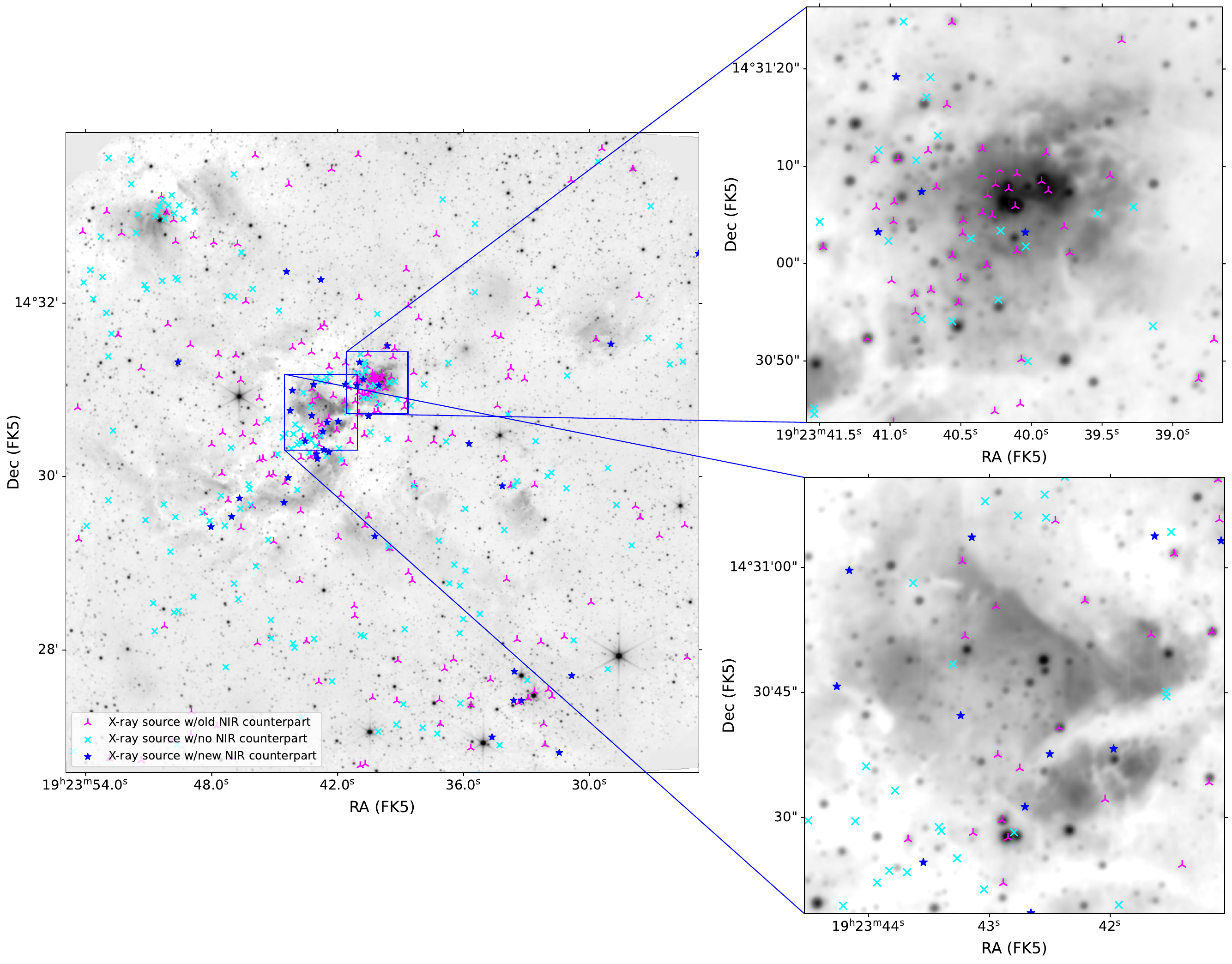}
    \caption{$K_{s}$-band image showing the locations of all the X-ray sources in the region from \citet{Townsley2014} with distinctions between the sources which have been previously-documented in the NIR by past surveys, sources which have been newly-detected in the NIR from our observations, and X-ray sources which have not been observed in the NIR.} 
    \label{xray_sources_fig}
\end{figure*}

We have identified the NIR counterparts to 194 X-ray sources from the \citet{Townsley2014} X-ray catalog, with 30 of these sources being newly-identified in the NIR from our observations. The proto-clusters and the surrounding regions have the highest concentration of newly-identified NIR counterparts. There is another small grouping of these newly-identified NIR counterparts located around the third region of high stellar density (See Section \ref{third_dense_region}). Combining the GTC observations presented here, the UKIDSS GPS, and our catalog of the NACO observations of W51 IRS2, we find that 60\% of the X-ray sources in our FOV have been identified in at least the $K_{s}$-band. We include Figure \ref{xray_sources_fig} to show the location of the X-ray sources in our FOV. 


\subsubsection{Extremely Luminous Sources}\label{SoI}
We have found three exceptionally luminous sources with $M_{K_S} < -7.37$ throughout the FOV ($M_{K_S} = -7.37$ is the brightest $M_{K_S}$ with an associated mass using our mass-luminosity relationship (MLR) (See Section \ref{Mass})). 
Despite its high luminosity, one source is not saturated in our images because of its high extinction. The other two sources (IRS2E and LS2) were saturated in our observations. We address the possible reasons for why these sources are saturated in their respective descriptions. 

IRS2E is the first source, with a measured $M_{K_S}$ range of $-7.45 < M_{K_S} < -11.58$, and the brightest $K_{s}$-band source within the proto-clusters (located at 19\textsuperscript{h}23\textsuperscript{m}40.11\textsuperscript{s}, +14\degree31\textsuperscript{'}5.88\textsuperscript{"}). IRS2E has been classified as an embedded, massive YSO through its spectra \citep{Barbosa2008}, and through SED fitting \citep{Lim2019}. We have estimated the range of extinction to IRS2E to be $30 < A_{V} < 67$. Our estimate for the extinction to this source is fairly consistent to what previous authors have measured; \citet{Barbosa2016} estimated the extinction to IRS2E to be $A_{V}\sim63$ by extrapolating mid-infrared flux measurements to $A_{V}$, and \citet{Lim2019} estimated the extinction to be $A_{V}\sim75$ using SED fitting and YSO radiative transfer models. 

IRS2E is saturated in our images despite the high extinction we have measured for this source. A potential explanation for this is that IRS2E is a small cluster of stars \citep{GoldaderWilliams} being detected as a single, bright source in our images. Although more recent $H$-band adaptive optics images from \citet{Figueredo2008} show that one source in this grouping is an extended clump of emission, and another is a likely foreground source due to its ``bluer" color in the $H$-band, but these would still contribute to the flux measured by the telescope. Another (likely compounding) contributor to IRS2E's saturation is nebular emission originating within close proximity to IRS2E which is shown by Fe and Kr emission lines in its $K$-band spectra using a 0.96" aperture \citep{Figueredo2008}. This emission is not seen in the $K$-band spectra taken of IRS2E by \citet{Barbosa2008} using a smaller aperture (0.4").



The second source is LS2 with a measured $M_{K_S} = -9.09$ (located at 19\textsuperscript{h}23\textsuperscript{m}32.66\textsuperscript{s}, +14\degree27\textsuperscript{'}28.64\textsuperscript{"}). A possible explanation for this star's extreme brightness is that we used an incorrect distance to calculate its $M_{K_S}$ (the distance to W51A). However, we find that the star's $M_{K_S}=-8.13$ (still exceptionally bright) if we use the distance approximation from GAIA instead ($3248\pm11516$ pc). \citet{Okumura2000} classified this star as a post-MS B- or B[e]-type supergiant based on its spectral features in the \textit{K}-band. We measure an $A_{V}\sim29$ for LS2, which is considerably higher than the average extinction of stars in the proto-clusters. The saturation of LS2 in our images could be explained by the existence of a surrounding reflection nebula (based on spectral features \citep{Okumura2000}), or that this star is very bright and suffers from less extinction due to its location away from the HII region of W51A. 

The third extremely luminous source is not associated with any region of high stellar density (located at 19\textsuperscript{h}23\textsuperscript{m}32.5\textsuperscript{s}, +14\degree33\textsuperscript{'}4.98\textsuperscript{"}). We measure an $M_{K_S} = -7.56$ for this source, and an $A_{V}\sim43$. A possible explanation for this star's brightness is that it is an intrinsically red source, which would make our extinction measurement an overestimate. 

Another source worth bringing attention to is located in W51 Main (specifically at 19\textsuperscript{h}23\textsuperscript{m}43.19\textsuperscript{s}, +14\degree30\textsuperscript{'}50.05\textsuperscript{"}). This source is likely the most extincted source in our observations having an $H-K_{s}$ color of 7.56 ($A_{V}\sim118$, using $H-K_{s}$ color to measure extinction, and assuming there is no $K_{s}$-band excess). \citet{Saral2017} classified this source as a Class I YSO, and \citet{Lim2019} suggests that it is a candidate massive YSO. We do not observe this source in the $J$-band, so we were only able to provide a much lower estimate to the extinction ($9 > A_{V} > 17$).

\subsubsection{Other Areas of High Stellar Density}\label{third_dense_region}
There is a third region in our observed field showing high stellar density which can be seen in the bottom right of Figure \ref{fig:sdm_inset}. This candidate cluster has an average stellar density of 161 stars/pc\textsuperscript{2}, and the approximate center is located at 19\textsuperscript{h}23\textsuperscript{m}32.62\textsuperscript{s} +14\degree27\textsuperscript{'}30.24\textsuperscript{"}. There is faint, diffuse emission associated with this candidate cluster which can be seen in the $H$, $K_{s}$, and $H_2(1-0)$ filters. We approximated the size of this candidate cluster with a circular region using the same justification we used when defining the proto-cluster boundaries (i.e. k-means clustering results, SDM. See Section \ref{pcb}). There are 198 stars in this defined region, and 85 of these stars are newly observed. The \textit{K\textsubscript{s}}-band magnitude of these new stars span the range $15 < m\textsubscript{\textit{K}} < 21.3$, and the average extinction of stars in this region is $A_{V}\sim17$. 


\subsection{Molecular Hydrogen Emission}
Comparisons between our narrow-band H\textsubscript{2}(1-0) and broad-band \textit{K\textsubscript{s}} observations allow us to identify excited 2.12 \um H\textsubscript{2} S(1) 1-0 emission.  This emission is likely produced by either protostellar outflows or ultraviolet radiation from massive stars. \citet{Hodapp2002} discovered five instances of shock-excited H\textsubscript{2} emission in our FOV. We confirm the existence of four of these shock fronts (W51H\textsubscript{2}J192339.7+143131, W51H\textsubscript{2}J192335.0+143028, W51H\textsubscript{2}J192336.6+143013, W51H\textsubscript{2}J192338.3+143047) in our observations, but the fifth (W51H\textsubscript{2}J192347.2+142944) appears to be a massive O-type star \citep{Figueredo2008,Bik2019}. \citet{Hodapp2002} concluded that the H\textsubscript{2} emission originates in the immediate vicinity of this star. However, we do not observe any H\textsubscript{2} emission near or surrounding this star.

\begin{figure*}
    \centering
    \includegraphics[width=1.0\linewidth]{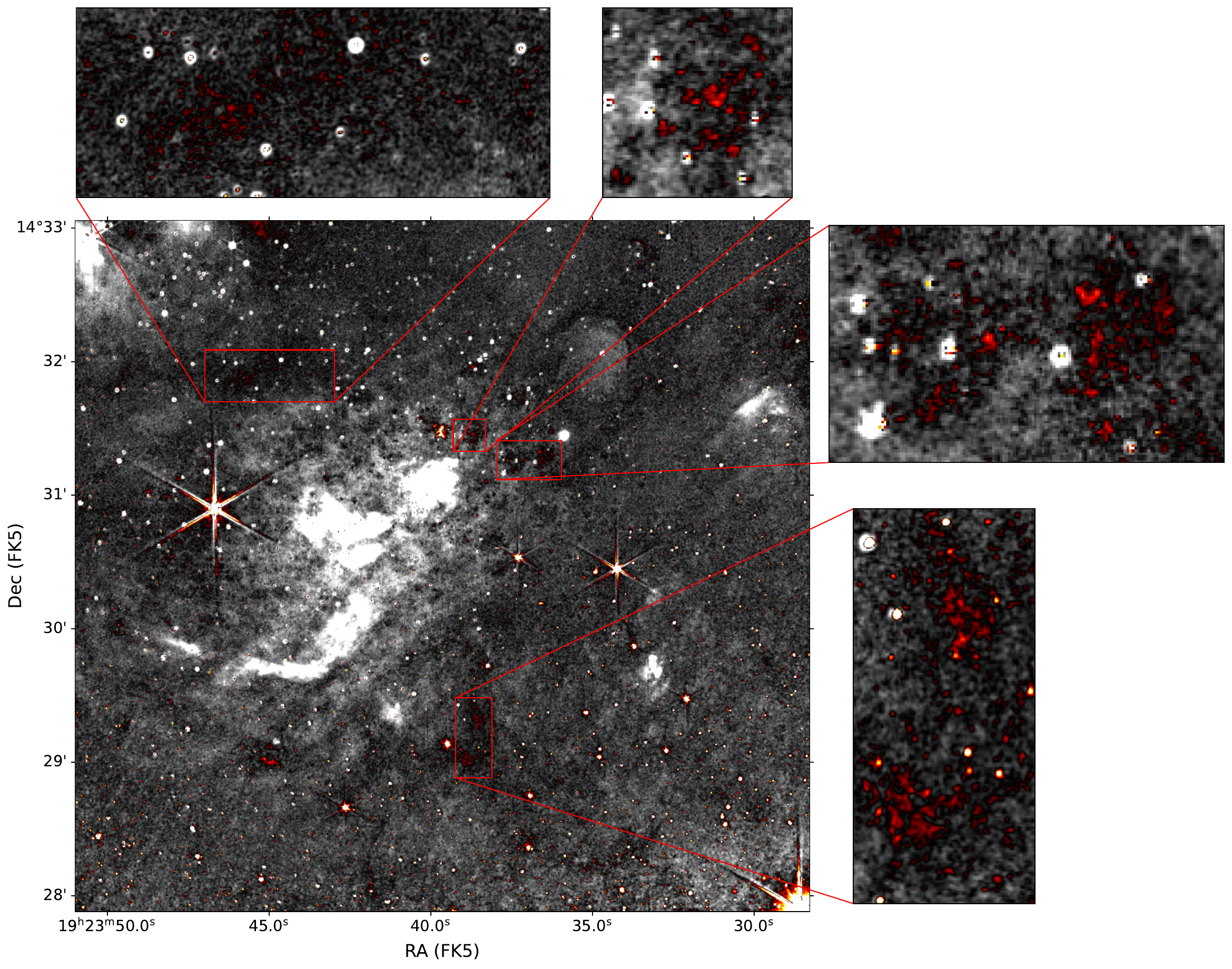}
    \caption{\textit{$H_{2}$(1-0)}-\textit{K\textsubscript{s}} image showing the location of the clustered H\textsubscript{2} emission seen throughout the observed field.}
    \label{distributed_h2_emission}
\end{figure*}

\begin{figure*}\centering
{\label{h2_emission_A}\includegraphics[scale=0.37]{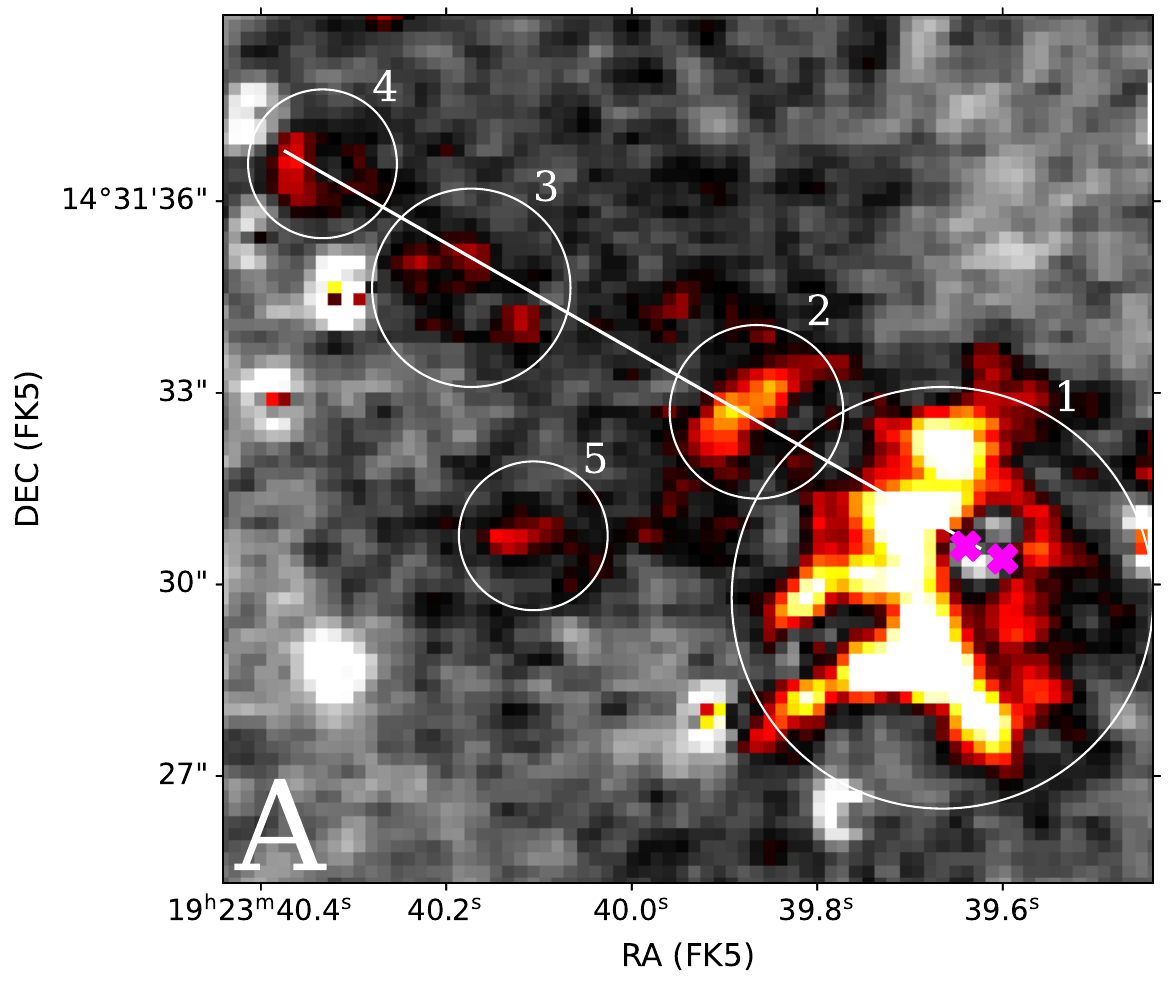}}
\hfill
{\label{h2_emission_B}\includegraphics[scale=0.37]{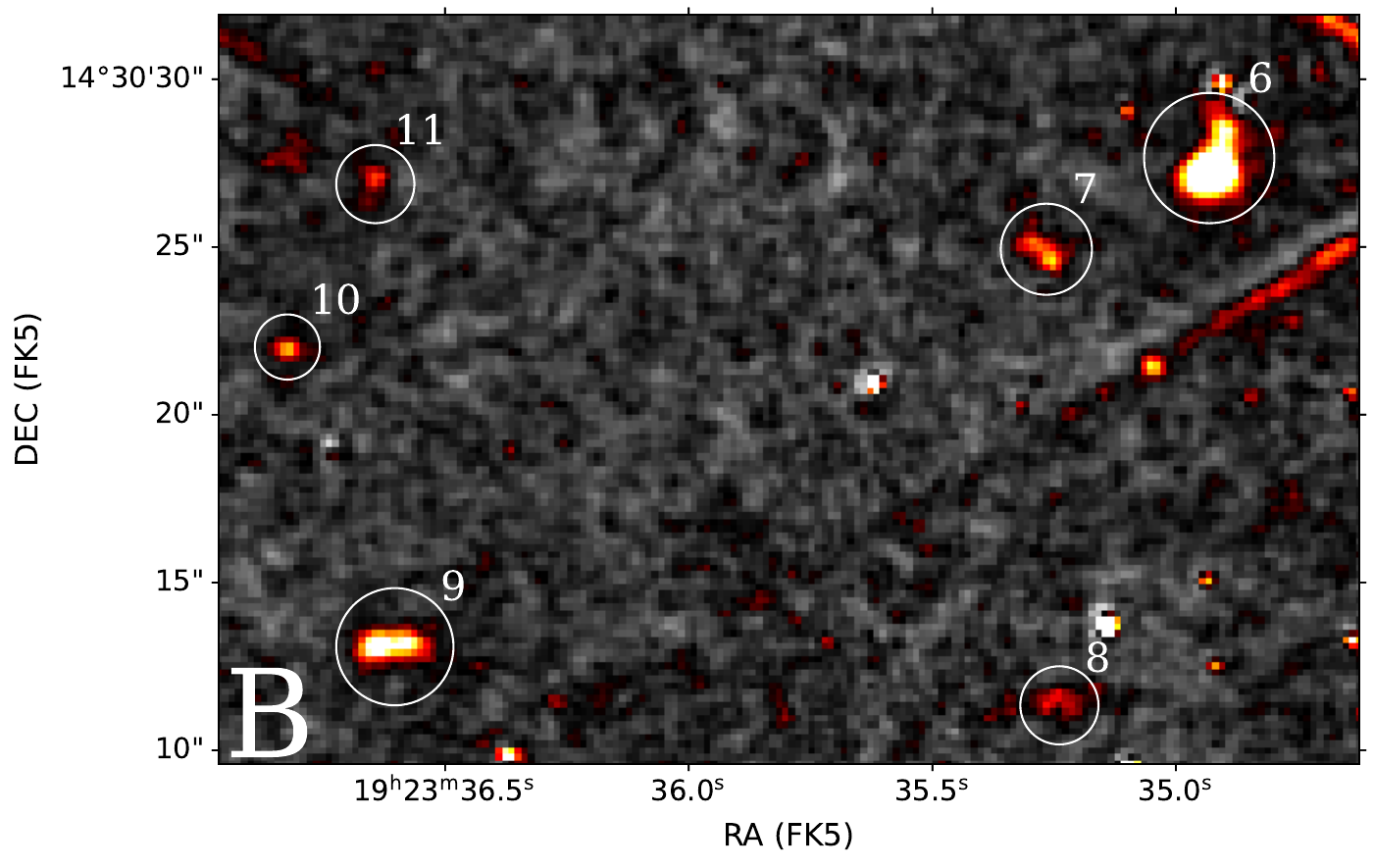}}
\vfill
{\label{h2_emission_C}\includegraphics[scale=0.37]{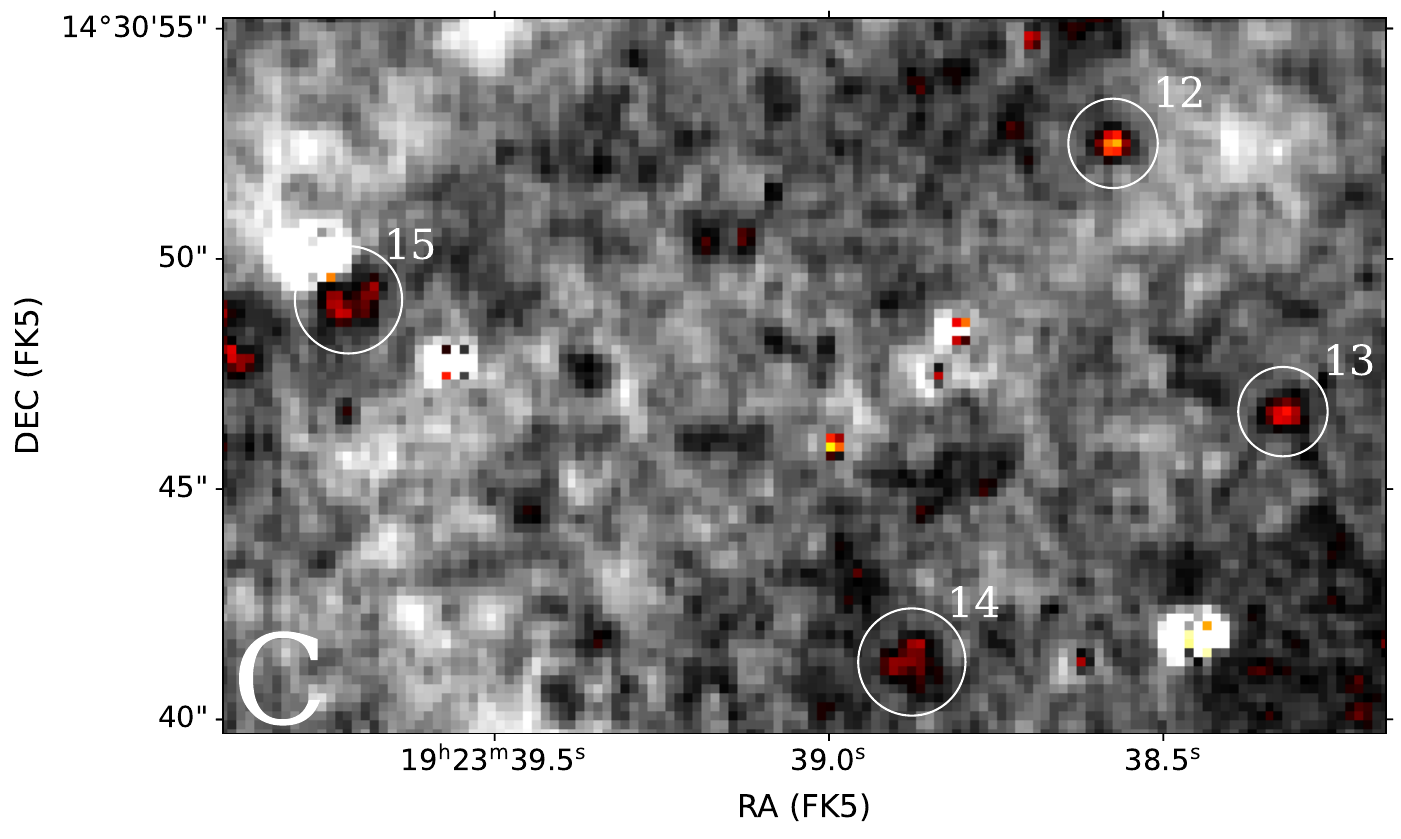}}
\hfill
{\label{h2_emission_D}\includegraphics[scale=0.37]{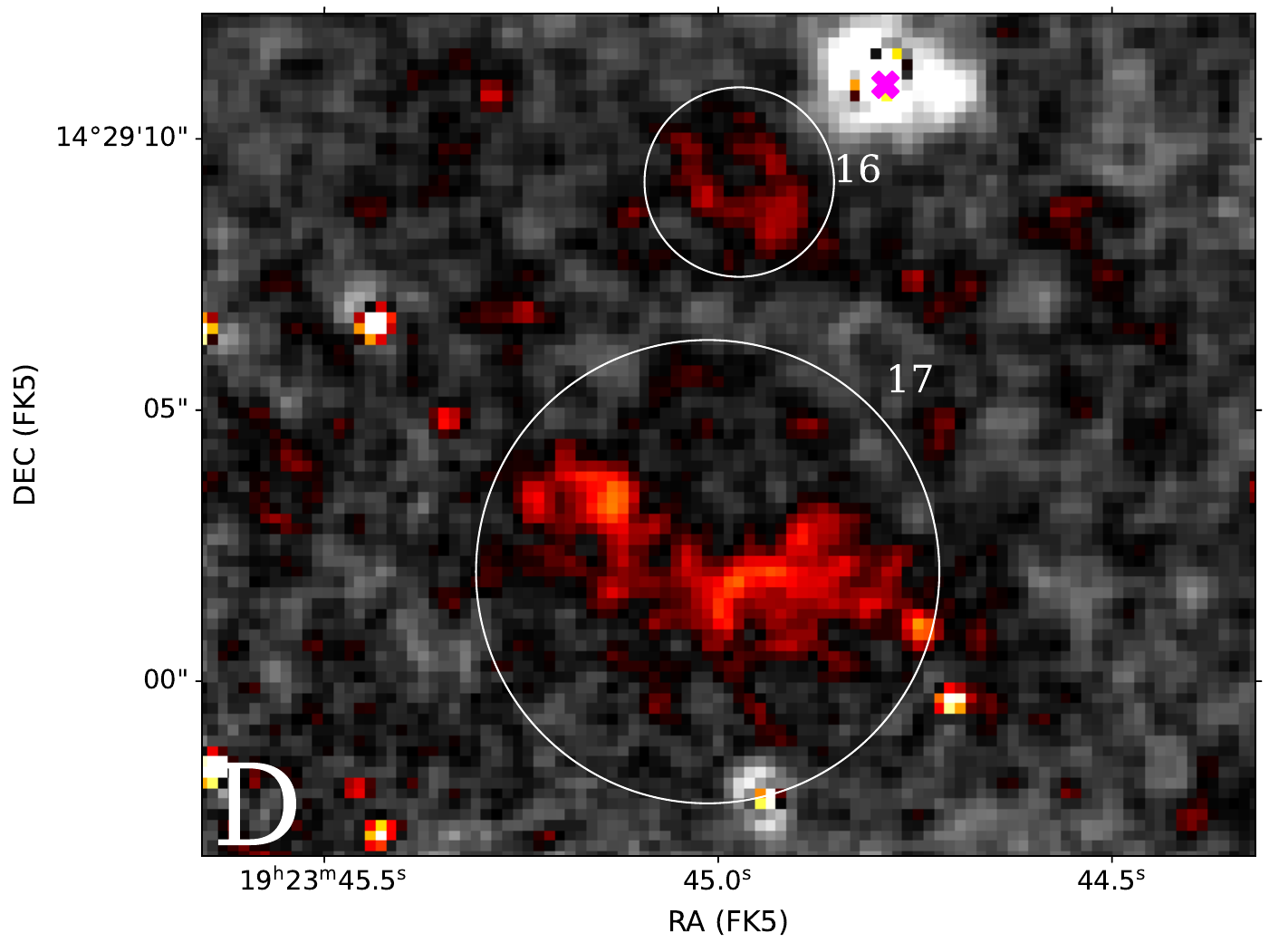}}
\vfill
{\label{h2_emission_E}\includegraphics[scale=0.37]{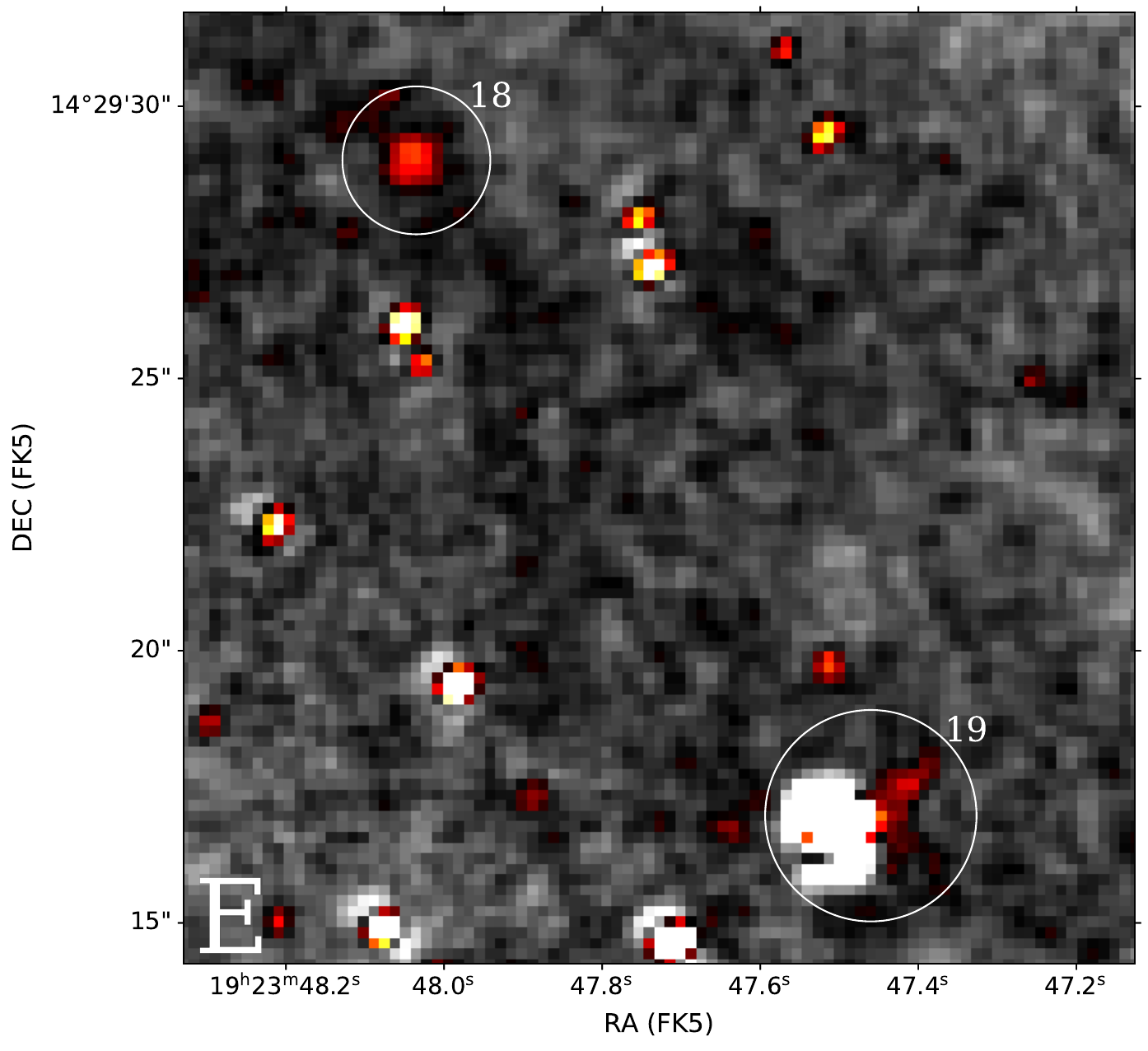}}
\hfill
{\label{h2_emission_F}\includegraphics[scale=0.4]{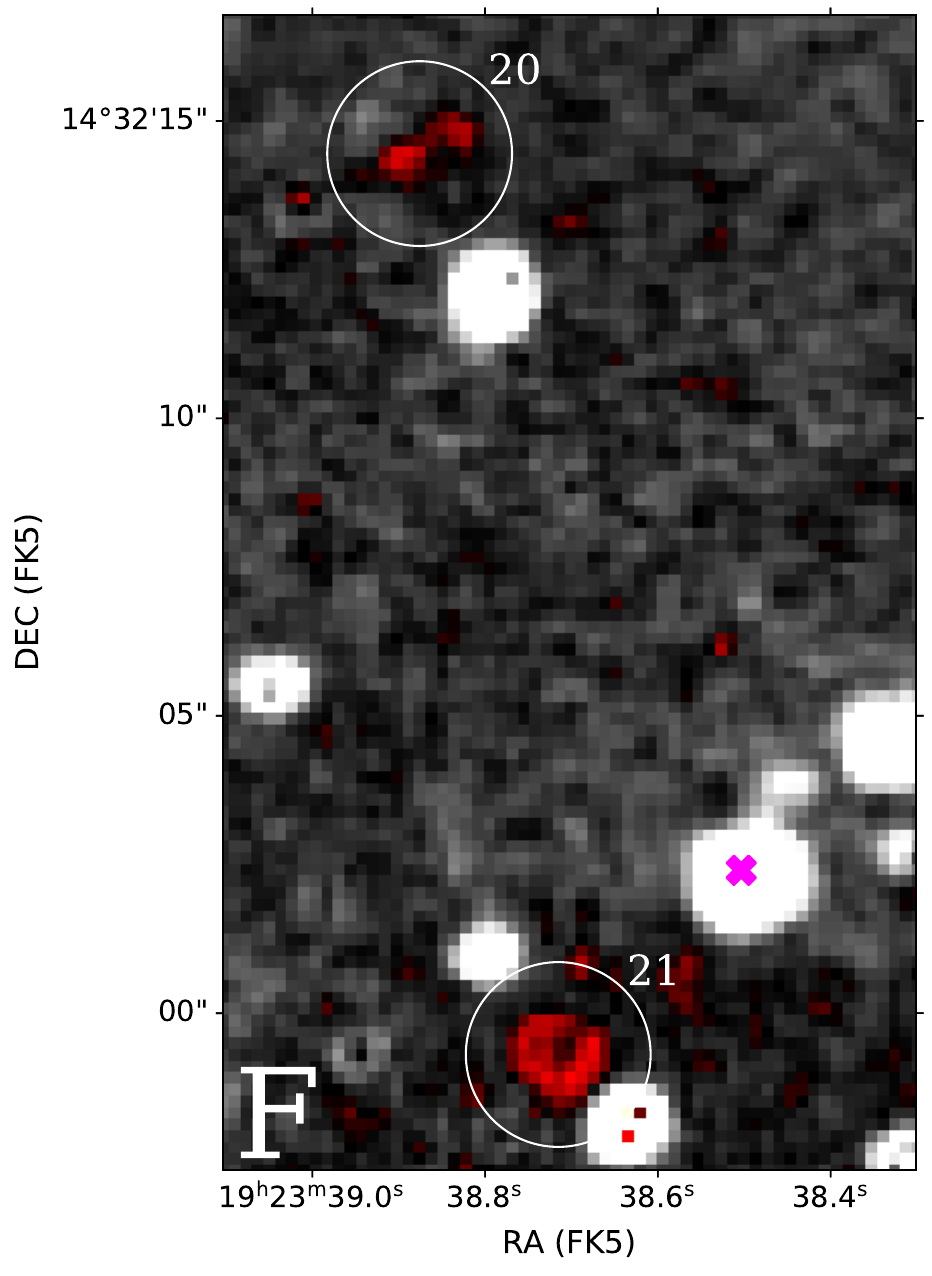}}
\caption{\textit{$H_{2}$(1-0)}-\textit{K\textsubscript{s}} image of isolated H\textsubscript{2} emission in the W51A region. The double star features seen in these panels (especially prominent in Panel E) are the result of imperfect \textit{$H_{2}$(1-0)}-\textit{K\textsubscript{s}} subtraction. Exact locations for these instances of H\textsubscript{2} are provided in Table \ref{h2_regions}. The filled, magenta X's mark the location of candidate driving sources of the nearby H$_{2}$ emission.}
\label{fig:h2_emission_all}
\end{figure*}

We compared the H\textsubscript{2} emission from \citet{Hodapp2002} to our observations, but we did not detect any significant morphological differences between the four shock fronts. However, our observations reveal a wealth of undocumented H\textsubscript{2} emission in the W51A region. We present Figure \ref{distributed_h2_emission} highlighting multiple regions in our FOV that show a clustering of H\textsubscript{2} emission, and Figure \ref{fig:h2_emission_all} which shows instances of more structured H\textsubscript{2} emission. For these figures, we subtracted the $K_{s}$-band image from the \textit{H$_2$(1-0)} narrow band image, resulting in regions with higher H$_{2}$ emission being shown in orange. The regions documented in Figure \ref{fig:h2_emission_all} are labeled 1 through 21 and have individual information about them in Table \ref{h2_regions}. Regions 1, 6, 9, and 13 are all outflows previously documented by \citet{Hodapp2002}, and the remaining 17 are all newly documented. 

A line drawn between regions 6 and 9 passes through a red star (m\textsubscript{K}=15.8, no other broad-band detection) at 19\textsuperscript{h}23\textsuperscript{m}32.65\textsuperscript{s}, +14\degree30\textsuperscript{'}47.23\textsuperscript{"} \citep{Hodapp2002} (we describe this alignment as being `collinear'). This star was also classified as a Class I young stellar object by \citet{Saral2017}. We find that Region 7 (found between Regions 6 and 9) also has a collinear alignment with this star, if it is indeed the source responsible for these emission knots. These regions are shown in Panel B of Figure \ref{fig:h2_emission_all}.

The outflow associated with Region 1 is the most thoroughly studied outflow in W51A due to its size and bow-shock structure. In addition to the H\textsubscript{2} emission, there exists [Fe II] emission and high-velocity Br$\gamma$ emission from this outflow. Based on the orientation of the bow-shocks, \citet{Hodapp2002} suggested that the driving source of this outflow is in the general direction of the W51 IRS2 proto-cluster, with a red star (m\textsubscript{K}=15.4, no other broad-band detection) at 19\textsuperscript{h}23\textsuperscript{m}39.76\textsuperscript{s} +14\degree31\textsuperscript{'}20.83\textsuperscript{"} being a plausible candidate. However, there are two candidate colliding-wind binaries (d4e and d4w) at the same on-sky location as this outflow \citep{Ginsburg2016}. Their positions are shown as filled, magenta X's in Panel A of Figure \ref{fig:h2_emission_all}. In the same panel, we include a line showing a potential collinear alignment between these two stars, parts of Region 1, and Regions 2 through 4.

Region 17 is the largest, continuous instance of H\textsubscript{2} emission observed in the FOV, spanning $\sim0.18$pc in the horizontal plane. Both Region 16 and 17 (Panel D in Figure \ref{fig:h2_emission_all}) are located to the south of a massive B-star at 19\textsuperscript{h}23\textsuperscript{m}44.79\textsuperscript{s} +14\degree29\textsuperscript{'}11.1\textsuperscript{"} \citep[][their source 7]{Bik2019}; a possible candidate responsible for this emission. This star's location is shown by a filled, magenta X in Panel D of Figure \ref{fig:h2_emission_all}.

Region 21 (Panel F of Figure \ref{fig:h2_emission_all}) is located southeast of a star classified as a YSO based on CO bandhead emission in its \textit{K}-band spectra \citep[][their source 16]{Bik2019}. The morphology of this emission suggests that the YSO is the driving source because the less intense H\textsubscript{2} emission is ``pointing" at this star, possibly being dissociated by a high-velocity shock from the YSO. This star's location is shown by a filled, magenta X in Panel F of Figure \ref{fig:h2_emission_all}.



\begin{table}
\centering
\caption{H$_{2}$ Emission in W51A}\label{h2_regions}
\begin{tabular}{l l l}
Region Number & R.A. & Dec. \\
\hline
1\textsuperscript{(\textit{a})} & 19 23 39.66 & +14 31 29.79 \\
2 & 19 23 39.87 & +14 31 32.7 \\
3 & 19 23 40.17 & +14 31 34.65 \\
4 & 19 23 40.33 & +14 31 36.59 \\
5 & 19 23 40.11 & +14 31 30.76 \\
6\textsuperscript{(\textit{a})} & 19 23 34.93 & +14 30 27.65 \\
7 & 19 23 35.27 & +14 30 24.93 \\
8 & 19 23 35.24 & +14 30 11.34 \\
9\textsuperscript{(\textit{a})} & 19 23 36.6 & +14 30 13.09 \\
10\textsuperscript{(\textit{b})} & 19 23 36.82 & +14 30 22.02 \\
11\textsuperscript{(\textit{b})} & 19 23 36.64 & +14 30 26.87 \\
12\textsuperscript{(\textit{b})} & 19 23 38.58 & +14 30 52.51 \\
13\textsuperscript{(\textit{a})}\textsuperscript{(\textit{b})} & 19 23 38.32 & +14 30 46.68 \\
14 & 19 23 38.88 & +14 30 41.25 \\
15\textsuperscript{(\textit{c})} & 19 23 39.72 & +14 30 49.11 \\
16 & 19 23 45.01 & +14 29 2.02 \\
17 & 19 23 44.97 & +14 29 9.21 \\
18 & 19 23 48.04 & +14 29 29.01 \\
19\textsuperscript{(\textit{c})} & 19 23 47.46 & +14 29 16.97 \\
20 & 19 23 38.88 & +14 32 14.45 \\
21 & 19 23 38.72 & +14 31 59.31 \\ 
\hline 
\end{tabular}
\textbf{Notes.}\textsuperscript{(\textit{a})}H\textsubscript{2} emission previously documented by \citet{Hodapp2002}. \textsuperscript{(\textit{b})}Star-like object only detectable in \textit{H$_2$(1-0)} narrow-band filter.\textsuperscript{(\textit{c})}Emission near foreground source.
\end{table}

\subsection{Age Estimation}\label{age}

Common methods of measuring cluster ages include isochrone fitting on a CMD, or taking spectra of individual cluster members to derive a spectral type or an evolutionary stage. In the case of W51A, isochrone fitting is an unreliable method due to the differential extinction the region suffers. Spectroscopic studies of the region have focused on the distributed OB population or HII regions \citep{Bik2019,Barbosa2022}. While previous works have derived estimated ages for those sources, there have been no spectra taken of embedded cluster members, with the exception of IRS2E and IRS2W \citep{Figueredo2008,Barbosa2008}.

\begin{figure*}
\gridline{\fig{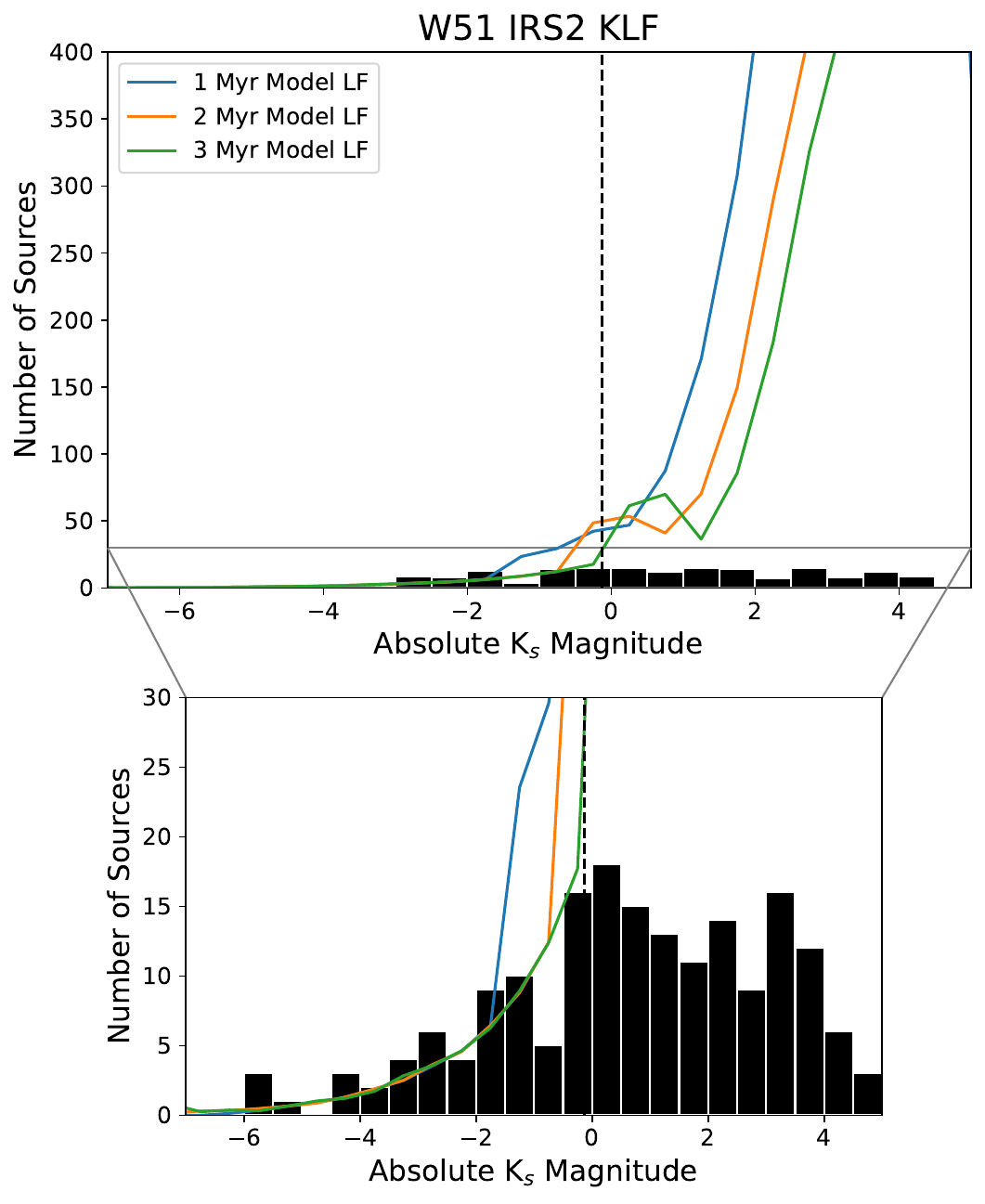}{0.4\textwidth}{(a)}
          \fig{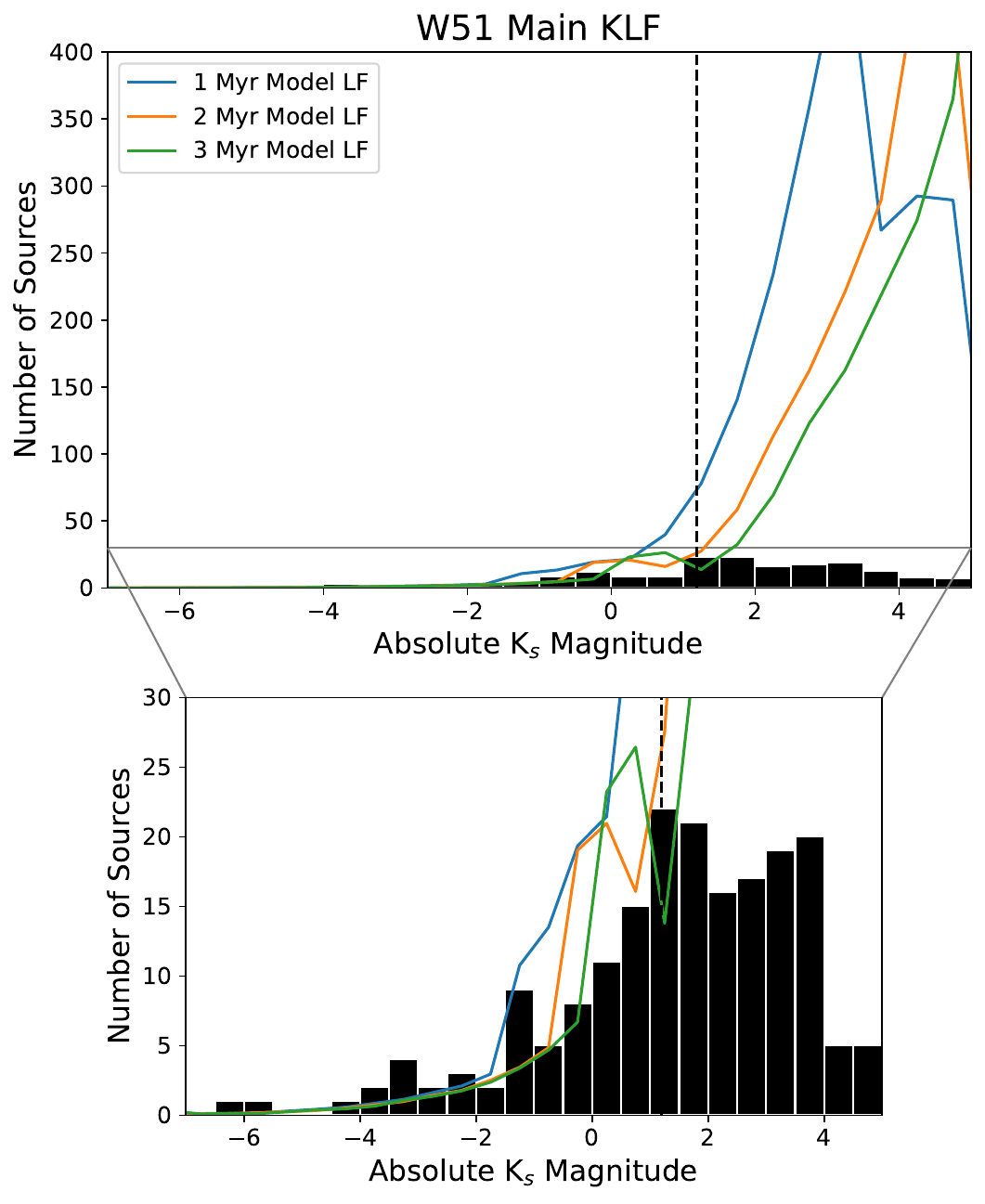}{0.4\textwidth}{(b)}}
\caption{Absolute KLFs for W51 IRS2 (a) and W51 Main (b) showing a possible sample population from our data (See Appendix \ref{sample_populations} for how these sample populations are generated). Included are the model KLFs of the assumed ages for the proto-clusters (1, 2, and 3 Myr). The dashed line is the 90\% completeness limit for each cluster.}
\label{fig:klf_both}
\end{figure*}

We attempted to measure the age of the proto-clusters by comparing the observed KLF to model KLFs of varying ages (0.5-100 Myr) using a two sample Kolmogorov-Smirnov test (K-S test), but this approach produced inconclusive results. This is because we cannot resolve enough of the stellar population in either proto-cluster to where their KLFs significantly differ in shape compared to the model KLFs. We include Figure \ref{fig:klf_both} to highlight how few sources we detect in the proto-clusters, assuming that the population is a simple, single-age stellar population. We provide a detailed description of our age estimation process, and the results of our K-S tests in Appendix \ref{age_estimation_app}. In brief, we are unable to provide age constraints for the proto-clusters using our photometric data alone. For the rest of the paper, we assume the age of the proto-clusters ranges from 1-3 Myr and allow for variation in that span to account for uncertainty in the cluster age. Also included in Appendix \ref{age_estimation_app} is our justification for why we think this age range is appropriate for these proto-clusters.

However, this method did return a plausible age range of 3-5 Myr for the candidate cluster (discussed in Section \ref{third_dense_region}), with all other tested ages in the 0.5-100 Myr range being rejected. The source LS2 is located near the center of this candidate cluster. \citet{Okumura2000} argues that, like LS1 (which is located north of our observed field), LS2 is a post-MS star in the W51 complex. Our age estimate, and the evolutionary stage estimate of LS2, suggest that this candidate cluster formed stars 3-5 Myr ago. The age estimate we provide for this cluster is slightly older than the distributed OB population of W51A ($<3$ Myr) \citep{Bik2019}, but is in agreement with the age estimate of LS1, which was estimated to be in the range of 3-6 Myr by \citet{Clark2009}.

\subsection{Stellar Masses}\label{Mass}
\subsubsection{Measuring the Masses}\label{measuring_mass}
One goal of our observations is to estimate the masses of stars within W51A and measure the IMF of W51 IRS2 and W51 Main. To accomplish this, we estimated the masses of the cluster members by using a mass-luminosity relation; the 1, 2, and 3 Myr isochrones of PARSEC-COLIBRI. We estimated the masses that correspond to each cluster member by interpolating the $M_{K_S}$ into the mass-luminosity relations. Sources with a range of possible $M_{K_S}$ values (See Section \ref{ext}) have their upper and lower limits interpolated into the mass-luminosity relation, yielding a corresponding range of possible masses.

These isochrones cover the range $-7.37 < M_{K_S} < 3.69$, but there are stars in the field with $M_{K_S} < -7.37$. These stars are unable to have their absolute magnitudes directly related to mass due to their brightness. For these sources, we assign them the maximum mass of the mass-luminosity relation (275$M_\odot$ using the 1 Myr isochrone, for example). These extremely bright stars were discussed in Section \ref{SoI}.

To convert from observed absolute magnitude to mass and propagate uncertainty, we took 1000 random samples from the measured $M_K$ assuming Gaussian random error $\sigma_{M_K}$ and converted them to mass by interpolating along the PARSEC-COLIBRI isochrones. Each mass estimate for the three assumed ages (1, 2, and 3 Myr) has its own uncertainty. We report the measured mass and uncertainty in Table \ref{tab:small_gtc_catalog}, where the uncertainty is the standard deviation of these 1000 samples.

To infer the total mass of the proto-clusters, we first calculated the integral of the completeness-corrected (CC) KLF above the chosen completeness limit. We vary the completeness limit from 70\% to 99\% to provide a range of possible total masses for the proto-clusters. Using the Kroupa IMF \citep{Kroupa2001} as a template, we calculated the average stellar mass above each limit and the corresponding mass fraction relative to the total cluster mass. The total mass was then derived by multiplying the average mass with the CC KLF integral and dividing by the mass fraction.


We estimated a total cluster mass range of $900-4700M_\odot$ for W51 IRS2, and a total cluster mass range of $500-2700M_\odot$ for W51 Main. These mass estimates were determined by only including stars with detections in more than one band when extrapolating the individual stellar masses to total cluster mass. The mass estimates change to $1100-4900M_\odot$ for W51 IRS2 and $400-2700M_\odot$ for W51 Main if we include stars that were only detected in the $K_{s}$-band. The mass estimate we present here is a lower limit on the proto-cluster masses. As mentioned previously, the NACO images of W51 IRS2 reveal a significant number of sources that we do not detect because they are confused with the extended emission from the HII region. There is also a cluster of X-ray sources found by \citet{Townsley2014} in both proto-clusters (discussed previously in Section \ref{xray}), and there are many sources from \citet{Ginsburg2016} that have yet to have their NIR counterparts identified. While some of these may be faint $K_{s}$-band sources, and therefore are accounted for by our population modeling, others may remain undetected despite being above our completeness limit because of confusion.

We also estimate a total mass of $330-1270M_\odot$ for the candidate cluster (discussed in Section \ref{third_dense_region}) assuming an age range of 3-5 Myr. The estimate we provide for the candidate cluster is a lower estimate as well. We find that this region has comparable completeness to the proto-clusters, so we are likely missing detections for many sources due to confusion from the higher stellar density.

\subsubsection{Comparison of the Total Mass to Previous Studies}
There have been two previous studies which have attempted to estimate the stellar mass in W51A using near-IR photometric data \citep{Okumura2000,Kumar2004}. \citet{Kumar2004} estimated the total stellar mass of the entire G49.5-0.4 region to be 9100$M_\odot$, assuming a distance of 6.5 kpc, and \citet{Okumura2000} estimated the total stellar mass of Region 3 (the region of W51A encompassing both proto-clusters and the surrounding area) to be 8200$M_\odot$, assuming a distance of 7 kpc. 

Both of these studies were conducted before the revised distance measurement to W51 was established using trigonometric parallax of methanol masers \citep{Xu2009} and H\textsubscript{2}O masers \citep{Sato2010}. To fairly compare the previous mass estimates to our own, we scaled down the previous mass estimates using a distance of 5.1 kpc \citep{Xu2009}. We did this by mimicking our methods in Section \ref{measuring_mass} using the authors measured magnitudes and extinction estimates, and then interpolating the $M_{K_{s}}$ along the 1 Myr mass-luminosity relationship. Doing so reduces both previous mass estimates by approximately 30\%, with the revised total mass from \citet{Kumar2004} being $\sim6400M_\odot$, and the revised total mass from \citet{Okumura2000} being $\sim5700M_\odot$.

The mass range we present here is consistent with the estimates from previous studies. If we sum both proto-cluster mass ranges ($900-4700M_\odot$ for W51 IRS2 and $400-2700M_\odot$ for W51 Main), we get a total cluster mass ($M_{tot}$) of $1300-7400M_\odot$. The similarity between our $M_{tot}$ and the (scaled-down) estimates from previous authors show how ground-based, non-adaptive optics observations are limiting the full analysis of this region. 

\subsection{The Initial Mass Function}
The high-mass end of the IMF for each proto-cluster can be described as a single power law of the form:
\[dN / d log M \propto M^{-\alpha}\]
where N is the number of stars, M is the stellar mass (in units of $M_\odot$), and $-\alpha$ is the IMF slope. 

We converted the proto-cluster KLFs to three mass distributions using the MLR for each assumed age (1, 2, and 3 Myr). We then re-sampled the proto-cluster mass distributions using bootstrapping, accounting for the mass uncertainties. The \texttt{powerlaw} Python package was used to measure the $\alpha$ (or slope) of the re-sampled mass distribution for each proto-cluster. This process was repeated 1000 times.

\begin{figure*}\centering
\gridline{\fig{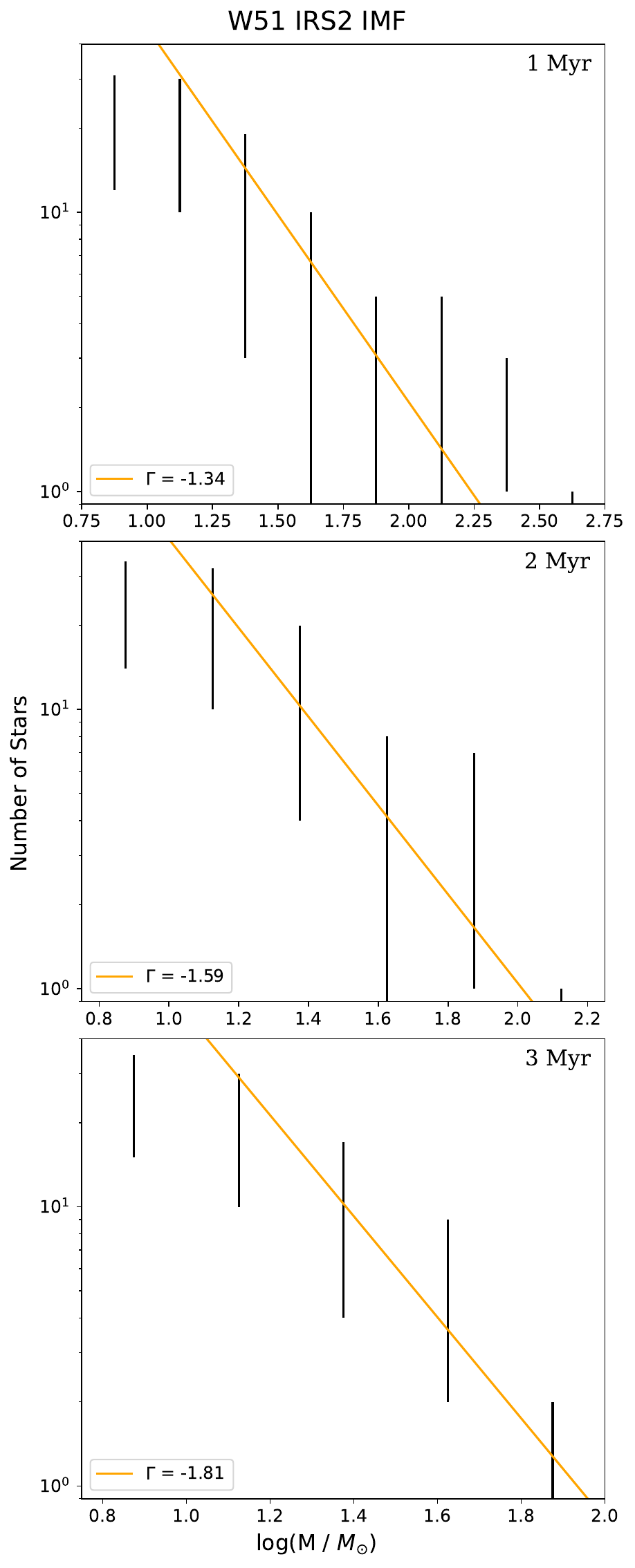}{0.45\textwidth}{(a)}
          \fig{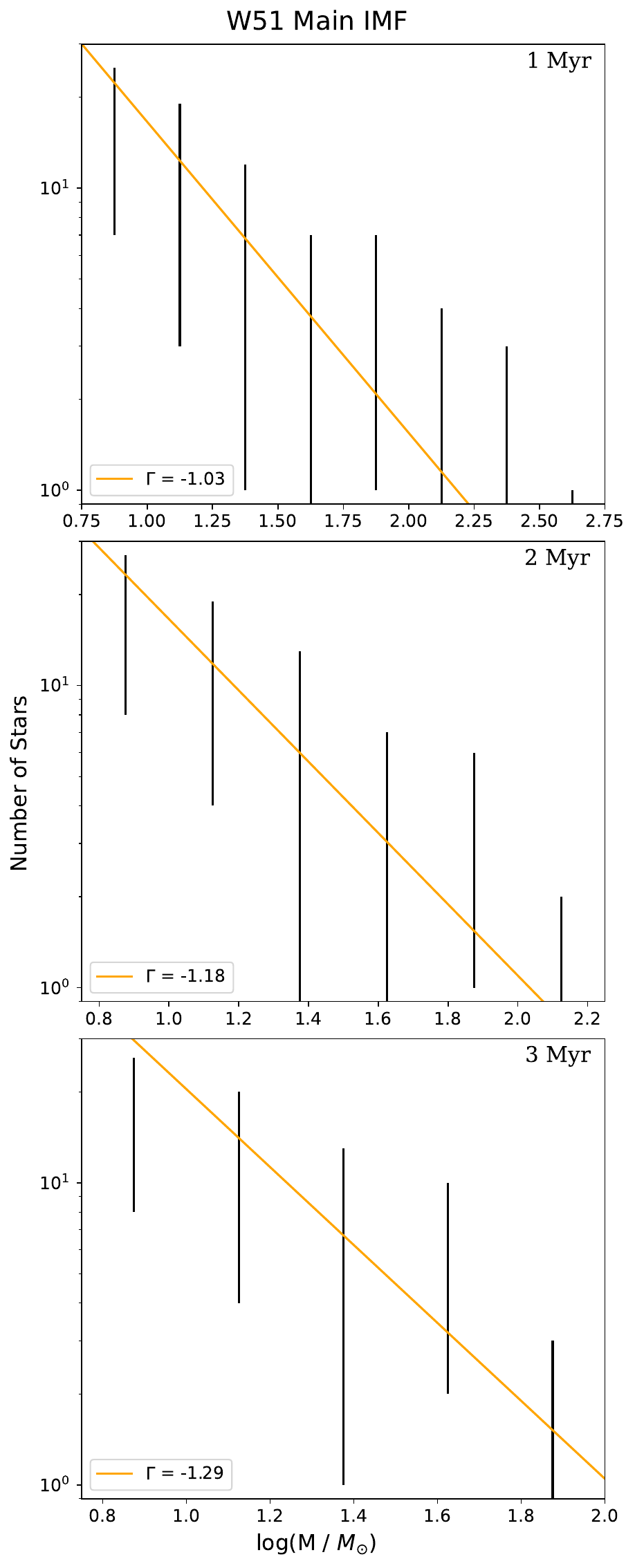}{0.45\textwidth}{(b)}}
\caption{High-mass IMFs for W51 IRS2 (a) and W51 Main (b). The panels, from top to bottom, show the IMFs assuming an age of 1, 2, and 3 Myr, respectively, for the proto-clusters. The orange line in each panel is the approximate slope of the IMF in that panel, and the value for each slope is included in the legend as well as in Table \ref{imf_slopes}.}
\label{fig:imf_both}
\end{figure*}

The reported IMF slope is the median $\Gamma$ ($\Gamma = -\alpha + 1$, the slope of the IMF in log-log space) over the 1000 re-samples, and the reported slope uncertainty is the standard deviation of the 1000 $\alpha$ values. Figure \ref{fig:imf_both} shows the IMFs for W51 IRS2 and W51 Main. The slopes for the IMFs are included in both the figure legend and in Table \ref{imf_slopes}. The IMF slopes we present here are limited to the high-mass end of the IMF ($M\gtrsim8M_\odot$) due to the limitations of our observations in these regions.


We find that all IMF slopes for both clusters at assumed ages of 1, 2, and 3 Myr are consistent with the \citet{Salpeter1955} IMF slope of $\Gamma = -1.35$ within $1\sigma$–$2\sigma$. The largest deviations occur for the W51 IRS2 slope at 3 Myr, which is $0.92\sigma$ steeper than Salpeter, and the W51 Main slope at 1 Myr, which is $1.39\sigma$ shallower.



The relative steepness of the W51 IRS2 IMFs are surprising, as high-mass clusters are the expected sites of the formation of the highest-mass stars.  Young massive clusters, with ages 1-5 Myr but without any remaining gas, exhibit shallower IMFs \citep{Hosek2019}.  There are several possibilities for this apparent discrepancy.  One is that the observations considered here have missed the top end of the IMF compared to those in other young clusters; as noted above, these observations are likely confusion-limited in the densest, brightest parts of the HII regions.  An alternative is that the mass function will grow shallower over time as the gas is evacuated, i.e., many of the massive stars are still growing to their final masses. 

Results from \citet{Okumura2000} show that the IMF slope of Region 3 (this is the region in their work that contains the proto-clusters) is fit well with $\Gamma = -1.8$ in the mass range of 10-30$M_\odot$. We attempted to define a region similar to Region 3 to compare to their results. We found that the IMF slope of our approximate Region 3 is -1.37 $\pm$ 0.17 only including stars with masses $>$10$M_\odot$. This IMF slope is much shallower the slope measured by \citet{Okumura2000} for this region, which can be explained by our observations being able to resolve many more O-type stars in our approximate Region 3 (31 compared to 20). We estimated the spectral type of these stars using their respective $M_{K_{s}}$ \citep{PecautMamajek2013}.

\begin{table}
\caption{IMF slopes for W51 IRS2 and W51 Main.}\label{imf_slopes}
\begin{tabular}{l l l}
Age (Myr) & $\Gamma_{IRS2}$ & $\Gamma_{Main}$  \\
\hline
1 & -1.34 $\pm$ 0.25 & -1.03 $\pm$ 0.23 \\
2 & -1.59 $\pm$ 0.36 & -1.18 $\pm$ 0.30 \\
3 & -1.81 $\pm$ 0.50 & -1.29 $\pm$ 0.34 \\
\end{tabular}

\textbf{Note.} Only sources with detections in two or more bands were included when determining the IMF slope.
\end{table}

\section{Conclusion}\label{conclusion}


We present NIR \textit{J}-, \textit{H}-, and \textit{K\textsubscript{s}}-band and narrow-band H$_2$(1-0) observations of W51A. We have provided photometric data to 3003 new sources discovered in the region, with 88 of these new sources being located in the embedded proto-clusters, W51 IRS2 and W51 Main. In W51 IRS2, the individual source $A_{V}$ ranges from 6-44 (with an average $A_{V}\sim19$), and in W51 Main it ranges from 5-53 (with an average $A_{V}\sim14$). 

By subtracting the \textit{K\textsubscript{s}}-band image from the \textit{H$_2$(1-0)} image, we document 17 new instances of isolated H\textsubscript{2} emission around W51A (21 in total). We also identify stars that are candidates for driving the observed H\textsubscript{2} outflows.

The stellar surface density map and the results of our k-means clustering analysis indicate a newly discovered candidate cluster near the source LS2. We found the age of the candidate cluster to be statistically consistent with 3-5 Myr based on the observed KLF of this region.

We were unable to estimate the ages of the proto-clusters only using photometric data due to our observations being confusion-limited in the central HII region. Instead, we assumed an age range between 1-3 Myr based on observations about the stellar population from previous authors and the physical structure of the HII region. This age range was used to provide three mass estimates for each cluster member using the absolute \textit{K\textsubscript{s}}-band magnitude via a mass-luminosity relationship (at 1, 2, and 3 Myr).

We estimated a total cluster mass ranging from  $900-4700M_\odot$ for W51 IRS2 and $500-2700M_\odot$ for W51 Main. Due to limitations in our observing, and the difficulty of resolving sources in the bright HII region, we were limited to measuring the high-mass end ($M\gtrsim8M_\odot$) of the IMF. The three IMF measurements for W51 IRS2 are $\Gamma_{1 Myr} = -1.34\pm0.25$, $\Gamma_{2 Myr} = -1.59\pm0.36$, $\Gamma_{3 Myr} = -1.81\pm0.50$, and the three IMF measurements for W51 Main are $\Gamma_{1 Myr} = -1.03\pm0.23$, $\Gamma_{2 Myr} = -1.18\pm0.30$, $\Gamma_{3 Myr} = -1.29\pm0.34$. We find that the IMF slopes for W51 IRS2 and W51 Main are consistent with the Salpeter IMF slope of $\Gamma = 1.35$ within $1\sigma-2\sigma$.


\begin{acknowledgments}
    
    A.G. acknowledges support from the NSF under AAG 2008101 and CAREER 2142300. This research made use of Astropy, a community-developed core Python package for Astronomy \citep{Astropy2022}, the PARSEC-COLIBRI evolutionary track web interface (\url{http://stev.oapd.inaf.it/cgi-bin/cmd_3.8}), NASA's Astrophysics Data System Bibliographic Services (ADS), Photutils, an Astropy package for detection and photometry of astronomical sources \citep{larry_bradley_2024_13989456}, and the SIMBAD database, operated at CDS, Strasbourg, France 
    \citep{Simbad}.
    
\end{acknowledgments}

\begin{longrotatetable}

\begin{deluxetable*}{c c c c c c c c c c}
\tabletypesize{\scriptsize}
\tablewidth{0pt} 
\tablecaption{Sample of sources from the GTC Catalog covering the proto-cluster region in W51A sorted by the estimated mass at 1 Myr.\label{tab:small_gtc_catalog}}
\tablehead{\colhead{R.A.} & \colhead{Dec.} & \colhead{$J$ (mag)} & \colhead{$H$ (mag)} & \colhead{$K_{s}$ (mag)} & \colhead{$A_{V}$} & \colhead{Mass ($M_{\odot}$, 1 Myr)} & \colhead{Mass ($M_{\odot}$, 2 Myr)} & \colhead{Mass ($M_{\odot}$, 3 Myr)} & \colhead{Notes}} 
\startdata 
19 23 40.1 & +14 31 5.7 & --- ± --- & 14.69 ± 0.005 & 9.44 ± 0.01 & 30.26-67.07 & 275.0-275.0 & 93.0-93.0 & 55.28-55.0 & B, D \\
19 23 42.78 & +14 30 27.66 & 21.18 ± 0.36 & 15.17 ± 0.01 & 12.43 ± 0.01 & 53.52 ± 3.38 & 200.64 ± 70.65 & 89.53 ± 10.28 & 54.59 ± 3.02 & C \\
19 23 42.86 & +14 30 27.67 & 18.04 ± 0.02 & 13.81 ± 0.004 & 11.12 ± 0.001 & 36.85 ± 0.2 & 124.53 ± 61.48 & 77.32 ± 11.88 & 51.73 ± 3.82 & C, E \\
19 23 39.93 & +14 31 8.62 & 15.48 ± 0.01 & 12.29 ± 0.001 & 10.11 ± 0.01 & 27.21 ± 0.05 & 115.79 ± 61.0 & 76.27 ± 11.95 & 51.22 ± 3.87 & C, D, F \\
19 23 40.18 & +14 31 7.78 & 17.78 ± 0.02 & 13.73 ± 0.01 & 11.71 ± 0.03 & 35.21 ± 0.17 & 92.99 ± 37.69 & 63.84 ± 12.59 & 46.65 ± 5.32 & C, D \\
19 23 42.89 & +14 30 29.58 & --- ± --- & 16.81 ± 0.01 & 13.11 ± 0.01 & 45.18-47.61 & 81.78-92.85 & 58.19-63.75 & 45.1-46.63 & B \\
19 23 41.53 & +14 30 49.56 & 19.83 ± 0.04 & 15.16 ± 0.01 & 12.7 ± 0.01 & 41.01 ± 0.36 & 79.35 ± 27.82 & 57.04 ± 12.21 & 44.54 ± 5.97 & C \\
19 23 40.95 & +14 31 10.73 & 21.26 ± 0.19 & 16.22 ± 0.01 & 13.44 ± 0.01 & 44.49 ± 1.75 & 66.54 ± 26.63 & 50.97 ± 12.56 & 41.09 ± 6.81 & C \\
19 23 40.04 & +14 31 7.74 & 19.04 ± 0.04 & 14.8 ± 0.003 & 13.25 ± 0.03 & 36.98 ± 0.33 & 47.72 ± 19.97 & 40.59 ± 11.62 & 33.89 ± 7.31 & C \\
19 23 40.2 & +14 31 6.34 & 14.81 ± 0.004 & 13.03 ± 0.04 & 11.17 ± 0.02 & 13.94 ± 0.37 & 36.88 ± 14.43 & 32.73 ± 9.74 & 28.93 ± 6.74 & C, D \\
19 23 40.25 & +14 31 7.8 & 16.93 ± 0.01 & 14.18 ± 0.02 & 12.62 ± 0.04 & 23.07 ± 0.19 & 29.41 ± 11.74 & 26.93 ± 8.71 & 24.74 ± 6.46 & C \\
19 23 43.67 & +14 30 48.81 & 20.16 ± 0.03 & 16.17 ± 0.03 & 14.08 ± 0.01 & 34.63 ± 0.35 & 26.96 ± 10.64 & 24.87 ± 8.18 & 23.04 ± 6.28 & C \\
19 23 40.13 & +14 31 2.52 & 17.77 ± 0.02 & 14.84 ± 0.01 & 13.14 ± 0.01 & 24.74 ± 0.2 & 24.7 ± 9.84 & 23.05 ± 7.83 & 21.51 ± 6.16 & C \\
19 23 40.51 & +14 31 5.56 & 19.81 ± 0.06 & 16.36 ± 0.06 & 14.0 ± 0.04 & 29.6 ± 0.78 & 21.26 ± 8.04 & 19.9 ± 6.69 & 18.74 ± 5.49 & C \\
19 23 40.32 & +14 30 59.8 & 20.34 ± 0.09 & 16.33 ± 0.01 & 14.61 ± 0.01 & 34.89 ± 0.83 & 21.11 ± 8.59 & 19.77 ± 7.12 & 18.62 ± 5.84 & C \\
19 23 40.53 & +14 30 56.01 & 20.77 ± 0.82 & 16.69 ± 0.02 & 14.7 ± 0.02 & 35.56 ± 7.71 & 20.91 ± 14.66 & 19.58 ± 11.13 & 18.45 ± 8.7 & C \\
19 23 40.92 & +14 31 6.71 & 18.53 ± 0.03 & 15.46 ± 0.01 & 13.68 ± 0.01 & 26.03 ± 0.25 & 20.39 ± 8.12 & 19.15 ± 6.73 & 18.07 ± 5.55 & C \\
19 23 40.53 & +14 30 53.43 & 16.11 ± 0.01 & 13.97 ± 0.01 & 12.72 ± 0.01 & 17.35 ± 0.15 & 20.28 ± 8.13 & 19.06 ± 6.78 & 17.99 ± 5.59 & C \\
19 23 39.83 & +14 31 5.48 & 18.16 ± 0.02 & 15.6 ± 0.03 & 13.2 ± 0.04 & 21.27 ± 0.31 & 19.76 ± 7.5 & 18.61 ± 6.38 & 17.63 ± 5.37 & C \\
19 23 42.35 & +14 30 48.62 & 19.01 ± 0.03 & 15.71 ± 0.01 & 13.98 ± 0.02 & 28.16 ± 0.34 & 19.71 ± 7.66 & 18.56 ± 6.47 & 17.59 ± 5.41 & C \\
19 23 40.69 & +14 30 59.99 & 20.72 ± 0.34 & 16.82 ± 0.01 & 14.63 ± 0.01 & 33.82 ± 3.14 & 19.54 ± 9.06 & 18.41 ± 7.48 & 17.47 ± 6.16 & C \\
19 23 42.55 & +14 30 47.54 & --- ± --- & 14.08 ± 0.03 & 12.98 ± 0.01 & 17.83-73.18 & 18.12-275.0 & 17.24-93.0 & 16.46-55.0 & B \\
19 23 43.07 & +14 30 36.6 & 20.52 ± 0.04 & 16.8 ± 0.01 & 14.6 ± 0.01 & 32.05 ± 0.36 & 17.78 ± 7.04 & 16.95 ± 6.04 & 16.2 ± 5.15 & C \\
19 23 40.86 & +14 31 4.09 & 21.09 ± 0.05 & 17.17 ± 0.02 & 14.87 ± 0.03 & 33.98 ± 0.48 & 17.36 ± 6.92 & 16.54 ± 5.98 & 15.86 ± 5.14 & C \\
19 23 40.85 & +14 31 3.39 & --- ± --- & 17.29 ± 0.02 & 14.88 ± 0.03 & 33.98-42.76 & 17.25-28.58 & 16.45-26.26 & 15.78-24.19 & B \\
19 23 39.89 & +14 31 7.41 & --- ± --- & 14.28 ± 0.04 & 14.41 ± 0.19 & 28.65-70.95 & 16.19-197.41 & 15.55-88.61 & 15.02-54.55 & B, D \\
19 23 40.26 & +14 31 4.39 & 17.86 ± 0.02 & 15.4 ± 0.03 & 13.64 ± 0.03 & 20.43 ± 0.32 & 15.13 ± 6.36 & 14.58 ± 5.54 & 14.08 ± 4.84 & C \\
19 23 42.44 & +14 30 44.86 & 21.34 ± 0.2 & 17.19 ± 0.01 & 15.44 ± 0.01 & 36.2 ± 1.87 & 14.87 ± 6.39 & 14.33 ± 5.56 & 13.82 ± 4.84 & C \\
19 23 39.93 & +14 31 4.01 & 20.26 ± 0.08 & 16.76 ± 0.02 & 14.81 ± 0.02 & 30.09 ± 0.8 & 14.51 ± 5.71 & 14.0 ± 5.06 & 13.47 ± 4.48 & C \\
19 23 40.11 & +14 31 1.23 & 19.38 ± 0.04 & 16.1 ± 0.01 & 14.62 ± 0.03 & 28.03 ± 0.42 & 14.21 ± 5.23 & 13.68 ± 4.7 & 13.2 ± 4.22 & C \\
19 23 40.35 & +14 31 11.77 & 20.21 ± 0.08 & 16.97 ± 0.02 & 14.6 ± 0.03 & 27.68 ± 0.77 & 14.06 ± 5.55 & 13.52 ± 4.95 & 13.07 ± 4.41 & C \\
19 23 43.19 & +14 30 50.05 & --- ± --- & 20.09 ± 1.62 & 12.48 ± 0.005 & 8.68-16.97 & 14.05-22.42 & 13.52-21.03 & 13.06-19.67 & B \\
19 23 40.49 & +14 31 4.39 & 19.58 ± 0.05 & 16.45 ± 0.03 & 14.54 ± 0.06 & 26.53 ± 0.56 & 13.62 ± 5.49 & 13.07 ± 4.9 & 12.68 ± 4.36 & C \\
19 23 40.27 & +14 31 4.88 & --- ± --- & --- ± --- & 13.39 ± 0.03 & 15.3-16.1 & 12.84-13.49 & 12.35-12.94 & 12.02-12.57 & B \\
19 23 42.96 & +14 30 55.34 & 20.81 ± 0.17 & 17.18 ± 0.01 & 15.22 ± 0.02 & 31.27 ± 1.62 & 12.51 ± 5.06 & 12.07 ± 4.57 & 11.79 ± 4.13 & C \\
19 23 40.58 & +14 31 11.77 & 19.53 ± 0.05 & 16.44 ± 0.02 & 14.71 ± 0.01 & 26.18 ± 0.49 & 12.12 ± 4.77 & 11.79 ± 4.31 & 11.54 ± 3.91 & C \\
19 23 40.37 & +14 30 59.58 & --- ± --- & 16.42 ± 0.01 & 14.89 ± 0.02 & 27.18-50.94 & 11.75-45.84 & 11.47-39.37 & 11.22-33.16 & B \\
19 23 42.42 & +14 30 40.73 & --- ± --- & 17.32 ± 0.02 & 13.31 ± 0.005 & 12.84-42.83 & 11.59-64.56 & 11.31-49.98 & 11.07-40.48 & B \\
19 23 40.48 & +14 30 58.23 & 19.78 ± 0.33 & 16.59 ± 0.01 & 14.97 ± 0.01 & 27.18 ± 3.12 & 11.35 ± 5.24 & 11.09 ± 4.71 & 10.84 ± 4.27 & C \\
19 23 39.74 & +14 31 1.05 & 20.81 ± 0.13 & 17.32 ± 0.03 & 15.29 ± 0.04 & 29.98 ± 1.24 & 11.32 ± 4.35 & 11.05 ± 3.96 & 10.8 ± 3.65 & C \\
19 23 41.76 & +14 30 35.83 & 19.23 ± 0.12 & 16.52 ± 0.01 & 14.48 ± 0.02 & 22.71 ± 1.1 & 11.28 ± 4.73 & 11.02 ± 4.31 & 10.77 ± 3.95 & C \\
19 23 40.83 & +14 30 56.79 & --- ± --- & 17.78 ± 0.02 & 15.73 ± 0.01 & 33.82-38.17 & 11.27-14.42 & 11.01-13.89 & 10.76-13.39 & B \\
19 23 43.24 & +14 31 2.09 & 21.66 ± 0.1 & 17.8 ± 0.01 & 15.73 ± 0.03 & 33.52 ± 0.91 & 11.07 ± 4.41 & 10.81 ± 4.04 & 10.55 ± 3.72 & C \\
19 23 40.32 & +14 31 7.03 & 17.64 ± 0.02 & 15.51 ± 0.05 & 13.95 ± 0.05 & 17.25 ± 0.52 & 10.89 ± 3.95 & 10.63 ± 3.63 & 10.38 ± 3.36 & C \\
19 23 43.14 & +14 30 36.92 & --- ± --- & 17.11 ± 0.01 & 15.69 ± 0.05 & 32.05-44.86 & 10.48-21.52 & 10.22-20.15 & 9.99-18.95 & B \\
19 23 39.8 & +14 31 13.2 & --- ± --- & 17.08 ± 0.03 & 14.39 ± 0.03 & 20.41-44.76 & 10.44-41.42 & 10.17-36.25 & 9.95-31.22 & B \\
19 23 41.97 & +14 30 36.91 & --- ± --- & 16.58 ± 0.01 & 13.19 ± 0.01 & 9.55-49.74 & 10.37-96.64 & 10.1-66.32 & 9.88-47.32 & B \\
19 23 40.98 & +14 31 3.34 & --- ± --- & 18.51 ± 0.04 & 15.71 ± 0.03 & 31.38-33.98 & 9.99-11.45 & 9.75-11.18 & 9.55-10.93 & B \\
19 23 40.7 & +14 31 9.85 & 20.0 ± 0.07 & 16.86 ± 0.03 & 15.19 ± 0.02 & 26.71 ± 0.69 & 9.95 ± 3.73 & 9.71 ± 3.44 & 9.52 ± 3.22 & C \\
19 23 39.7 & +14 31 0.29 & --- ± --- & 17.22 ± 0.03 & 14.73 ± 0.02 & 22.32-43.46 & 9.83-32.31 & 9.6-29.2 & 9.41-26.49 & B \\
\enddata
\tablecomments{A: Foreground source. B: Extinction estimated using surrounding sources (See Section \ref{ext}). C: Extinction measured using the NICE method. D: NACO $K_{s}$-band magnitude adopted. E: UKIDSS $K_{s}$-band magnitude adopted. F: UKIDSS $H$-band magnitude adopted. G: UKIDSS $J$-band magnitude adopted. H: 2MASS $H$- and $K_{s}$-band magnitudes adopted.}
\end{deluxetable*}
\end{longrotatetable}

\appendix
\renewcommand{\thefigure}{A\arabic{figure}}
\setcounter{figure}{0}

\renewcommand{\thetable}{A\arabic{table}}
\setcounter{table}{0}

\section{WCS Correction Using GAIA}\label{app_wcs}
Correcting the WCS information proceeded as follows. First, we did a by-eye comparison of the on-sky locations of the GAIA sources and the GTC sources; shifting the WCS to provide better initial positions for the cross-matching. We then cross-matched the GAIA and GTC catalogs and recorded the R.A. and Dec. of the GAIA sources with GTC matches. Included in this table are the on-sky separations between the matched sources given in arcseconds and the source ID of the GAIA source. A new WCS object was created using Astropy's \texttt{fit\_wcs\_from\_points} with the XY coordinates from the GTC detector and the sky coordinates from the GTC-matched GAIA sources. A second degree simple imaging polynomial (SIP) was used to correct for geometric distortion. The new WCS object was used to create new R.A. and Dec. coordinates for the GTC sources. The residuals between the newly created GTC sky coordinates and the GTC-matched GAIA sky coordinates were calculated, and if the residual for either the R.A. or Dec. were $<$ -1" or $>$ 1" the match would be considered bad. We found the good matches by inverting the ``bad match" mask. The final WCS object was created using \texttt{fit\_wcs\_from\_points} strictly with the XY GTC coordinates and the R.A. and Dec. from GAIA that were considered good matches. An abridged catalog of GTC sources with good GAIA matches that were used to create the final WCS object can be found in Table \ref{tab:gaia_cm_tab}. The original table consists of 2194 sources and the first 100 sources are shown in the table.

With the new WCS, we then calculated the astrometric accuracy of the observed field. We chose the UKIDSS GPS catalog for comparison. We need to compare pixel coordinates, so the R.A. and Dec. from the UKIDSS catalog is converted to detector coordinates using \texttt{all\_world2pix} and the newly determined WCS information. We down-selected to only sources with on-sky separations $<$ 0.5". The difference between the GTC and UKIDSS X and Y coordinates ($X_{GTC} - X_{UKIDSS}$; $Y_{GTC} - Y_{UKIDSS}$) was calculated and then squared.

\section{Aperture Photometry on the NACO images}
\label{naco_photometry}
In this appendix, we present the results of our aperture photometry routine which was performed on $K_{s}$-band, adaptive optics images centered on W51 IRS2 (ra = 290.917360 deg, dec = 14.51946 deg). These images were taken using the NAOS+CONICA instrument \citep{Lenzen2003,Rousset2003} at the ESO VLT UT4. We obtained the data from \citet{Figueredo2008}. To begin, we used the same PSF photometry routine that was used for the $K_{s}$-band GTC images for source detection with two differences: the detection threshold, and the ePSF. The ePSF was created using bright, isolated stars from the NACO image. Since the observations were carried out using adaptive optics, the ePSF was not representative of the varying PSFs found in the field. This caused many false detections to be returned from the photometry routine, so we placed regions around the real detections, and documented their XY positions.

\begin{figure}
    \centering
    \includegraphics[width=0.5\textwidth]{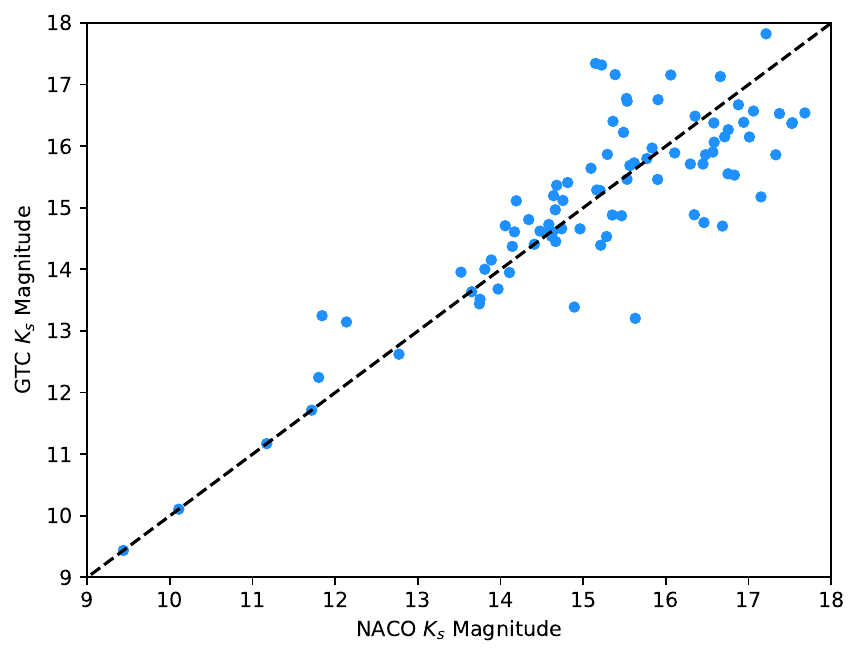}
    \caption{Plot showing the comparison between the calibrated GTC and NACO magnitudes of cross-matched sources. The dashed line shows the 1:1 magnitude correlation.}
    \label{fig:naco_gtc_comparison}
\end{figure}

We placed circular apertures (radius = 3 pixels) and annulus apertures (inner radius = 3.5 pix, outer radius = 4.5 pix) centered on the star’s XY position, and recovered the flux within. The mean per-pixel background (calculated using ApertureStats from photutils.aperture) was used to subtract off the background flux in the aperture, and we converted the background-subtracted flux to magnitude. We calibrated the recovered magnitudes using a calibration factor which was obtained by cross-matching the NACO catalog to the GTC catalog and taking the median difference between the recovered NACO magnitudes and the calibrated GTC magnitudes of matched sources. We include Figure \ref{fig:naco_gtc_comparison} to show how the calibrated magnitudes compare between the NACO sources and the GTC sources. Table \ref{naco_table} shows the catalog of sources we recovered from the $K_{s}$-band NACO image of W51 IRS2.

\section{Extinction Law Comparisons and Justification}\label{ext_law_comp}
\begin{figure*}
    \centering
    \includegraphics[width=1.0\linewidth]{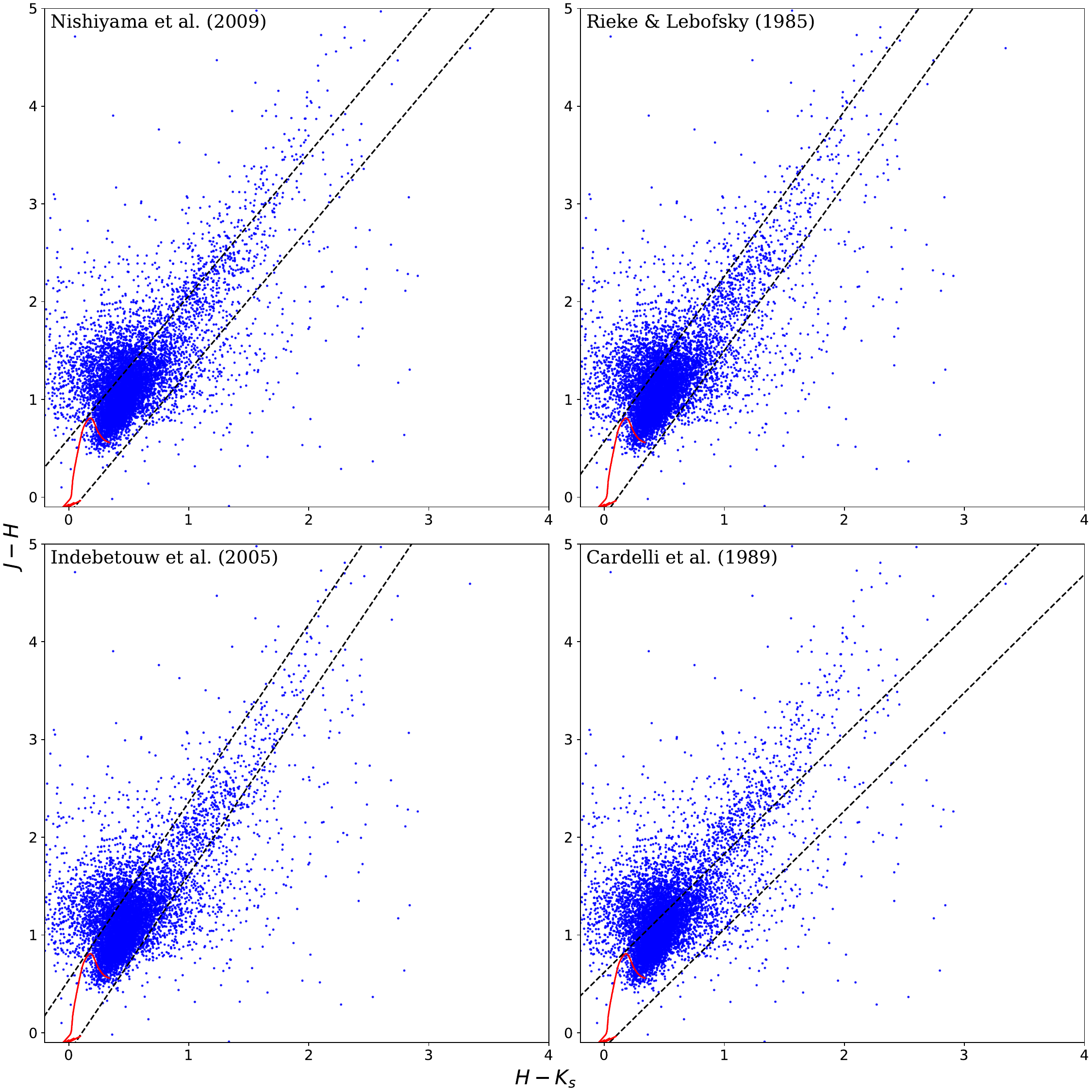}
    \caption{$H-K_{s}$ vs. $J-H$ CCDs including all sources in the observed field that were used for visual inspection of the 4 extinction laws. A 1 Myr isochrone is plotted in red, and the reddening zone of the isochrone is shown by the black, dashed lines. Each reddening zone has a different slope depending on the reddening slope of the extinction law used.}
    \label{extinction_law_comparison}
\end{figure*}

Presented here are the CCDs showing the 4 tested extinction laws as well as further justification as to why we chose the extinction law that we did. The first test was a visual comparison of each extinction law reddening slope and how it compared to the reddening slope we observed in the CCD. An extinction law was determined to be a good fit based on how the reddening zone around the plotted isochrone encapsulated the reddened stellar population. These plots are shown in Figure \ref{extinction_law_comparison}. The stars we trusted to make this determination are bounded by the inequalities: $J-H > 1.6 \times H-K_{s} + 0.9$ and $J-H < 1.6 \times H-K_{s} - 0.7$. The stars in these bounds do not exhibit filter-specific excess and would redden normally down the reddening vector.

We determined that the \citet{Cardelli1989} and \citet{Nishiyama2009} extinction laws were too shallow for our data. The extinction laws of \citet{RL85} and \citet{Indebetouw2005} both appear to be good fits, so more context was needed to pick between the two. We calculated the extinction of 9 sources, which had extinctions reported by \citet{Bik2019}, using the NICE method. We did this with all extinction laws listed prior and found that extinctions calculated with the extinction law of \citet{RL85} had the lowest average difference when compared to the reported extinctions.

\section{Generating Sample Populations}\label{sample_populations}
During the analysis (and for some figures, like Figure \ref{fig:comparison_between_new_old} (b) and Figure \ref{fig:klf_both}), we generate possible sample populations with the following method: For sources which have a range of possible values (for $A_{V}$, $M_{K_{s}}$, Mass) we draw from a uniform distribution over the given interval (a minimum value and a maximum value). If we measured the $A_{V}$ to a source (i.e. the source has a propagated, Gaussian uncertainty for all physical properties), we draw from a normal distribution with the mean being the quantity we measured and the $1\sigma$ being the associated error on that quantity. We draw 1000 samples for each source from their respective distributions and use element-wise combination to create a sample population.



\section{Age Estimation Process and Age Justification}\label{age_estimation_app}
We attempted to estimate the age of the proto-clusters by comparing the absolute \textit{K\textsubscript{s}}-band luminosity function (KLF) to model KLFs of varying ages. We used the PARSEC-COLIBRI isochrone web interface found here\footnote{http://stev.oapd.inaf.it/cgi-bin/cmd} to generate the model KLFs. The ages of the model KLFs range from 0.5-100 Myr and they are all generated with an underlying IMF from \citet{Kroupa2001}. Before the models could be used for comparison, they needed to be scaled appropriately to the observed KLF. We did this by comparing the integrals of the observed and model KLFs within the 90\% completeness limit of the respective proto-cluster, and scaling the models by a value that would equate the two integrals. 

We created synthetic stellar populations from the scaled model KLFs, and then conducted a two-sample Kolmogorov-Smirnov test (K-S test) to compare the distribution of the $M_{K_S}$ of the synthetic stars to the observed absolute KLF. A significance level of 0.1 was established prior to running the test. We conducted the K-S test 1000 times to account for variations within the randomly-generated, synthetic populations, and the test-statistic, D, and p-values were averaged over all iterations. The null hypothesis (H\textsubscript{o}) was rejected if the average p-value was less than the established significance level of 0.1. For the model KLFs where we fail to reject H\textsubscript{o}, we conclude that the age of the model KLF is a plausible age for the proto-cluster, based on the observed population. In Section \ref{age}, we mention that we do not observe enough of the stellar population in either proto-cluster to where their KLFs significantly differ in shape compared to the model KLFs. This results in a very wide age range not being rejected for W51 Main (7-15 Myr), and the age range of 3-6 Myr to not be rejected for W51 IRS2. The full results from our K-S tests are shown in Table \ref{kstest_results}.


The 90\% completeness limit was chosen because we wanted to include as much of the proto-cluster stellar population as possible before the $M_{K_{s}}$ distribution deviated too far from the models (due to an incomplete sample), which would result in poor fits for every model KLF. For example, H\textsubscript{o} for all model KLFs would be rejected for both proto-clusters if we use the 50\% completeness limit instead. When comparing how the scale factor changes with respect to the completeness limit, we find that the scale factors are similar for completeness fractions from 85-95\%. The measurements are Poisson-noise-dominated at completeness limits $>95\%$. For completeness limits $<85\%$, the number of stars per $M_{K_{s}}$ ($\frac{dN}{dM_{K_{s}}}$) is no longer increasing, while the integrals of the model KLFs rapidly increase (See Figure \ref{fig:klf_both} for a visual representation of this.). This causes the scale factor for these completeness limits to be too small to accurately represent the observed KLF.

\begin{table*}
\centering
\caption{Kolmogorov–Smirnov test results for both proto-clusters.\label{kstest_results}}
\begin{tabular}{c c c c c c c}
& \multicolumn{3}{c}{W51 IRS2} & \multicolumn{3}{c}{W51 Main} \\
\hline
Model KLF Age (Myr) & D & p-value & H\textsubscript{o} rejected? & D & p-value & H\textsubscript{o} rejected?\\
\hline
0.5 & 0.359 & 0.002 & Yes & 0.287 & 0.009 & Yes \\
1 & 0.407 & 0.0 & Yes & 0.236 & 0.043 & Yes \\
1.5 & 0.383 & 0.001 & Yes & 0.237 & 0.046 & Yes \\
2 & 0.378 & 0.002 & Yes & 0.251 & 0.031 & Yes \\
3 & 0.202 & 0.21 & No & 0.296 & 0.007 & Yes \\
4 & 0.234 & 0.111 & No & 0.256 & 0.028 & Yes \\
5 & 0.22 & 0.14 & No & 0.249 & 0.034 & Yes \\
6 & 0.227 & 0.124 & No & 0.229 & 0.062 & Yes \\
7 & 0.244 & 0.083 & Yes & 0.192 & 0.155 & No \\
8 & 0.241 & 0.088 & Yes & 0.176 & 0.213 & No \\
9 & 0.263 & 0.05 & Yes & 0.17 & 0.241 & No \\
10 & 0.287 & 0.027 & Yes & 0.183 & 0.184 & No \\
11 & 0.3 & 0.018 & Yes & 0.183 & 0.185 & No \\
12 & 0.299 & 0.02 & Yes & 0.202 & 0.119 & No \\
13 & 0.291 & 0.022 & Yes & 0.2 & 0.123 & No \\
14 & 0.312 & 0.013 & Yes & 0.203 & 0.116 & No \\
15 & 0.303 & 0.016 & Yes & 0.203 & 0.113 & No \\
20 & 0.306 & 0.014 & Yes & 0.228 & 0.058 & Yes \\
40 & 0.329 & 0.007 & Yes & 0.254 & 0.025 & Yes \\
60 & 0.362 & 0.002 & Yes & 0.269 & 0.016 & Yes \\
100 & 0.377 & 0.001 & Yes & 0.271 & 0.015 & Yes \\
\hline
\end{tabular}
\end{table*}

Here we provide our justification for why we believe that the age range of 1-3 Myr is an appropriate assumption for the age of the proto-clusters:
\begin{itemize}
    \item The physical structure of the W51A HII region is indicative of a very young SFR because of the abundance of dust and bright nebulosity, which are characteristics of an active star forming region. When comparing to other young stellar clusters, such as Westerlund 2 (Wd2) with an estimated age of $\sim$2 Myr \citep{Ascenso2007}, it can be seen that the core of Wd2 is devoid of dust and bright nebulosity. As star formation progresses, feedback from protostars and YSOs will expel the surrounding gas and dust causing the molecular cloud to slowly dissipate. 
    
    \item W51A hosts a significant population of X-ray emitting sources (discussed prior in Section \ref{xray}). Pre-main sequence (PMS) stars frequently emit X-rays due to their strong magnetic fields \citep{Feigelson1999}, or through accretion processes \citep{Calvet1998}. \citet{Nunez2021} show that if the hard X-ray luminosity (L\textsubscript{X,H}) of a star exceeds $10^{31.1}$ erg s$^{-1}$, it is likely an intermediate-mass PMS star, or a T-Tauri star, with an age $<$3 Myr. We used the X-ray catalog from \citet{Townsley2014} to calculate the L\textsubscript{X,H} of X-ray sources in W51A. There are 14 and 7 sources with log(L\textsubscript{X,H}) $>$ 31.1 in W51 IRS2 and W51 Main, respectively, and 20 additional sources within close proximity to the central HII region, indicating that these X-ray sources have ages $<$ 3 Myr.
    
    \item There are also several sources in the proto-clusters providing evidence for a very young age. W51 IRS2 hosts three sources which have had their spectral type/evolutionary stage estimated: IRS2E was spectroscopically classified as a massive YSO \citep{Figueredo2008,Barbosa2008}, \citet{Lim2019} classified a source they call IRS2/\#10 as a massive YSO using SOFIA photometry, and \citet{Ginsburg2016} classified the source W51d2 as a hyper-compact HII region. W51 Main hosts two sources that have had their spectral type/evolutionary stage estimated. The first star is an extremely massive (95$\begin{array}{c} +30 \\ -20 \end{array}$M$_\odot$ or 200$\begin{array}{c} +100 \\ -80 \end{array}$M$_\odot$ , depending on the extinction law used), young star ($<$ 2 Myr) which was spectroscopically classified as an O3V-O8V star by \citet{Bik2019}. The authors note that the reasoning behind this wide range in specrtal type is attributed to conflicting pieces of evidence in the star's spectra that point to either an O5V type star or earlier, or as late as an O8V type star. The second star is a source commonly referred to as IRS3 due to its extreme brightness in the \textit{K\textsubscript{s}}-band. This star was classified as a massive YSO using SOFIA photometry by \citet{Lim2019}, and a Class I source using the IRAC colors and the SED slope by \citet{Saral2017}. The location of these sources can be seen in Figure \ref{fig:new_stars_protocluster}. The existence of these massive YSOs in the system show that this is a site of active star formation with a very young age. 
\end{itemize}


\startlongtable
\begin{deluxetable*}{l l l l l l l}
\tabletypesize{\small}
\tablecaption{Catalog of cross-matched GAIA sources used to correct the WCS offset \label{tab:gaia_cm_tab}}
\tablehead{\colhead{R.A.} & \colhead{Dec.} & \colhead{$J$ (mag)} & \colhead{$H$ (mag)} & \colhead{$K_{s}$ (mag)} & \colhead{GAIA Offset (arcsec)} & \colhead{GAIA Source ID}} 
\startdata 
19 23 40.15 & +14 26 36.25 & --- & --- & 11.07 & 0.19 & 4319842513926127360 \\
19 23 36.92 & +14 26 36.03 & --- & --- & 15.51 & 0.46 & 4319842406550146304 \\
19 23 35.26 & +14 26 36.33 & --- & --- & 16.04 & 0.37 & 4319842475270283008 \\
19 23 37.09 & +14 26 36.67 & --- & --- & 13.94 & 0.15 & 4319842406563671040 \\
19 23 33.6 & +14 26 36.54 & --- & --- & 17.16 & 0.4 & 4319842445206690816 \\
19 23 41.74 & +14 26 38.82 & 14.73 & 14.28 & 13.94 & 0.04 & 4319842303470932224 \\
19 23 29.21 & +14 26 41.28 & 16.5 & 15.62 & 16.06 & 0.26 & 4319845395860723200 \\
19 23 29.21 & +14 26 41.3 & --- & --- & 16.05 & 0.29 & 4319845395860723200 \\
19 23 35.34 & +14 26 39.15 & 17.16 & 16.6 & 16.53 & 0.18 & 4319842479566418176 \\
19 23 34.72 & +14 26 39.61 & 15.52 & 14.98 & 14.91 & 0.3 & 4319842479579880320 \\
19 23 34.72 & +14 26 39.58 & --- & --- & 14.92 & 0.33 & 4319842479579880320 \\
19 23 28.35 & +14 26 39.59 & --- & 15.93 & 15.95 & 0.22 & 4319845395860701184 \\
19 23 38.3 & +14 26 42.55 & 17.04 & 16.14 & 15.81 & 0.12 & 4319842410846927616 \\
19 23 30.73 & +14 26 40.11 & 16.02 & 15.16 & 14.81 & 0.28 & 4319845395848044416 \\
19 23 43.25 & +14 26 41.24 & 14.51 & 13.41 & 12.78 & 0.01 & 4319842303471591424 \\
19 23 43.29 & +14 26 41.35 & --- & --- & 13.19 & 0.66 & 4319842303471591424 \\
19 23 29.94 & +14 26 40.97 & 17.38 & 16.6 & 16.36 & 0.29 & 4319845400146097024 \\
19 23 41.21 & +14 26 41.43 & 14.12 & 13.19 & 12.56 & 0.05 & 4319842509630022016 \\
19 23 37.46 & +14 26 42.12 & 15.99 & 15.46 & 15.05 & 0.13 & 4319842410846931968 \\
19 23 34.63 & +14 26 44.65 & --- & --- & 17.16 & 0.99 & 4319842475285318528 \\
19 23 34.67 & +14 26 45.66 & 16.37 & 15.55 & 15.31 & 0.17 & 4319842475285318528 \\
19 23 31.15 & +14 26 42.59 & 16.27 & 15.42 & 14.9 & 0.25 & 4319845395860776576 \\
19 23 37.81 & +14 26 44.31 & 16.59 & 15.73 & 15.09 & 0.15 & 4319842406563687040 \\
19 23 29.03 & +14 26 43.06 & 16.78 & 16.04 & 16.42 & 0.31 & 4319845395860719232 \\
19 23 37.59 & +14 26 45.08 & 16.36 & 15.47 & 14.91 & 0.12 & 4319842406563682176 \\
19 23 31.59 & +14 26 45.37 & 16.7 & 15.74 & 15.45 & 0.23 & 4319845395848047104 \\
19 23 35.57 & +14 26 47.48 & 17.14 & 16.44 & 16.26 & 0.16 & 4319842479567342848 \\
19 23 32.18 & +14 26 46.38 & 16.6 & 16.1 & 16.12 & 0.2 & 4319842440910548096 \\
19 23 35.88 & +14 26 49.09 & 16.71 & 15.97 & 15.49 & 0.16 & 4319842479566422400 \\
19 23 39.77 & +14 26 47.48 & 16.34 & 15.33 & 14.88 & 0.12 & 4319842509642938880 \\
19 23 27.51 & +14 26 49.72 & 15.21 & 14.65 & 14.65 & 0.22 & 4319845017890265344 \\
19 23 32.54 & +14 26 50.45 & 17.03 & 16.16 & 15.69 & 0.18 & 4319842445208108416 \\
19 23 31.56 & +14 26 50.95 & --- & --- & 16.48 & 1.04 & 4319845400144213632 \\
19 23 31.6 & +14 26 51.97 & 16.47 & 15.6 & 15.18 & 0.25 & 4319845400144213632 \\
19 23 34.83 & +14 26 54.53 & 13.69 & 12.56 & 12.1 & 0.17 & 4319842479568333696 \\
19 23 29.91 & +14 26 51.73 & 15.21 & 14.64 & 14.5 & 0.29 & 4319845400157652736 \\
19 23 29.81 & +14 26 54.06 & 15.53 & 14.94 & 14.68 & 0.28 & 4319845395860740992 \\
19 23 43.38 & +14 26 52.87 & 16.67 & 16.07 & 15.77 & 0.02 & 4319842337831335936 \\
19 23 43.41 & +14 26 52.93 & --- & --- & 15.95 & 0.53 & 4319842337831335936 \\
19 23 36.09 & +14 26 55.06 & 16.72 & 15.9 & 15.44 & 0.14 & 4319842475285319808 \\
19 23 31.15 & +14 26 53.08 & 16.39 & 15.59 & 15.07 & 0.22 & 4319845395860776704 \\
19 23 42.83 & +14 26 53.47 & 16.71 & 16.04 & 15.72 & 0.03 & 4319842303484566016 \\
19 23 39.56 & +14 26 53.57 & 15.22 & 14.6 & 14.17 & 0.29 & 4319842513926141824 \\
19 23 38.47 & +14 26 53.46 & 16.49 & 15.71 & 15.28 & 0.13 & 4319842509642914816 \\
19 23 38.99 & +14 26 53.65 & 15.86 & 15.29 & 14.87 & 0.13 & 4319842509630027904 \\
19 23 36.78 & +14 26 54.62 & 17.56 & 16.83 & 16.43 & 0.16 & 4319842479566419072 \\
19 23 40.33 & +14 26 57.19 & 15.91 & 15.0 & 14.48 & 0.07 & 4319842513926139392 \\
19 23 38.05 & +14 26 57.41 & 17.2 & 16.35 & 15.94 & 0.07 & 4319842513927572992 \\
19 23 37.97 & +14 26 58.89 & 14.17 & 13.1 & 12.45 & 0.13 & 4319842509630030208 \\
19 23 42.1 & +14 26 55.44 & 16.31 & 15.51 & 15.02 & 0.05 & 4319842509630028416 \\
19 23 35.59 & +14 26 56.49 & 12.94 & 11.93 & 11.35 & 0.19 & 4319842479579881856 \\
19 23 42.64 & +14 26 57.05 & 15.44 & 14.94 & 14.63 & 0.02 & 4319842509629366272 \\
19 23 37.11 & +14 26 57.37 & 17.46 & 16.52 & 16.26 & 0.12 & 4319842479566418560 \\
19 23 42.81 & +14 26 57.76 & 16.54 & 15.82 & 15.41 & 0.01 & 4319842544006830848 \\
19 23 42.85 & +14 26 57.84 & --- & --- & 15.62 & 0.56 & 4319842544006830848 \\
19 23 35.8 & +14 26 57.53 & 16.25 & 15.47 & 15.07 & 0.16 & 4319842479566428672 \\
19 23 34.74 & +14 26 58.54 & 15.88 & 15.07 & 14.77 & 0.27 & 4319842479567344512 \\
19 23 38.97 & +14 26 58.62 & 15.13 & 13.97 & 13.2 & 0.16 & 4319842509642924672 \\
19 23 33.13 & +14 26 58.59 & 16.26 & 15.56 & 15.23 & 0.18 & 4319842475270291072 \\
19 23 36.95 & +14 26 59.07 & 13.72 & 13.12 & 12.87 & 0.2 & 4319842479579879680 \\
19 23 40.61 & +14 27 0.02 & 15.58 & 14.97 & 14.61 & 0.08 & 4319842513939610368 \\
19 23 39.81 & +14 27 1.35 & 16.64 & 15.97 & 15.62 & 0.13 & 4319842513926144512 \\
19 23 43.98 & +14 27 1.48 & 15.62 & 14.62 & 14.36 & 0.04 & 4319842337831339520 \\
19 23 39.13 & +14 27 3.64 & 18.08 & 17.18 & 16.6 & 1.07 & 4319842513927570560 \\
19 23 39.16 & +14 27 4.61 & 17.32 & 16.45 & 16.03 & 0.06 & 4319842513927570560 \\
19 23 38.99 & +14 27 6.31 & 15.12 & 14.54 & 14.22 & 0.12 & 4319842509629368192 \\
19 23 42.08 & +14 27 3.33 & 16.44 & 15.83 & 15.48 & 0.05 & 4319842509630032512 \\
19 23 31.58 & +14 27 2.34 & 16.49 & 15.45 & 14.97 & 0.26 & 4319845395860788480 \\
19 23 38.53 & +14 27 2.57 & 16.43 & 15.75 & 15.3 & 0.12 & 4319842513939615232 \\
19 23 39.85 & +14 27 3.86 & 15.91 & 15.22 & 14.82 & 0.05 & 4319842513926146432 \\
19 23 43.41 & +14 27 4.52 & 16.95 & 16.2 & 15.71 & 0.03 & 4319842544002745088 \\
19 23 43.45 & +14 27 4.61 & --- & --- & 16.06 & 0.62 & 4319842544002745088 \\
19 23 34.74 & +14 27 5.15 & 16.35 & 15.66 & 15.37 & 0.04 & 4319842475283100544 \\
19 23 31.3 & +14 27 6.39 & 15.89 & 15.21 & 14.83 & 0.22 & 4319845400161135360 \\
19 23 31.86 & +14 27 13.86 & 15.92 & 15.09 & 14.66 & 0.22 & 4319845434517391872 \\
19 23 39.86 & +14 27 7.89 & 16.85 & 16.22 & 15.76 & 0.12 & 4319842513939613184 \\
19 23 36.74 & +14 27 8.19 & 16.95 & 16.26 & 15.94 & 0.12 & 4319842582646559744 \\
19 23 32.2 & +14 27 8.73 & 16.23 & 15.2 & 15.1 & 0.23 & 4319845430220542592 \\
19 23 41.38 & +14 27 6.69 & 16.29 & 15.61 & 15.11 & 0.06 & 4319842509630033024 \\
19 23 37.12 & +14 27 8.57 & 12.52 & 12.1 & 11.93 & 0.19 & 4319842582659096448 \\
19 23 37.13 & +14 27 10.64 & 16.29 & 15.58 & 15.36 & 0.13 & 4319842582659446528 \\
19 23 26.95 & +14 27 8.65 & 17.28 & 16.4 & 16.08 & 0.26 & 4319845468877138688 \\
19 23 40.31 & +14 27 8.92 & 16.26 & 15.56 & 15.18 & 0.07 & 4319842513926146560 \\
19 23 40.24 & +14 27 10.29 & 16.92 & 16.08 & 15.75 & 0.06 & 4319842513926147200 \\
19 23 35.88 & +14 27 10.97 & 18.06 & 17.35 & 17.17 & 1.09 & 4319842479567654272 \\
19 23 35.85 & +14 27 11.93 & 17.3 & 16.52 & 16.25 & 0.14 & 4319842479567654272 \\
19 23 35.62 & +14 27 9.63 & 16.25 & 15.55 & 15.19 & 0.15 & 4319842479579884928 \\
19 23 40.8 & +14 27 9.77 & 16.96 & 16.26 & 15.91 & 0.05 & 4319842513926144128 \\
19 23 37.54 & +14 27 9.94 & 16.63 & 15.85 & 15.52 & 0.16 & 4319842582645641088 \\
19 23 37.49 & +14 27 10.63 & 16.15 & 15.54 & 15.3 & 0.14 & 4319842578362371712 \\
19 23 33.67 & +14 27 9.89 & 16.17 & 15.52 & 15.32 & 0.21 & 4319842475270297472 \\
19 23 25.7 & +14 27 9.87 & 20.52 & 18.87 & 15.97 & 1.24 & 4319845086610413568 \\
19 23 25.79 & +14 27 10.76 & 16.74 & 16.02 & 16.33 & 0.26 & 4319845086610413568 \\
19 23 24.97 & +14 27 10.12 & 17.09 & 16.3 & 16.16 & 0.18 & 4319845086610412160 \\
19 23 39.05 & +14 27 10.66 & 15.49 & 14.91 & 14.56 & 0.11 & 4319842513939615872 \\
19 23 31.35 & +14 27 10.55 & 12.75 & 12.44 & 12.28 & 0.23 & 4319845434517392256 \\
19 23 33.23 & +14 27 10.81 & 15.52 & 14.81 & 14.41 & 0.16 & 4319845430207796608 \\
19 23 29.61 & +14 27 11.09 & 14.93 & 14.34 & 14.07 & 0.25 & 4319845395847253120 \\
19 23 38.32 & +14 27 11.94 & 17.06 & 16.25 & 15.9 & 0.12 & 4319842582645635840 \\
19 23 42.13 & +14 27 13.5 & 15.56 & 15.08 & 14.67 & 0.04 & 4319842543989775872 \\
\enddata
\end{deluxetable*}

\startlongtable
\begin{deluxetable*}{l l l l l}
\tabletypesize{\small}
\tablecaption{NACO $K_{s}$-band source catalog of W51 IRS2 \label{naco_table}}
\tablehead{\colhead{R.A.} & \colhead{Dec.} & \colhead{$K_{s}$ (mag)} & \colhead{GTC Offset (arcsec)} & \colhead{GTC Matched?}} 
\startdata 
19 23 40.22 & +14 30 55.31 & 13.75 ± 0.08 & 0.33 & Yes \\
19 23 40.19 & +14 30 55.87 & 17.28 ± 0.7 & 0.8 & No \\
19 23 39.9 & +14 30 55.83 & 17.37 ± 0.75 & 0.15 & Yes \\
19 23 40.31 & +14 30 59.82 & 14.17 ± 0.09 & 0.1 & Yes \\
19 23 40.3 & +14 30 58.48 & 15.2 ± 0.16 & 0.07 & Yes \\
19 23 40.38 & +14 30 59.65 & 16.34 ± 0.34 & 0.07 & Yes \\
19 23 40.37 & +14 30 59.47 & 15.44 ± 0.18 & 0.16 & No \\
19 23 40.47 & +14 30 58.22 & 14.66 ± 0.12 & 0.11 & Yes \\
19 23 40.52 & +14 30 55.97 & 16.68 ± 0.44 & 0.12 & Yes \\
19 23 40.5 & +14 30 58.51 & 18.5 ± 1.99 & 0.48 & No \\
19 23 40.46 & +14 30 58.7 & 17.62 ± 0.93 & 0.52 & No \\
19 23 40.34 & +14 31 0.69 & 16.83 ± 0.48 & 0.14 & Yes \\
19 23 40.54 & +14 31 0.6 & 15.68 ± 0.21 & 0.44 & No \\
19 23 40.64 & +14 30 59.59 & 17.33 ± 0.72 & 0.13 & Yes \\
19 23 39.71 & +14 30 58.88 & 17.06 ± 0.58 & 0.09 & Yes \\
19 23 39.72 & +14 30 59.55 & 17.34 ± 0.73 & 0.11 & Yes \\
19 23 39.6 & +14 31 0.31 & 17.09 ± 0.6 & 0.17 & Yes \\
19 23 39.69 & +14 31 0.33 & 14.58 ± 0.11 & 0.13 & Yes \\
19 23 39.73 & +14 31 1.12 & 15.16 ± 0.16 & 0.13 & Yes \\
19 23 39.72 & +14 31 0.86 & 18.18 ± 1.55 & 0.28 & No \\
19 23 39.68 & +14 31 0.78 & 17.66 ± 0.96 & 0.58 & No \\
19 23 39.85 & +14 31 0.77 & 17.68 ± 0.98 & 0.05 & Yes \\
19 23 40.06 & +14 31 0.79 & 16.43 ± 0.37 & 0.79 & No \\
19 23 39.94 & +14 31 2.03 & 16.58 ± 0.42 & 0.12 & Yes \\
19 23 40.09 & +14 31 0.98 & 16.76 ± 0.5 & 0.33 & No \\
19 23 40.1 & +14 31 1.3 & 14.48 ± 0.11 & 0.08 & Yes \\
19 23 40.21 & +14 31 1.21 & 16.35 ± 0.38 & 0.84 & No \\
19 23 40.14 & +14 31 1.82 & 14.47 ± 0.11 & 0.71 & No \\
19 23 40.26 & +14 31 1.79 & 15.28 ± 0.18 & 0.21 & Yes \\
19 23 40.36 & +14 31 1.94 & 16.06 ± 0.28 & 0.14 & Yes \\
19 23 40.36 & +14 31 2.86 & 15.52 ± 0.19 & 0.1 & Yes \\
19 23 40.26 & +14 31 2.21 & 16.78 ± 0.49 & 0.62 & No \\
19 23 40.12 & +14 31 2.56 & 12.13 ± 0.03 & 0.1 & Yes \\
19 23 40.04 & +14 31 3.24 & 15.15 ± 0.16 & 0.06 & Yes \\
19 23 40.01 & +14 31 3.26 & 16.78 ± 0.51 & 0.47 & No \\
19 23 39.92 & +14 31 4.08 & 14.34 ± 0.11 & 0.14 & Yes \\
19 23 40.33 & +14 31 3.77 & 14.81 ± 0.13 & 0.07 & Yes \\
19 23 40.25 & +14 31 4.41 & 13.65 ± 0.07 & 0.11 & Yes \\
19 23 40.13 & +14 31 3.64 & 17.41 ± 0.8 & 1.12 & No \\
19 23 40.06 & +14 31 4.31 & 16.09 ± 0.31 & 1.11 & No \\
19 23 40.14 & +14 31 4.35 & 16.69 ± 0.53 & 1.46 & No \\
19 23 40.19 & +14 31 4.6 & 14.8 ± 0.14 & 0.98 & No \\
19 23 40.32 & +14 31 4.32 & 17.59 ± 1.09 & 0.58 & No \\
19 23 40.31 & +14 31 4.39 & 16.24 ± 0.35 & 0.67 & No \\
19 23 40.35 & +14 31 5.25 & 14.67 ± 0.13 & 0.04 & Yes \\
19 23 40.28 & +14 31 4.95 & 14.7 ± 0.15 & 0.12 & No \\
19 23 40.27 & +14 31 4.95 & 14.89 ± 0.18 & 0.07 & Yes \\
19 23 40.11 & +14 31 5.22 & 15.69 ± 0.35 & 0.53 & No \\
19 23 40.07 & +14 31 4.72 & 16.99 ± 0.65 & 1.08 & No \\
19 23 40.02 & +14 31 5.1 & 16.77 ± 0.58 & 1.32 & No \\
19 23 39.6 & +14 31 4.3 & 16.29 ± 0.34 & 3.51 & No \\
19 23 39.66 & +14 31 4.14 & 17.0 ± 0.57 & 2.81 & No \\
19 23 39.77 & +14 31 3.78 & 16.97 ± 0.56 & 1.91 & No \\
19 23 40.69 & +14 30 59.99 & 14.6 ± 0.11 & 0.09 & Yes \\
19 23 40.56 & +14 31 0.84 & 14.96 ± 0.14 & 0.07 & Yes \\
19 23 40.57 & +14 31 0.98 & 16.68 ± 0.48 & 0.2 & No \\
19 23 40.55 & +14 31 1.76 & 17.18 ± 0.64 & 0.97 & No \\
19 23 40.82 & +14 31 1.41 & 16.66 ± 0.4 & 0.06 & Yes \\
19 23 40.97 & +14 31 3.32 & 16.45 ± 0.36 & 0.07 & Yes \\
19 23 40.84 & +14 31 3.41 & 15.35 ± 0.18 & 0.09 & Yes \\
19 23 40.79 & +14 31 4.13 & 16.48 ± 0.37 & 0.11 & Yes \\
19 23 40.83 & +14 31 3.95 & 16.4 ± 0.34 & 0.38 & No \\
19 23 40.86 & +14 31 4.17 & 15.46 ± 0.2 & 0.08 & Yes \\
19 23 40.97 & +14 31 6.32 & 16.44 ± 0.35 & 0.81 & No \\
19 23 40.92 & +14 31 6.75 & 13.97 ± 0.09 & 0.05 & Yes \\
19 23 40.49 & +14 31 3.11 & 15.9 ± 0.24 & 0.17 & Yes \\
19 23 40.51 & +14 31 3.97 & 15.49 ± 0.19 & 0.47 & No \\
19 23 40.48 & +14 31 4.44 & 14.62 ± 0.12 & 0.15 & Yes \\
19 23 40.51 & +14 31 5.63 & 13.81 ± 0.08 & 0.07 & Yes \\
19 23 40.47 & +14 31 5.4 & 15.76 ± 0.25 & 0.64 & No \\
19 23 40.45 & +14 31 5.49 & 16.03 ± 0.3 & 0.87 & No \\
19 23 40.49 & +14 31 6.67 & 15.39 ± 0.18 & 0.1 & Yes \\
19 23 40.44 & +14 31 6.1 & 16.77 ± 0.52 & 0.8 & No \\
19 23 40.39 & +14 31 6.09 & 15.29 ± 0.18 & 0.21 & Yes \\
19 23 40.3 & +14 31 5.47 & 16.8 ± 0.6 & 0.72 & No \\
19 23 40.25 & +14 31 5.52 & 16.29 ± 0.47 & 0.69 & No \\
19 23 40.23 & +14 31 5.44 & 16.59 ± 0.69 & 0.79 & No \\
19 23 40.21 & +14 31 5.67 & 12.42 ± 0.04 & 0.68 & No \\
19 23 40.11 & +14 31 5.9 & 9.44 ± 0.01 & 0.31 & Yes \\
19 23 40.14 & +14 31 5.78 & 15.32 ± 0.47 & 0.68 & No \\
19 23 40.2 & +14 31 6.36 & 11.17 ± 0.02 & 0.07 & Yes \\
19 23 40.3 & +14 31 5.92 & 17.1 ± 0.79 & 0.99 & No \\
19 23 40.22 & +14 31 6.06 & 16.84 ± 1.2 & 0.33 & No \\
19 23 40.25 & +14 31 6.18 & 17.14 ± 1.11 & 0.76 & No \\
19 23 40.31 & +14 31 6.59 & 17.56 ± 1.22 & 0.46 & No \\
19 23 40.32 & +14 31 6.82 & 17.18 ± 0.92 & 0.23 & No \\
19 23 40.31 & +14 31 7.04 & 14.11 ± 0.1 & 0.12 & Yes \\
19 23 40.35 & +14 31 6.32 & 17.18 ± 0.73 & 0.65 & No \\
19 23 40.41 & +14 31 7.01 & 16.81 ± 0.51 & 0.63 & No \\
19 23 40.46 & +14 31 7.11 & 17.12 ± 0.63 & 0.76 & No \\
19 23 40.39 & +14 31 7.71 & 15.56 ± 0.22 & 0.2 & Yes \\
19 23 40.45 & +14 31 7.98 & 17.61 ± 1.0 & 0.71 & No \\
19 23 40.47 & +14 31 8.13 & 16.75 ± 0.49 & 0.91 & Yes \\
19 23 40.49 & +14 31 7.65 & 16.7 ± 0.48 & 1.06 & No \\
19 23 40.79 & +14 31 7.11 & 16.75 ± 0.46 & 0.11 & Yes \\
19 23 40.78 & +14 31 7.58 & 16.71 ± 0.46 & 0.2 & Yes \\
19 23 40.67 & +14 31 7.84 & 15.09 ± 0.15 & 0.1 & Yes \\
19 23 40.61 & +14 31 8.39 & 16.29 ± 0.32 & 0.28 & Yes \\
19 23 40.73 & +14 31 8.55 & 15.53 ± 0.19 & 0.18 & Yes \\
19 23 40.83 & +14 31 8.37 & 17.01 ± 0.55 & 0.15 & Yes \\
19 23 40.88 & +14 31 8.74 & 15.77 ± 0.22 & 0.09 & Yes \\
19 23 40.95 & +14 31 10.77 & 13.74 ± 0.08 & 0.05 & Yes \\
19 23 40.77 & +14 31 9.15 & 16.35 ± 0.34 & 0.11 & Yes \\
19 23 40.7 & +14 31 9.82 & 14.64 ± 0.12 & 0.04 & Yes \\
19 23 40.69 & +14 31 10.06 & 15.74 ± 0.23 & 0.25 & No \\
19 23 40.76 & +14 31 10.01 & 17.05 ± 0.56 & 0.91 & No \\
19 23 40.62 & +14 31 10.17 & 16.11 ± 0.31 & 0.04 & Yes \\
19 23 40.61 & +14 31 10.15 & 16.81 ± 0.57 & 0.06 & No \\
19 23 40.66 & +14 31 10.32 & 17.18 ± 0.64 & 0.65 & No \\
19 23 40.73 & +14 31 11.6 & 16.85 ± 0.47 & 0.96 & No \\
19 23 40.66 & +14 31 11.84 & 14.68 ± 0.12 & 0.08 & Yes \\
19 23 40.61 & +14 31 10.69 & 15.75 ± 0.22 & 0.52 & No \\
19 23 40.58 & +14 31 11.83 & 14.06 ± 0.09 & 0.08 & Yes \\
19 23 40.5 & +14 31 10.46 & 15.36 ± 0.17 & 0.05 & Yes \\
19 23 40.44 & +14 31 11.77 & 16.76 ± 0.43 & 1.19 & No \\
19 23 40.37 & +14 31 12.0 & 15.67 ± 0.22 & 0.37 & No \\
19 23 40.35 & +14 31 11.76 & 14.63 ± 0.12 & 0.1 & Yes \\
19 23 40.35 & +14 31 11.2 & 16.58 ± 0.43 & 0.57 & No \\
19 23 40.37 & +14 31 9.16 & 16.22 ± 0.34 & 0.21 & No \\
19 23 40.35 & +14 31 9.04 & 15.9 ± 0.27 & 0.02 & Yes \\
19 23 40.29 & +14 31 10.29 & 16.09 ± 0.32 & 0.2 & No \\
19 23 40.3 & +14 31 11.36 & 17.09 ± 0.63 & 0.94 & No \\
19 23 40.3 & +14 31 8.74 & 16.57 ± 0.52 & 0.88 & No \\
19 23 40.29 & +14 31 9.96 & 17.15 ± 0.78 & 0.15 & Yes \\
19 23 40.27 & +14 31 9.82 & 15.5 ± 0.22 & 0.41 & No \\
19 23 40.27 & +14 31 7.3 & 14.78 ± 0.15 & 0.53 & No \\
19 23 40.25 & +14 31 8.14 & 14.03 ± 0.1 & 0.35 & No \\
19 23 40.25 & +14 31 7.84 & 12.77 ± 0.05 & 0.11 & Yes \\
19 23 40.24 & +14 31 10.19 & 17.07 ± 0.72 & 0.76 & No \\
19 23 40.23 & +14 31 9.63 & 15.3 ± 0.2 & 1.03 & No \\
19 23 40.2 & +14 31 8.09 & 14.95 ± 0.2 & 0.47 & No \\
19 23 40.17 & +14 31 7.83 & 11.71 ± 0.03 & 0.13 & Yes \\
19 23 40.18 & +14 31 7.5 & 16.63 ± 1.1 & 0.28 & No \\
19 23 40.16 & +14 31 7.7 & 17.36 ± 3.24 & 0.23 & No \\
19 23 40.18 & +14 31 8.01 & 14.5 ± 0.21 & 0.23 & No \\
19 23 40.16 & +14 31 7.99 & 13.29 ± 0.09 & 0.3 & No \\
19 23 40.2 & +14 31 9.33 & 15.63 ± 0.26 & 1.17 & No \\
19 23 40.16 & +14 31 8.73 & 15.27 ± 0.23 & 0.84 & No \\
19 23 40.14 & +14 31 7.42 & 14.32 ± 0.13 & 0.69 & No \\
19 23 40.18 & +14 31 11.36 & 16.48 ± 0.4 & 0.99 & No \\
19 23 40.16 & +14 31 10.77 & 16.8 ± 0.53 & 0.63 & No \\
19 23 40.12 & +14 31 11.03 & 16.36 ± 0.39 & 0.24 & No \\
19 23 40.13 & +14 31 10.77 & 16.88 ± 0.58 & 0.16 & Yes \\
19 23 40.12 & +14 31 10.51 & 17.11 ± 0.75 & 0.29 & No \\
19 23 40.19 & +14 31 9.8 & 17.11 ± 0.8 & 1.13 & No \\
19 23 40.12 & +14 31 9.59 & 15.88 ± 0.33 & 0.29 & No \\
19 23 40.1 & +14 31 9.24 & 16.1 ± 0.41 & 0.21 & Yes \\
19 23 40.14 & +14 31 9.33 & 17.02 ± 0.9 & 0.35 & No \\
19 23 40.19 & +14 31 8.84 & 17.21 ± 1.03 & 1.07 & No \\
19 23 40.1 & +14 31 8.31 & 15.15 ± 0.21 & 1.03 & No \\
19 23 40.11 & +14 31 8.07 & 16.03 ± 0.42 & 1.03 & No \\
19 23 40.09 & +14 31 7.71 & 14.84 ± 0.17 & 0.84 & No \\
19 23 40.06 & +14 31 7.88 & 15.94 ± 0.43 & 0.3 & No \\
19 23 40.07 & +14 31 7.65 & 14.57 ± 0.15 & 0.45 & No \\
19 23 40.09 & +14 31 7.54 & 16.88 ± 1.0 & 0.86 & No \\
19 23 40.09 & +14 31 6.83 & 16.1 ± 0.52 & 1.14 & No \\
19 23 40.24 & +14 31 9.36 & 17.38 ± 1.06 & 1.01 & No \\
19 23 40.87 & +14 31 12.7 & 15.83 ± 0.23 & 0.06 & Yes \\
19 23 40.7 & +14 31 14.17 & 17.21 ± 0.61 & 0.09 & Yes \\
19 23 40.17 & +14 31 12.51 & 14.19 ± 0.09 & 0.11 & Yes \\
19 23 40.62 & +14 31 15.77 & 17.52 ± 0.81 & 0.35 & Yes \\
19 23 40.6 & +14 31 16.31 & 17.53 ± 0.81 & 0.14 & Yes \\
19 23 40.76 & +14 31 16.29 & 13.52 ± 0.07 & 0.07 & Yes \\
19 23 40.53 & +14 31 18.04 & 14.75 ± 0.12 & 0.08 & Yes \\
19 23 40.42 & +14 31 19.02 & 15.53 ± 0.2 & 0.07 & Yes \\
19 23 40.31 & +14 31 18.53 & 16.58 ± 0.41 & 0.12 & Yes \\
19 23 40.25 & +14 31 13.32 & 17.47 ± 0.78 & 1.33 & No \\
19 23 40.21 & +14 31 12.82 & 17.44 ± 0.79 & 0.59 & No \\
19 23 40.01 & +14 31 5.84 & 15.67 ± 0.27 & 1.27 & No \\
19 23 39.96 & +14 31 5.89 & 16.26 ± 0.4 & 1.79 & No \\
19 23 39.83 & +14 31 5.64 & 15.63 ± 0.32 & 0.16 & Yes \\
19 23 39.82 & +14 31 5.77 & 16.59 ± 0.75 & 0.32 & No \\
19 23 39.81 & +14 31 5.58 & 15.85 ± 0.38 & 0.28 & No \\
19 23 39.87 & +14 31 7.46 & 15.37 ± 0.4 & 0.37 & No \\
19 23 39.88 & +14 31 7.5 & 14.41 ± 0.19 & 0.19 & Yes \\
19 23 39.97 & +14 31 7.8 & 16.08 ± 0.46 & 0.96 & No \\
19 23 39.92 & +14 31 8.57 & 10.11 ± 0.01 & 0.14 & Yes \\
19 23 40.03 & +14 31 7.85 & 11.84 ± 0.03 & 0.17 & Yes \\
19 23 39.98 & +14 31 8.91 & 16.38 ± 0.53 & 0.78 & No \\
19 23 40.04 & +14 31 8.45 & 18.16 ± 2.73 & 0.71 & No \\
19 23 40.03 & +14 31 8.32 & 16.55 ± 0.65 & 0.58 & No \\
19 23 39.74 & +14 31 7.24 & 11.8 ± 0.03 & 0.12 & Yes \\
19 23 39.65 & +14 31 9.6 & 15.7 ± 0.23 & 2.82 & No \\
19 23 39.83 & +14 31 9.88 & 15.22 ± 0.18 & 0.13 & Yes \\
19 23 39.85 & +14 31 9.88 & 16.64 ± 0.56 & 0.15 & No \\
19 23 40.0 & +14 31 12.88 & 16.46 ± 0.41 & 0.27 & Yes \\
19 23 39.9 & +14 31 11.37 & 16.7 ± 0.49 & 1.62 & No \\
19 23 39.79 & +14 31 13.34 & 15.21 ± 0.18 & 0.24 & Yes \\
19 23 39.71 & +14 31 14.09 & 14.74 ± 0.13 & 0.11 & Yes \\
19 23 39.54 & +14 31 14.11 & 15.49 ± 0.2 & 0.14 & Yes \\
19 23 39.45 & +14 31 14.51 & 13.89 ± 0.08 & 0.11 & Yes \\
19 23 39.76 & +14 31 20.93 & 15.61 ± 0.21 & 0.11 & Yes \\
19 23 39.58 & +14 31 17.35 & 16.94 ± 0.52 & 0.02 & Yes \\
19 23 39.19 & +14 31 15.55 & 16.57 ± 0.41 & 0.18 & Yes \\
19 23 39.44 & +14 31 9.05 & 15.99 ± 0.28 & 4.48 & No \\
19 23 39.53 & +14 31 7.92 & 16.98 ± 0.61 & 3.13 & No \\
19 23 39.49 & +14 31 6.72 & 16.81 ± 0.54 & 3.68 & No \\
19 23 39.14 & +14 31 8.16 & 14.14 ± 0.1 & 0.13 & Yes \\
\enddata
\end{deluxetable*}

\bibliography{bib}{}
\bibliographystyle{aasjournal}

\end{document}